\documentclass[american,a4paper]{report}

\usepackage{amssymb}
\usepackage{amsthm}

\usepackage{epsf}
\usepackage{graphicx}

\usepackage[dvips,usenames]{color}
\usepackage[american]{babel}
\usepackage[latin1]{inputenc}
\usepackage[T1]{fontenc}
\usepackage{ae}
\usepackage{aecompl}
\newtheorem{theorem}{Theorem}[chapter]
\newtheorem{lemma}[theorem]{Lemma}
\newtheorem{corollary}[theorem]{Corollary}
\newtheorem{prop}[theorem]{Proposition}

\theoremstyle{definition}
\newtheorem{example}[theorem]{Example}
\newtheorem{defn}[theorem]{Definition}
\newtheorem{claim}[theorem]{Claim}
\newtheorem{note}[theorem]{Remark}
\newtheorem{red}[theorem]{Reduction}

\newtheorem{algorithm}[theorem]{Algorithm}

\newenvironment{postulate}[1]{\vspace{3mm}{\bf Postulate #1: }}

\newcommand{\comment}[1]{} 
\newcommand{\suppress}[1]{}

\newcommand{\complex}{{\mathbb C}}
\newcommand{\reals}{{\mathbb R}}
\newcommand{\N}{{\mathbb N}}

\newcommand{\tensor}{\otimes}

\newcommand{\ket}[1]{|#1\rangle}
\newcommand{\bra}[1]{\langle #1|}
\newcommand{\braket}[2]{\langle #1 | #2\rangle}    

\newcommand{\identity}{1 \hspace{-2.5pt} \mathrm{l}}
\newcommand{\hilbert}{{\cal H}}         

\newcommand{\set}[1]{{\left\{#1\right\}}}    
\newcommand{\size}[1]{\left|#1\right|}             

\newcommand{\trace}{{\rm tr}}
\newcommand{\transpose}{{\mathrm t}}
\newcommand{\weight}{{\rm wt}}

\newcommand{\floor}[1]{\left\lfloor#1\right\rfloor}
\newcommand{\ceil}[1]{\left\lceil#1\right\rceil}

\newcommand{\res}[2]{\mathrm{res}_{#1} \left( #2 \right)}
\newcommand{\Diff}{{\rm Diff}}
\newcommand{\F}{\mathbb{F}}
\newcommand{\places}{\mathbb{P}}
\newcommand{\divgroup}{{\cal D}_F}
\newcommand{\adelespace}{{\cal{A}}_F}
\newcommand{\ff}{{\rm F}}
\newcommand{\K}{{K}}
\newcommand{\chara}{{\rm char}}
\newcommand{\codeL}{{\cal L}}
\newcommand{\mod}{{\rm mod}}
\newcommand{\diag}{{\rm diag}}
\newcommand{\Quot}{{\rm Quot}}
\newcommand{\Gal}{{\rm Gal}}
\newcommand{\supp}{{\rm supp}}
\newcommand{\code}{{\cal C}}
\newcommand{\Cperp}{C^{\perp}}
\newcommand{\Cperps}{C^{\perp_s}}

\newcommand{\G}{{\cal G}}
\newcommand{\qecc}[3]{\left[ \left[ #1,\, #2,\, #3 \right] \right]}
\newcommand{\ecc}[3]{\left[ #1,\, #2,\, #3 \right]}
\newcommand{\pauli}{{\cal P}}
\newcommand{\error}{{\cal E}}
\newcommand{\GL}{{\rm GL}}

\newcommand{\matrixip}[2]{\langle #1,#2 \rangle}
\newcommand{\ip}[2]{\langle #1,#2 \rangle}
\newcommand{\ipa}[2]{\langle #1,#2 \rangle^a}
\newcommand{\ipb}[2]{\langle #1,#2 \rangle^b}
\newcommand{\sip}[2]{\langle #1,#2 \rangle_s}
\newcommand{\sipa}[2]{\langle #1,#2 \rangle_s^a}
\newcommand{\sipb}[2]{\langle #1,#2 \rangle_s^b}
\newcommand{\sipp}[2]{\langle #1,#2 \rangle_s^p}

\begin{document}

\pagenumbering{arabic}

\begin{titlepage}
  \centering
  
  \vspace{70mm}

  	 {\Huge \bf Quantum Goppa Codes\\ \vspace{1mm}
	    over\\ \vspace{5mm}
	    Hyperelliptic Curves}
	 
	 \vspace{40mm}
	 
		{\huge Diplomarbeit}
		
		\vspace{60mm}

		    {\Large

		  {\bf Annika Niehage}\\ \par

		  \vspace{25mm}
		  November 2004\\
			 }
		  \vspace{15mm}
		  {\large
		  Institute for Quantum Computing, University of Waterloo
		  
		  - Prof. Dr. Raymond Laflamme, Dr. Martin R\"otteler -
		  \vspace{1mm}

		  Institut f\"ur Mathematik, Universit\"at Mannheim
		  
		  - Prof. Dr. Wolfgang K. Seiler -
		  }	       
			       
\end{titlepage}

\cleardoublepage
\pagenumbering{roman}

\begin{abstract}

  This thesis provides an explicit construction of a quantum Goppa code for
  any hyperelliptic curve over a non-binary field. Hyperelliptic curves have
  conjugate pairs of rational places. We use these pairs to construct
  self-orthogonal classical Goppa codes with respect to a weighted inner
  product. These codes are also self-orthogonal with respect to a symplectic
  inner product and therefore define quantum stabilizer codes. A final
  transformation leads to a quantum Goppa code with respect to the standard
  symplectic inner product. Some examples illustrate the described
  construction.

  Furthermore we present a projection of a higher dimensional code onto the
  base field and a special case when the projected code is again weighted
  self-orthogonal and symmetric.

\end{abstract}

\tableofcontents
\cleardoublepage

\pagenumbering{arabic}
\parskip2mm

\chapter*{Introduction}


This diploma thesis uses algebraic geometry, coding theory, and quantum error
correction to construct some classes of new quantum Goppa codes. In particular
we show that every hyperelliptic curve which has at least one pair of rational
places can be used to construct a code. The presented constructions lead to
explicit descriptions of quantum error correcting codes. We illustrate all
constructions by concrete computations which have been carried out with the
help of the computer algebra system Magma \cite{magma}. In order to be able to
prove properties of these quantum Goppa codes, some basic theory has to be
introduced. This thesis does not assume the reader to be familiar with all
three different topics and gives brief introductions. It is structured in the
following way:

In Chapter 1 the reader gets a short introduction to classical coding
theory. The classical Hamming code will serve as an example in which we
explain what it means to encode, correct errors, and finally decode codewords.

The basics of quantum mechanics are introduced in Chapter 2. This chapter
includes the ideas of quantum information and quantum errors. We will see some
general quantum error correcting codes before looking at stabilizer codes, a
special kind of quantum codes that can use classical codes for quantum
computing. Finally, this chapter shows a way how to transform stabilizer codes
in order to make them orthogonal with respect to various symplectic inner
products.

Chapter 3 is an introduction to algebraic curves, function fields, divisors,
and differentials. The main theorems are the theorem of Riemann-Roch and the
strong approximation theorem. It also introduces towers of function fields in
order to be able to study the asymptotics of codes. Finally, Goppa codes are
defined and we give special properties concerning self-orthogonality.

In Chapter 4, a construction by R. Matsumoto for asymptotically good quantum
codes over fields of characteristic two will be presented in detail. His ideas
will play an important part in the constructions of the subsequent chapter.

Chapter 5 is the main part of this thesis and gives some new constructions of
quantum Goppa codes. We will use hyperelliptic function fields to construct
weighted self-orthogonal Goppa codes that can be transformed to symplectic
self-orthogonal quantum codes. One method to construct these codes is to use
the well known CSS construction. The other method will use a direct
construction similar to the one of Chapter 4. Some examples shall help to
understand the ideas better.

Finally, Chapter 6 concludes the thesis.


\section*{Table of Notations}

This section gives an overview over the used symbols and notations.

\paragraph{Notation for Codes}$ $\\
\begin{tabular}{ll}
  &\\
  $C$ & linear code\\
  $d(x,y)$ & Hamming distance\\
  $\weight(x)$ & weight of $x$\\
  $\dim C$ & dimension of code $C$\\
  $\ecc{n}{k}{d}$ & classical linear code with parameters $n,\, k, \, d$\\
  $d(C)$ & minimum distance of $C$\\
  $\ip{x}{y}$ & canonical inner product on $\F_q^n$\\
  $\ipa{x}{y}$ & inner product with weights $a_i$ on $\F_q^n$\\
  $\G$ & generator matrix of a code $C$\\
  $H$ & parity check matrix of a code $C$\\
  $s(x)$ & syndrome of $x$ corresponding to a code $C$\\
  $C^{\perp}$ & dual code of $C$\\
  $C^{\perp^a}$ & dual code of $C$ with respect to $\ipa{\;}{\;}$\\
  $C_{\cal L}(D,G)$ & geometric Goppa code associated with $D$ and $G$\\
  $ev_D$ & evaluation map\\
  $C_{\Omega}(D,G)$ & Goppa code over differentials associated with $D$ and
  $G$\\
  $\qecc{n}{k}{d}$ & quantum stabilizer code with parameters $n,\, k, \, d$\\
  $\sip{x}{y}$ & symplectic inner product on $\mathbb{F}_q^{2n}$\\
  $\sipa{x}{y}$ & symplectic inner product with weights $a_i$ on $\F_q^{2n}$\\
  $C^{\perp_s}$ & symplectic dual code of $C$\\
  $C^{\perp_s^a}$ & symplectic dual code of $C$ with respect to
  $\sipa{\;}{\;}$\\

\end{tabular}

\paragraph{Quantum Information Language}$ $\\
\begin{tabular}{ll}
  &\\
  $\ket{\psi}$ & qudit in ket notation\\
  $\ket{b_1\ldots b_n}$ & quantum register in ket notation\\
  $\pauli$ & Pauli group on single qudits\\
  $\pauli_n$ & Pauli group on quantum registers of length $n$\\
  $X_j$, $Z_j$ & generators of the Pauli group\\
  $\identity$ & identity operator\\
  $H$ & Hadamard gate\\
  $C-NOT$ & controlled-not gate\\
  $C-U$ & controlled unitary operation $U$\\
  Toffoli & Toffoli gate\\
  $N(S)$ & Normalizer of the stabilizer $S$\\
 
\end{tabular}

\paragraph{Used Algebraic Geometry Terminology}
$ $\\
\begin{tabular}{ll}
  &\\
  $\F_q$ & finite field with $q$ elements\\
  $\F_q^n$ & $n$-dimensional vector space over $\mathbb{F}_q$\\
  $\mathbb{P}^2$ & projective space of dimension $2$\\
  $[F^{\prime}:F]$ & degree of a field extension $F^{\prime}/F$\\
  $char \, K$ & characteristic of $K$\\
  $K[T]$ & polynomial ring in one variable over $K$\\
  $K(x)$ & rational function field\\
  $(f(X,Y,Z))$ & ideal spanned by $f(X,Y,Z)$\\
  $F/K$ & algebraic function field of one variable\\
  ${\cal O}$ & valuation ring of $F/K$\\
  $P$ & place of $F/K$\\
  $\mathbb{P}_F$ & set of places of $F/K$\\
  ${\cal O}_P$ & valuation ring of a place $P$\\
  $v_P$ & discrete valuation corresponding to $P$\\
  $F_P$ & residue class field of $P$\\
  $x(P)$ & residue class of an element $x \in {\cal O}$\\
  $\deg P$ & degree of a place $P$\\
  $P_{\infty}$ & infinite place of $K(x)$\\
  $\divgroup$ & divisor group of $F/K$\\
  $\supp \, D$ & support of the divisor $D$\\
  $D_1 \le D_2$ & ordering of divisors\\
  $\deg D$ & degree of a divisor $D$\\
  $(x)$ & principal divisor of $x$\\
  $(x)_{\infty}$ & pole divisor of $x$\\
  ${\cal L}(A)$ & space of functions associated with the divisor $A$\\
  $\dim A$ & dimension of a divisor $A$\\
  $g$ & genus of $F/K$\\
  ${\cal A}_F$ & adele space of $F/K$\\
  ${\cal A}_F(A)$ & space of adeles associated with the divisor $A$\\
  $\Omega_F$ & module of Weil differentials of $F/K$\\
  $\Omega_F(A)$ & module of Weil differentials associated with the divisor
  $A$\\
  $(\eta)$ & divisor of a Weil differential $\eta \ne 0$\\
  $dx,\; u \, dx$ & differentials of $F/K$\\
  $\res{P}{\eta}$ & residue of a differential $\eta$ at a place $P$\\
  $F^{\sigma}$ & fixed field of $\sigma$\\
  $e(P^{\prime}|P)$ & ramification index of $P^{\prime}$ over $P$\\
  $f(P^{\prime}|P)$ & relative degree of $P^{\prime}$ over $P$\\  
  $d(P^{\prime}|P)$ & different exponent of $P^{\prime}$ over $P$\\
  $d(P^{\prime})$ & different exponent of $P^{\prime}$ if $d(P^{\prime}|P)$ is
  independent of $P$\\
  $\Diff(F^{\prime}|F)$ & different of $F^{\prime}/F$\\
  $\Gal(F^{\prime}|F)$ & Galois group of $F^{\prime}/F$\\    
  \end{tabular}

\vspace{10mm}

\section*{Acknowledgements}

\vspace{5mm}

There are a lot of people who supported me in this work and who have to be
thanked.

First, I thank Martin R\"otteler for the weekly discussions that helped so
much and gave so many ideas. Then I have to thank Raymond Laflamme who gave me
the opportunity to work at the Institute for Quantum Computing (IQC) at the
University of Waterloo. He, Michele Mosca and the whole institute helped me a
lot to integrate in Waterloo and to get an idea of what research means. I
thank my office mates, especially Pranab, Niel, and Donny for the mathematical
discussions, but also all the others especially for their mental support and
motivating words.

Also thanks to Wolfgang K. Seiler and Hans-Peter Butzmann at the Universit\"at
Mannheim that they accepted my thesis written in Waterloo and helped me to
arrange all formalities in Germany. Thanks to Claus Hertling who proof-read
especially the mathematical part of this thesis and gave me helpful hints. And
last but not least thanks to Wolfgang Effelsberg who sent me on the exchange
program Mannheim - Waterloo and therefore made me meet the people from IQC.

\chapter{Classical Coding Theory}
\label{cec}
Classical codes are used for error correction. Transmission of information
over a noisy channel will always cause bit flip errors with a certain
probability. When designing a code we try to maximise the number of
correctable errors in a codeword and minimise the number of bits we have to
send. Interesting math problems arise from the question how to construct
families of codes that reach special bounds in their asymptotics. An example
for this is the tower of quantum codes constructed in Chapter
\ref{qechyper}. In general, there are lots of different ideas, how to
construct good codes, we will mostly look at codes coming from algebraic
geometry.

This chapter gives a short overview over the basic ideas of classical linear
codes to introduce the notation used later on. For more information about
classical coding theory we refer to \cite{macwil}.

\section{Linear Codes}

Codes do not have to be linear, but if they are linear, many things become
easier. Hence we will only use linear codes over finite fields $\F_q$ where
$q = p^m$ is a prime power.

\subsection{Basic Definitions}

\begin{defn}
  A {\bf linear code {\it C}} (over the alphabet $\F_q$) is a linear subspace
  of $\F_q^n$; the elements of $C$ are called {\bf codewords}. We call $n$ the
  {\bf length} of $C$ and $\dim C$ (as $\F_q$-vector space) the {\bf
  dimension} of $C$. 
\end{defn}

\begin{defn}
  For $ a = (a_1, \ldots , a_n)$ and $ b = (b_1, \ldots , b_n) \in
  \mathbb{F}_q^n$ let
  \begin{displaymath}
    d(a,b) := \size{\set{i \; | \; a_i \neq b_i}}.
  \end{displaymath}
  This function $d$ is called the {\bf Hamming distance} on
  $\F_q^n$ and defines a metric. The {\bf weight} of an element $ a \in
  \F_q^n$ is defined as
  \begin{displaymath}
    \weight(a) := d(a,0) = \size{\set{ i \; | \; a_i \neq 0}}
  \end{displaymath}
\end{defn}

\begin{defn}
\label{mindistance}
 The {\bf minimum distance} $d(C)$ of a linear code $C \neq \set{0}$ is
 defined as
 \begin{eqnarray*}
   d(C) &:= & \min \set{d(a,b) \; | \; a,b \in C \;\rm{and}\; a \neq b } = \min
   \set{\weight(c) \; | \; 0 \neq c \in C}.
 \end{eqnarray*}
 The second equality holds, because the code is linear and therefore for $a,b
 \in C$
 \begin{eqnarray*}
   d(a,b) & = & d(a-b,0) = \weight(a-b)
 \end{eqnarray*}
 and $a-b \in C$ because of linearity.
\end{defn}

\begin{defn}
  An {\bf $\ecc{n}{k}{d}$ code} is a code of length $n$ and dimension $k$ with
  minimum distance $d$ (see Definition \ref{mindistance}) detecting $d-1$ and
  correcting $t = \floor{\frac{d-1}{2}}$ errors.
\end{defn}

\subsection{Encoding}

\begin{defn}
 Let $C$ be an
$[n,k]$ code over $\F_q$. A {\bf generator matrix} $\G$ of $C$ is a
$ k \times n$ matrix whose rows form a basis of $C$, i.e. if
 $\set{c_1,\ldots,c_k}$ is a basis of $C$, then
 \begin{eqnarray*}
   \G & = & \left(
   \begin{array}{c}
     c_1\\ \vdots \\ c_k
   \end{array}
   \right)
   \in \F_q^{k \times n}.
 \end{eqnarray*}
\end{defn}

\begin{defn}
 The {\bf canonical inner product} on $\F_q^n$ is defined by
 \begin{displaymath}
   \ip{a}{b} := \sum_{i=1}^n a_i b_i
 \end{displaymath}
 for $ a = ( a_1, \ldots , a_n)$ and $ b = ( b_1, \ldots , b_n) \in
 \F_q^n$. This is a {\it symmetric bilinear form} on $\F_q^n$.
\end{defn}

\begin{defn}
  If $ C \subseteq \F_q^n$ is a code, then
  \begin{eqnarray*}
    \Cperp & := & \set{u \in \F_q^n \; | \; \ip{u}{c} = 0 \; \rm{for \; all}
    \; c \in C}
  \end{eqnarray*}
  is called the {\bf dual} of $C$. A code $C$ is called {\bf self-dual}
  (respectively. {\bf self-orthogonal}) if $C = \Cperp$ (respectively. $C \subseteq \Cperp$).
\end{defn}

\begin{defn}
  A generator matrix $H$ of $\Cperp$ is said to be a {\bf parity check matrix}
  for $C$.

  Clearly, a parity check matrix of an $\ecc{n}{k}{d}$ code $C$ is an $(n-k)
  \times n$ matrix $H$ of rank $n-k$, and we have
  \begin{eqnarray*}
    C & = & \set{u \in \F_q^n \; | \; H \cdot u^\transpose = 0}
  \end{eqnarray*}
\end{defn}

\subsection{Decoding}

We can decode linear codes by so-called {\it syndromes}. What syndromes are
and how to decode with their help will be explained in this section.

\begin{defn}
  The {\bf syndrome} $s(u)$ of a vector $u \in \F_q^n$ with respect to
  a code $C$ is defined by
  \begin{eqnarray*}
    s(u) & = & u \cdot H^\transpose
  \end{eqnarray*}
  where $H^\transpose$ is the transpose of the parity check matrix of $C$.
\end{defn}

By definition of the parity check matrix, we have $ \forall c \in C:
s(c) = 0$, so that we can characterize $C$ as $C = \{u \in \F_q^n \; | \;
s(u) = 0 \}$.

\begin{note}
  \label{mostlikelyerror}
  We can divide $\F_q^n$ into cosets according to the syndromes:
  \begin{eqnarray*}
    s(u) = s(v) & \Longleftrightarrow & u + C = v + C
  \end{eqnarray*}
  This property holds because as seen above, $C$ is the kernel of the linear
  map that maps every vector to its syndrome with respect to $H$. We call the
  vector with the smallest weight in every coset the {\bf leader}. The
  leader of a coset need not be unique.
\end{note}

\begin{lemma}
  \label{cosetleader}
  For a $t$-error-correcting code $C$, every vector, whose weight is at most
  $t$, is the leader of a coset. This vector is unique and describes the error
  which has happened during the transmission.
\end{lemma}

\begin{algorithm}[Decoding]
  \begin{enumerate}
  \item For a vector $x \in \F_q^n$ determine its coset by syndrome
    calculations.
  \item Find a leader $y$ of its coset. Note that if the coset leader is
  unique, the decoded vector will be equal to the original one. This is the
  case if less than $t+1$ errors occurred in a $t$-error-correcting code (see
  Lemma \ref{cosetleader}).
  \item Decode $x$ by calculating $x-y$, which is the wanted vector.
  \end{enumerate}

  It is known that decision problems corresponding to the general decoding
  problem are NP-hard \cite{berlekamp}, also to determine the minimum weight
  is known to be NP-hard \cite{vardy}. An (inefficient) decoding strategy is
  to use look-up tables, especially if many codewords have to be decoded. An
  overview of possible strategies is given in \cite{barg}.
\end{algorithm}

\section{Example: Hamming Code}

The following example shows how the whole procedure of encoding and decoding
including error correction described in the previous section works.

\begin{defn}
  The {\bf Hamming code} $C$ is a $\ecc{7}{4}{3}$ classical linear code over
  the field $\F_2$. It has dimension 7, encodes 4 bits and has a minimum
  distance of 3. That means it can detect 2 errors and correct 1 error.
\end{defn}

\subsection{Encoding}

As we have seen above, the {\it generator matrix} is a $k \times n$
matrix, so in this example we obtain a $4 \times 7$ matrix. It is defined by
\begin{eqnarray*}
  \G & = &
  \left(
  \begin{array}{rrrrrrr}
    1 & 0 & 1 & 0 & 1 & 0 & 1\\
    0 & 1 & 1 & 0 & 0 & 1 & 1\\
    0 & 0 & 0 & 1 & 1 & 1 & 1\\
    1 & 1 & 1 & 0 & 0 & 0 & 0\\
  \end{array}
  \right)
\end{eqnarray*}

This means that our code $C$ is generated by the vectors 
\begin{displaymath}
  (1,0,1,0,1,0,1),\;(0,1,1,0,0,1,1),\;(0,0,0,1,1,1,1),\;
  \mathrm{and}\;(1,1,1,0,0,0,0),
\end{displaymath}
i.e. it consists of all linear combinations of those 4 vectors over $\F_2$.

Now, from the generator matrix $G$, we can construct the {\it parity check
matrix}. This is an $(n-k) \times n$ matrix which consists of the generators
of the space orthogonal to $C$. The rows have to be
orthogonal to every row of the generator matrix ($ H \cdot \G^t = 0$). For our
code $C$ the following matrix is one example that satisfies this condition:

\begin{displaymath}
  H =
  \left(
  \begin{array}{rrrrrrr}
    1 & 0 & 1 & 0 & 1 & 0 & 1\\
    0 & 1 & 1 & 0 & 0 & 1 & 1\\
    0 & 0 & 0 & 1 & 1 & 1 & 1\\
  \end{array}
  \right)
\end{displaymath}
If we calculate the matrix product, we get:

\begin{eqnarray*}
  H \cdot \G^{\transpose} & = &
  \left(
  \begin{array}{rrrrrrr}
    1 & 0 & 1 & 0 & 1 & 0 & 1\\
    0 & 1 & 1 & 0 & 0 & 1 & 1\\
    0 & 0 & 0 & 1 & 1 & 1 & 1\\
  \end{array}
  \right)
  \cdot
  \left(
  \begin{array}{rrrr}
    1 & 0 & 0 & 1\\
    0 & 1 & 0 & 1\\
    1 & 1 & 0 & 1\\
    0 & 0 & 1 & 0\\
    1 & 0 & 1 & 0\\
    0 & 1 & 1 & 0\\
    1 & 1 & 1 & 0
  \end{array}
  \right)
   = 
   {\bf 0}
\end{eqnarray*}

Transmitting a vector, e.g. $( 1,0,1,0)$, we get its encoding:
\begin{eqnarray*}
  (1,0,1,0) \cdot \G & = & (1,0,1,1,0,1,0)
\end{eqnarray*}

\subsection{Decoding}

Suppose that a one bit error occurs during the transmission:
\begin{displaymath}
  (1,0,1,1,0,1,0) \longrightarrow (1,1,1,1,0,1,0)
\end{displaymath}
For error correction we have to calculate the {\bf syndrome} of the vector:
\begin{displaymath}
  (1,1,1,1,0,1,0) \cdot H^t = (0,1,1)
\end{displaymath}
Now we have to find the leader of the coset $(1,1,1,1,0,1,0) +
C$. To do this, we use the equation:
\begin{displaymath}
  (a,b,c,d,e,f,g) \cdot H^t = (0,1,1) 
\end{displaymath}
This equation has the solution $(0,1,0,0,0,0,0)$. In general we will calculate
the leaders once and store them in a look up table. As we know from Lemma 15,
a vector is the unique leader of a $t$ correcting code, if it has at most
weight $t$. We have a one error correcting code and $(0,1,0,0,0,0,0)$ has
weight one, so it is the unique leader. Then we can correct by calculating:
\begin{displaymath}
  (1,1,1,1,0,1,0) - (0,1,0,0,0,0,0) = (1,0,1,1,0,1,0)
\end{displaymath}
This gives us back our original codeword. We only have to calculate
the linear combination of the rows of $G$ to decode $(1,0,1,1,0,1,0)$, and we
get $(1,0,1,0)$.

\chapter{Quantum Error Correcting Codes}
\label{qecc}
This chapter gives a brief introduction to quantum information and quantum
codes. It introduces qudits in general, operations on quantum states, quantum
error correcting codes, and stabilizer codes. The relation of this theory to
physical models can be found in Appendix \ref{appendix} in form of the
postulates of quantum mechanics.

\section{Concepts of Quantum Information}


The mathematical foundation of quantum information theory is linear
algebra. This section gives the relation between common notations in algebra
and its corresponding ones in quantum information.

\begin{defn}
  A {\bf qubit} is the short form for quantum bit and the analog of a bit in
  classical computation. The possible states of a quantum bit are the states
  of the vector space $\complex^2 = \hilbert$. The standard basis of this
  state space is denoted by
  \begin{displaymath}
    \ket{0} = \left( {1 \atop  0} \right), \quad 
    \ket{1} = \left( {0 \atop  1} \right),
  \end{displaymath}
  where $\ket{\cdot}$ is called the {\bf ket notation}. The right side denotes
  the usual vector notation.

  The general state of a qubit is a linear combination
  \begin{eqnarray*}
    \ket{\psi} & = & \alpha \ket{0} + \beta \ket{1}
  \end{eqnarray*}
  with $\alpha, \, \beta \in \complex$ and $\size{\alpha}^2 + \size{\beta}^2 =
  1$.
\end{defn}

This definition is too restrictive for the following chapters, because the
codes we are going to construct will be defined over larger alphabet size than
two. Therefore we have to introduce qudits as the generalisation to non-binary
systems.

\begin{defn}
  A {\bf qudit} $\ket{\psi}$ is the generalisation of a qubit. Possible states
  are the elements of the Hilbert space $\hilbert = \complex^d$ The standard
  basis for this state space consists of the elements
  $\ket{0},\ldots,\ket{d-1}$ and every state can be represented as
  \begin{eqnarray*}
    \ket{\psi} & = & \sum_{i=0}^{d-1} \alpha_i \ket{i}
  \end{eqnarray*}
  with $\sum_{i=0}^{d-1} \size{\alpha_i}^2 = 1$. In usual vector notation,
  every qudit $\ket{i}$ corresponds to the $i$-th vector of unity with one in
  the $i$-th component, all other components are zero.
\end{defn}

To get consistent definitions and good properties, from now on, we restrict
ourselves to $d=p^m$ for $m \in \N$.

\begin{defn}
  \label{paulimatrices}
  Let $\omega_p =  e^{\frac{2\pi i}{p}}$ be a primitive $p$-th root of unity,
  then we define the {\bf Pauli matrices} in the $p$-ary case by
  \begin{eqnarray*}
    X_j \ket{k} = X^j \ket{k}  & = & \ket{k+j},\\
    Z_j \ket{k} = Z^j \ket{k}  & = & \omega_p^{k \cdot j} \ket{k},\\
    \identity \ket{k} & = & \ket{k} 
  \end{eqnarray*}
  for $j,k=0,\ldots,d-1$.

  For $d=p^m$ with $m > 2$, we have the more general definition
  \begin{eqnarray*}
	X_j \ket{k} & = & \ket{k+j},\\
	Z_j \ket{k} & = & \omega_p^{\trace(k\cdot j)} \ket{k},\\
	\identity \ket{k} & = & \ket{k} 
      \end{eqnarray*}
  with  $\omega_p =  e^{\frac{2\pi i}{p}}$ for $j,k \in \F_{p^m}$ and $\trace:
  \F_{p^m} \rightarrow \F_p$ the trace map from the finite field onto its base
  field.
\end{defn}

\begin{defn}
  The {\bf Pauli group} for qudits is defined as
  \begin{displaymath}
    \pauli = \braket{X_i Z_j}{i,j \in \F_{p^m}}.
  \end{displaymath}

  Definition and properties of this group can be found in
  \cite{gottesmanheisenberg, grasslroetteler}.
\end{defn}

\begin{example}
  For the case of qubits we get the well known Pauli matrices

  \begin{displaymath}
    \identity = \left( 
    \begin{array}{rr}
      1&0\\0&1
    \end{array}
    \right), \quad
    X = \left( 
    \begin{array}{rr}
      0&1\\1&0
    \end{array}
    \right), \quad
    Z =  \left( 
    \begin{array}{rr}
     1&0\\0&-1
    \end{array}
    \right), \quad
    Y =  \left( 
    \begin{array}{rr}
      0&-1\\1&0
    \end{array}
    \right)
  \end{displaymath}
  and the Pauli group consists of the span of these four matrices. It is easy
  to check that these matrices are all self-inverse.
  \footnote{Note that in physics literature instead of $Y$ the matrix $iY$ is
  used which has the advantage of being hermitian. However, the vector space
  spanned over the complex numbers is the same.}
\end{example}

If we want to generalise these definitions to systems of more than one qubit
respectively. qudit, we have to use tensor products of Hilbert spaces in order to get
quantum registers.

\begin{defn}
  \label{qregister}
  A {\bf quantum register} is a multiple qudit system. Its state $\ket{b_1
  \cdots b_n}$ is a tensor product of qudits $\ket{b_i}$, $b_i \in
  \F_{p^m}$, i.e.
  \begin{eqnarray*}
    \ket{b_1 \cdots b_n} & := & \ket{b_1} \tensor \cdots \tensor \ket{b_n}.
  \end{eqnarray*}
  The state of a quantum register is in the space $\hilbert^{\tensor n} =
  (\complex^d)^{\tensor n} \cong \complex^{d^n}$ and can be written as
  \begin{eqnarray*}
    \ket{\psi} & = & \sum_{x \in \F_{p^m}^n} c_x \ket{x}
  \end{eqnarray*}
  where $c_x \in \complex$ and $\sum_{x \in \F_{p^m}^n} \size{c_x}^2
  = 1$.
\end{defn}

To be able to use quantum registers, we also have to generalise the operators
acting on them.

\begin{defn}
  The {\bf Pauli group} on a quantum register of length $n$ is the tensor
  product of the Pauli group on single qubits
  \begin{eqnarray*}
    \pauli_n & := & \braket{ U_1 \tensor \cdots \tensor U_n}{U_i \in \pauli \;
    \mathrm{for} \; i = 1,\ldots,n}
  \end{eqnarray*}
  Every operator acts on the corresponding qudit.
\end{defn}

With help of the Pauli group we have already some possible operators that act
on qudits respectively. quantum registers, but this will not be sufficient for
circuits that we need for quantum error correction. Therefore we will
introduce some more important gates. These are defined over $\F_2$.

\begin{defn}$ $
  \begin{itemize}
  \item The {\bf Hadamard} gate $H$ is given by the matrix
    \begin{eqnarray*}
      H & = & \frac{1}{\sqrt{2}} \left( 
      \begin{array}{rr}
	1&1\\1&-1
      \end{array}
      \right)
    \end{eqnarray*}
    and drawn in the Feynman circuit notation \cite{nielsen} for a qubit
    $\ket{\psi}$
    \begin{center}
      \includegraphics[scale=1]{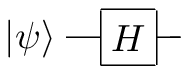}
    \end{center}

  \item The {\bf controlled-not} gate $C-NOT$ is given by the matrix
    \begin{eqnarray*}
      C-NOT & = & \left(
      \begin{array}{rrrr}
	1&0&0&0\\0&1&0&0\\0&0&0&1\\0&0&1&0
      \end{array}
      \right)
    \end{eqnarray*}
    and the conventional circuit is denoted by
    \begin{center}
      \includegraphics[scale=1]{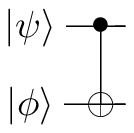}
    \end{center}
    i.e. the $C-NOT$ is a controlled-X gate.

  \item The {\bf controlled-U} gate $C-U$ is the generalisation of the $C-NOT$
  gate for any unitary transformation $U = \left( 
  \begin{array}{cc}
    u_1&u_2\\u_3&u_4
  \end{array} \right)$
  and given by the matrix
    \begin{eqnarray*}
      C-U & = & \left(
      \begin{array}{rrrr}
	1&0&0&0\\0&1&0&0\\0&0&u_1&u_2\\0&0&u_3&u_4
      \end{array}
      \right)
    \end{eqnarray*}
    and the conventional circuit notation is
    \begin{center}
      \includegraphics[scale=1]{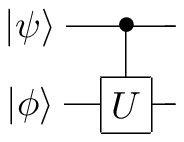}
    \end{center}

  \item The {\bf Toffoli} gate is a controlled gate with two control and one
  target bit. This three qubit gate is given by the matrix
  \begin{eqnarray*}
    {\rm Toffoli} & = & \left(
    \begin{array}{rrrrrrrr}
      1&0&0&0&0&0&0&0\\
      0&1&0&0&0&0&0&0\\
      0&0&1&0&0&0&0&0\\
      0&0&0&1&0&0&0&0\\
      0&0&0&0&1&0&0&0\\
      0&0&0&0&0&1&0&0\\
      0&0&0&0&0&0&0&1\\
      0&0&0&0&0&0&1&0\\
    \end{array}
    \right)
  \end{eqnarray*}
  and its circuit is given by
  \begin{center}
    \includegraphics[scale=1]{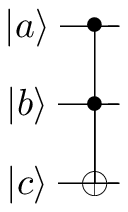}
  \end{center}

  \end{itemize}
\end{defn}

\section{Error Model and Quantum Codes in General}


In order to construct quantum error correcting codes, we first have to define
 an error model. An error model tells us what kind of errors appear and what
 kind of errors we would like to detect or correct.

In the following we will assume that there are no correlated errors, but just
errors on single qudits. Errors on single qubits in our error model can be any
linear combination of Pauli group elements. We also assume that every qudit is
independently affected with the same probability $p$ \cite{laflamme}.

We can look at errors as the interaction of a state in a state space $S$ with
the environment $Env$ by looking at the total Hilbert space $\hilbert =
\hilbert_S \tensor \hilbert_{Env}$ where $\hilbert_S$ is the Hilbert
space of our system and $\hilbert_{Env}$ is the Hilbert space of the
environment. Then an error in its most general form is an operator
\begin{displaymath}
  E: \ket{\psi}_S \ket{0}_{Env} \mapsto \sum_{a \in I} E_a
  \ket{\psi}_S \ket{a}_{Env}
\end{displaymath}
where $\ket{a}_{\mathrm Env}$ are some states of the environment and $E_a$
acts on our state $\ket{\psi}_S$ $\cite{nielsen}$.

\begin{defn}
  A $\qecc{n}{k}{d}$ {\bf quantum error correcting code} (QECC) is a linear
  code that encodes $k$ qudits into $n$ qudits and is able to detect errors
  affecting $d-1$ qudits and to correct errors on $\floor{\frac{d-1}{2}}$
  qudits.
\end{defn}

We cite the following theorem from \cite{knilafl}

\begin{theorem}
  Let $C$ be a subspace of the state space with orthonormal basis
  $\set{\ket{c_1},\ldots,\ket{c_K}}$. Then $C$ is a quantum error correcting
  code for the error operators $\error = \set{E_1,\ldots,E_N}$ iff there are
  constants $\alpha_{k,l} \in \complex$ such that for all
  $\ket{c_i},\,\ket{c_j}$ and for all $E_k,\,E_l \in \error$:
  \begin{eqnarray*}
    \bra{c_i} E_k^{\dagger} E_l \ket{c_j} & = & \delta_{i,j} \alpha_{k,l}.
  \end{eqnarray*}
\end{theorem}

This theorem tells us that errors have to act on the states by mapping them
into orthogonal subspaces of the state space in order to be able to recover
the state.

For our error model it suffices to correct so called dit and phase errors. Dit
errors are bit flip errors in the binary case and given by the acting of the
$X$ gate on the state. Phase flip errors occur when the $Z$ operator is
applied to our state.

\begin{defn}
  A bit flip error is given by the following mapping:
  \begin{eqnarray*}
    \ket{0} & \mapsto & \ket{1}\\
    \ket{1} & \mapsto & \ket{0}
  \end{eqnarray*}
  Similarly the phase error acts on a qubit:
  \begin{eqnarray*}
    \ket{0} & \mapsto & \ket{0}\\
    \ket{1} & \mapsto & - \ket{1}   
  \end{eqnarray*}
  More generally dit errors are given by the operators $X_{\alpha}$ and phase
  errors by the operators $Z_{\beta}$ as in Definition \ref{paulimatrices}.
\end{defn}

\begin{theorem}
  For correction of independent errors on single qudits, it suffices to
  correct dit and phase flip errors, i.e. the set
  \begin{displaymath}
    \pauli = \set{X_{\alpha} Z_{\beta} \; | \; \alpha, \beta \in \F_{p^m}}
  \end{displaymath}
  forms an orthonormal basis for the set of matrices acting on one qudit,
 $\complex^{p^m \times p^m}$ with respect to the matrix inner product
 $\matrixip{A}{B} = \frac{1}{p^m} \, \trace(A^{\dagger} B)$.
\end{theorem}

\begin{proof}
  If we can show that for all $\alpha_i, \beta_j \in \F_{p^m}$
  \begin{eqnarray*}
    \matrixip{X_{\alpha_1}Z_{\beta_1}}{X_{\alpha_2}Z_{\beta_2}} & = &
    \delta_{(\alpha_1,\beta_1),(\alpha_2,\beta_2)},
  \end{eqnarray*}
  these operators have to form a basis, because 
  \begin{eqnarray*}
    \size{\pauli} & = & (p^m)^2 = \dim \complex^{p^m \times p^m}
  \end{eqnarray*}
  
  Let $X_{\alpha_1}Z_{\beta_1}$ and $X_{\alpha_2}Z_{\beta_2}$ be two arbitrary
  elements of $\pauli$, then

  \begin{eqnarray*}
    \matrixip{X_{\alpha_1}Z_{\beta_1}}{X_{\alpha_2}Z_{\beta_2}}
    & = & \frac{1}{p^m}\trace ((X_{\alpha_2} Z_{\beta_2})^{\dagger}
    X_{\alpha_1}Z_{\beta_1})\\
    & = & \frac{1}{p^m}\trace (Z_{\beta_2}^{\dagger} X_{\alpha_2}^{\dagger}
    X_{\alpha_1}Z_{\beta_1})\\
    & = & \frac{1}{p^m} \trace (Z_{-\beta_2} X_{-\alpha_2}
    X_{\alpha_1}Z_{\beta_1})\\
    & = & \frac{1}{p^m} \trace (Z_{-\beta_2} X_{\alpha_1-\alpha_2}
    Z_{\beta_1})\\
    & = & \frac{1}{p^m}
    \trace (X_{\alpha_1-\alpha_2} Z_{\beta_1-\beta_2})
  \end{eqnarray*}
  Note that $Z_{\alpha}$ has nothing but diagonal terms for any $\alpha$ and
  $X_{\alpha}$ has no diagonal terms if $\alpha \ne 0$. So we can conclude
  \begin{itemize}
  \item if $\alpha_1 \ne \alpha_2$, $\trace
    (X_{\alpha_1-\alpha_2} Z_{\beta_1-\beta_2}) = 0$ and the two elements are
    orthogonal
  \item if $\alpha_1 = \alpha_2$,
    \begin{eqnarray*}
       \matrixip{X_{\alpha_1}Z_{\beta_1}}{X_{\alpha_2}Z_{\beta_2}}
       & = & \frac{1}{p^m} \trace(Z_{\beta_1-\beta_2})
    \end{eqnarray*}
    \begin{itemize}
    \item if $\beta_1 = \beta_2$,
	$\matrixip{X_{\alpha_1}Z_{\beta_1}}{X_{\alpha_2}Z_{\beta_2}}
	 =  \frac{1}{p^m} \trace(\identity)
	 =  \frac{1}{p^m} \cdot p^m = 1$
    \item if $\beta_1 \ne \beta_2$,
    \begin{eqnarray*}
       \matrixip{X_{\alpha_1}Z_{\beta_1}}{X_{\alpha_2}Z_{\beta_2}}
       & = & \frac{1}{p^m} \trace(Z_{\beta_1-\beta_2})
        =  \frac{1}{p^m} \sum_{z \in \F_{p^m}}
       \omega_p^{\trace((\beta_1-\beta_2)z)}
       = 0
    \end{eqnarray*}
    Therefore the two operators are orthogonal.
    \end{itemize}
  \end{itemize}
\end{proof}

In the following we will give an example of a simple quantum code which is
also known as repetition code. The difference to the classical repetition code
is that we cannot copy states:

\begin{theorem}[No-Cloning Theorem \cite{nielsen}]
  \label{nocloning}
  It is not possible to copy arbitrary quantum states.
\end{theorem}

\begin{proof}
  Assume there exists a unitary operation $U$ that copies arbitrary quantum
  states. Let $\ket{\psi}$ and $\ket{\phi}$ be arbitrary states. Then
  \begin{eqnarray*}
    U (\ket{\psi} \tensor \ket{s}) & = & \ket{\psi} \tensor \ket{\psi}\\
    U (\ket{\phi} \tensor \ket{s}) & = & \ket{\phi} \tensor \ket{\phi}.    
  \end{eqnarray*}
  The inner product of these two equations gives
  \begin{eqnarray*}
    \braket{\psi}{\phi} & = & \braket{\psi}{\phi}^2.
  \end{eqnarray*}
  This equation can only hold if $\braket{\psi}{\phi}=1$ or
  $\braket{\psi}{\phi}=0$, i.e. if $\ket{\psi} = \ket{\phi}$ or if
  $\ket{\psi}$ and $\ket{\phi}$ are orthogonal.

  Hence cloning of unknown states is not possible.
\end{proof}

Therefore we have to use the properties of linear algebra. These examples can
for example be reviewed in \cite{qeclecture,laflamme}.

\begin{example}[Repetition Code for Bit Flips]
  \label{exbitflip}
  
  The easiest way to correct one error in classical coding theory is the
  repetition code, that maps
  \begin{eqnarray*}
    0 & \mapsto & 000,\\
    1 & \mapsto & 111.
  \end{eqnarray*}
  Decoding is performed by correcting the most likely error. In this
  case, we will decode with the majority principle, e.g.
  \begin{eqnarray*}
    001 & \stackrel{decoding}{\Longrightarrow} & 000,\\
    101 & \stackrel{decoding}{\Longrightarrow} & 111.\\
  \end{eqnarray*}
  The problem with this code in quantum error correction is that we cannot
  copy arbitrary quantum states (see Theorem \ref{nocloning}). However, we can
  do something similar to the classical repetition code. First, we will look at
  bit flips and imitate the classical code. Then in the next example, we will
  transform the code such that it can correct phase errors.

  Let us encode one qubit into three, i.e.
  \begin{eqnarray*}
    \ket{0} & \mapsto & \ket{000},\\
    \ket{1} & \mapsto & \ket{111}.    
  \end{eqnarray*}
  Then an arbitrary superposition will be encoded as
  \begin{eqnarray*}
    \alpha \ket{0} + \beta \ket{1} & \mapsto & \alpha \ket{000} + \beta
    \ket{111}.
  \end{eqnarray*}
  In the following we will assume that a bit flip error occurred on the first
  qubit. This is sufficient, because of the symmetry of the code, all
  calculations will be the same for an error on one of the other qubits.

  An error on the first qubit will map the state $\alpha \ket{000} + \beta
  \ket{111}$ to $\alpha \ket{100} + \beta \ket{011}$. This error can be
  detected with two ancilla qubits by parity check. Let the ancilla qubits be
  in the state $\ket{0}$, then we use two C-NOT gates on every ancilla qubit
  like in Figure \ref{bitflip} and get
  \begin{eqnarray*}
    \ket{000}\ket{00} & \mapsto & \ket{000}\ket{00},\\
    \ket{100}\ket{00} & \mapsto & \ket{100}\ket{10},\\
    \ket{010}\ket{00} & \mapsto & \ket{010}\ket{01},\\
    \ket{001}\ket{00} & \mapsto & \ket{001}\ket{11},\\
    \ket{111}\ket{00} & \mapsto & \ket{111}\ket{00},\\
    \ket{011}\ket{00} & \mapsto & \ket{011}\ket{10},\\
    \ket{101}\ket{00} & \mapsto & \ket{101}\ket{01},\\
    \ket{110}\ket{00} & \mapsto & \ket{110}\ket{11}.\\
  \end{eqnarray*}
  If we measure the ancilla bits, we will deterministically get syndrome bits
  0 or 1 and can uniquely assign it to the single bit flip error that has
  occurred.
  \begin{figure}
    \centering
    \includegraphics[scale=1]{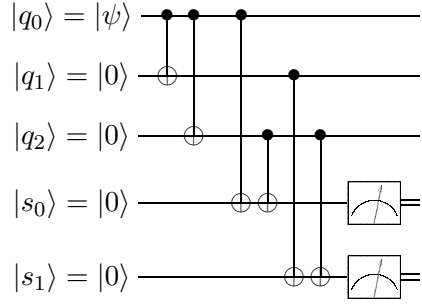}
  \caption{Encoding and syndrome measurement circuit for the repetition bit
  flip code with code bits $\ket{q_0} = \ket{\psi}, \; \ket{q_1}, \;
  \ket{q_2}$ and syndrome bits $\ket{s_0}, \; \ket{s_1}$.}
  \label{bitflip}
\end{figure}

\end{example}

\begin{example}[Repetition Code for Phase Errors]
  \label{exphaseflip}

  The important observation for phase correction is that bit flip errors
  transform to phase errors under the Hadamard transformation and
  vice versa. So we encode
  \begin{eqnarray*}
    \ket{0} & \mapsto & \frac{1}{\sqrt{2^3}}
    (\ket{0}+\ket{1})(\ket{0}+\ket{1})(\ket{0}+\ket{1}) =: \ket{{+++}}\\
    \ket{1} & \mapsto & \frac{1}{\sqrt{2^3}}
    (\ket{0}-\ket{1})(\ket{0}-\ket{1})(\ket{0}-\ket{1}) =: \ket{{---}}\\
  \end{eqnarray*}
  and a phase error on the first qubit will map
  \begin{eqnarray*}
    \ket{{+++}} & \mapsto & \ket{{-++}}\\
    \ket{{---}} & \mapsto & \ket{{+--}}
  \end{eqnarray*}
  We can detect these errors by Hadamard transformation to the standard basis
  and the same syndrome calculations as before. Then we apply another Hadamard
  and perform the necessary error corrections. The complete circuit of this
  encoding and decoding process is given in Figure \ref{phaseflip}.
  \begin{figure}
    \centering
    \includegraphics[scale=1]{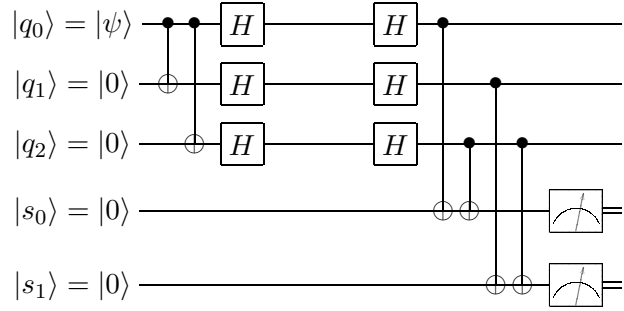}
  \caption{Encoding and syndrome measurement circuit for the repetition phase
  flip code with code bits $\ket{q_0} = \ket{\psi}, \; \ket{q_1}, \;
  \ket{q_2}$ and syndrome bits $\ket{s_0}, \; \ket{s_1}$.}
  \label{phaseflip}
\end{figure}
 
\end{example}

The combination of these two codes gives our first quantum code, known as {\bf
  Shor code} \cite{shor}, encoding one qubit into nine and being able to
  correct bit and phase flip errors and therefore any error on single
  qubits. This code is a $\qecc{9}{1}{3}$ code.

\begin{example}[9-qubit code]
  
  If we concatenate the codes given in Examples \ref{exbitflip} and
  \ref{exphaseflip}, we get a code that can correct bit and phase flip errors
  on one qubit. Therefore we encode
  \begin{eqnarray*}
    \ket{0} 
    & \mapsto & \frac{1}{\sqrt{2^3}} (\ket{000} + \ket{111})(\ket{000} +
    \ket{111})(\ket{000} + \ket{111})\\
    & = & \frac{1}{\sqrt{2^3}} (\ket{000000000} + \ket{000000111} +
    \ket{000111000} + \ket{000111111}\\
    & & + \ket{111000000} + \ket{111000111} +
    \ket{111111000} + \ket{111111111})\\
    & = & \ket{\bar{0}}\\
    \ket{1} 
    & \mapsto & \frac{1}{\sqrt{2^3}} (\ket{000} - \ket{111})(\ket{000} -
    \ket{111})(\ket{000} - \ket{111})\\
    & = & \frac{1}{\sqrt{2^3}} (\ket{000000000} - \ket{000000111} -
    \ket{000111000} + \ket{000111111}\\
    & & - \ket{111000000} + \ket{111000111} +
    \ket{111111000} - \ket{111111111})\\
    & = & \ket{\bar{1}}\\
  \end{eqnarray*}
  Then bit flip errors can be corrected in blocks of three qubits, phase flip
  errors in the comparison of the three qubit blocks. The circuit for
  this encoding is given in Figure \ref{IXqubitcode}.
  \begin{figure}
    \centering
    \includegraphics[scale=1]{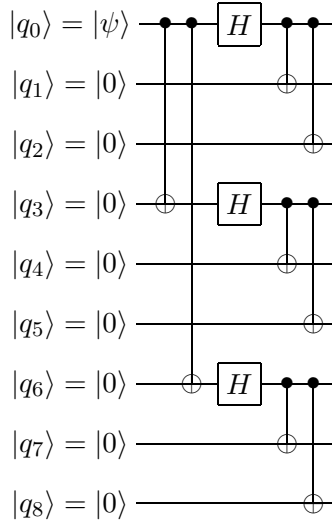}
    \caption{Encoding circuit for $\qecc{9}{1}{3}$-Code}
    \label{IXqubitcode}
  \end{figure}

\end{example}

\section{Stabilizer Codes}

Stabilizer codes were introduced by Daniel Gottesman
\cite{gottesman}. This type of quantum error correction is very useful since
it allows us to reuse results from classical coding theory and to describe a
large class of quantum codes.

\subsection{Basic Definitions}

In the last section we saw that it is sufficient to correct dit and phase
errors. The idea of a stabilizer code is to define a commuting set of Pauli
operators and use the set of states that is invariant under these operators as
code space, i.e. the code is the +1 eigenspace of all the
operators.

Recall that two operators $E,F$ {\bf commute} iff $EF=FE$.

\begin{defn}
  A {\bf Stabilizer code} $S$ of length $n$ is an Abelian subgroup of the
  Pauli group $\pauli_n$. Codewords are the states of the state space that are
  invariant under all elements of $S$.
\end{defn}

\begin{note}
  \label{commute}
  An important feature of stabilizer codes is that the stabilizer forms an
  Abelian group. Error detection and correction takes place by measuring
  commutation and anti-commutation of operators.
 
  First we will look at the commutation behaviour of single qudit operators in
  the $p$-ary case.  It is easy to show that $X^{\alpha}$ commutes with
  $X^{\beta}$. Similarly $Z^{\alpha}$ commutes with $Z^{\beta}$.

  The problem is what happens with $X^{\alpha}$ and $Z^{\beta}$. It suffices
  to show what happens with $X$ and $Z$, because we can split $X^{\alpha}$
  into $X \cdot X \cdots X$. In general, in case $p$ prime $X$ is given by
  the matrix
  \begin{eqnarray*}
    X & = & \left(
    \begin{array}{ccccc}
      0 & 0 & \cdots & 0 & 1\\
      1 & 0 & \cdots & 0 & 0\\
      0 & 1 & \cdots & 0 & 0\\
      \vdots & \vdots & \ddots & \vdots & \vdots\\
      0 & 0 & \cdots & 1 & 0\\
    \end{array}
    \right).
  \end{eqnarray*}
  The matrix for $Z$ looks like
  \begin{eqnarray*}
    Z & = & \left(
    \begin{array}{cccc}
      1 & 0 & \cdots & 0\\
      0 & \omega_p & \cdots & 0\\
      \vdots & \vdots & \ddots & \vdots\\
      0 & 0 & \cdots & \omega_p^{p-1}\\
    \end{array}
    \right).
   \end{eqnarray*}
   Then
   \begin{eqnarray*}
     XZ & = & \left(
     \begin{array}{ccccc}
       0 & 0 & \cdots & 0 & \omega_p^{p-1}\\
       1 & 0 & \cdots & 0 & 0\\
       0 & \omega_p & \cdots & 0 & 0\\
       \vdots & \vdots & \ddots & \vdots & \vdots\\
       0 & 0 & \cdots & \omega_p^{p-2} & 0\\
     \end{array}
     \right)
   \end{eqnarray*}
   and
   \begin{eqnarray*}
     ZX & = & \left(
     \begin{array}{ccccc}
       0 & 0 & \cdots & 0 & 1\\
       \omega_p & 0 & \cdots & 0 & 0\\
       0 & \omega_p^2 & \cdots & 0 & 0\\
       \vdots & \vdots & \ddots & \vdots & \vdots\\
       0 & 0 & \cdots & \omega_p^{p-1} & 0\\
     \end{array}
     \right)
     = \omega_p XZ.
   \end{eqnarray*}
  So we get for $X^{\alpha}$ and $Z^{\beta}$, $\alpha, \beta \in \F_p$
  \begin{eqnarray*}
    Z^{\beta} X^{\alpha} & = & \omega_p^{\alpha \cdot \beta} X^{\alpha}
    Z^{\beta}.
  \end{eqnarray*}
  In the case of a field $\F_{p^m}$ we get for $\alpha, \beta \in \F_{p^m}$
  \begin{eqnarray*}
    Z_{\beta} X_{\alpha} & = & \omega_p^{\trace(\alpha \cdot \beta)} X_{\alpha}
    Z_{\beta}.
  \end{eqnarray*} 


  Since we have defined the commutation rules on a single qudit operator, we
  can generalise this to operators on quantum registers. Hence for two
  operators
  \begin{displaymath}
    U_1 = X_{\alpha_1} Z_{\beta_1} \tensor \cdots \tensor  X_{\alpha_n}
    Z_{\beta_n}, \quad
    U_2 = X_{\mu_1} Z_{\nu_1} \tensor \cdots \tensor  X_{\mu_n} Z_{\nu_n}  
  \end{displaymath}
  we have that
  \begin{eqnarray*}
    U_2 \cdot U_1 & = & \omega_p^{\sum_{i=1}^n \alpha_i \nu_i - \mu_i \beta_i}
    U_1 \cdot U_2
  \end{eqnarray*}
  in case $p$ prime and therefore two operators commute iff
  \begin{eqnarray*}
    \sum_{i=1}^n \alpha_i \nu_i - \mu_i \beta_i & = & 0.
  \end{eqnarray*}

  In the case of a field $\F_{p^m}$ with $m \ge 2$, we get
  \begin{eqnarray*}
    U_2 \cdot U_1 & = & \omega_p^{\sum_{i=1}^n \trace(\alpha_i \nu_i - \mu_i
    \beta_i)} U_1 \cdot U_2
  \end{eqnarray*}
  and therefore two operators commute iff
  \begin{eqnarray*}
    \sum_{i=1}^n \trace(\alpha_i \nu_i - \mu_i \beta_i) & = & 0.
  \end{eqnarray*}

\end{note}

\begin{corollary}
  Let $S$ be a stabilizer, i.e. an Abelian subgroup of $\pauli_n$. If a state
  $\ket{\psi}$ is in the +1 eigenspace of a set of generators
  $\set{G_1,\ldots,G_l}$ of $S$, it is an
  eigenstate of all elements in $S$.
\end{corollary}

\begin{proof}
  An arbitrary element $F \in S$ is a linear combination of the set of
  generators:
  \begin{eqnarray*}
    F \ket{\psi} 
    & = & (G_1^{\alpha_1} \cdots G_l^{\alpha_l})\ket{\psi}\\
    & = & (G_1^{\alpha_1} \cdots G_{l-1}^{\alpha_{l-1}})
    \, G_l^{\alpha_l}\, \ket{\psi}\\
    & = & (G_1^{\alpha_1} \cdots G_{l-1}^{\alpha_{l-1}}) \ket{\psi}\\
    & = & \cdots\\
    & = & \ket{\psi}
  \end{eqnarray*}
  for any $\ket{\psi}$ that is in the +1 eigenspace of the set of
  generators and $\alpha_i \in \F_{p^m}$ for all
  $i$. Therefore $\ket{\psi}$ is in the +1 eigenspace of $F$.
\end{proof}

\begin{lemma}
  The stabilizer $S$ of a stabilizer code is an Abelian group, i.e. all
  stabilizer elements commute.
\end{lemma}

\begin{proof}
  Assume that not all of them commute, then there exist $E,F \in S$ with $EF =
  \omega_p^{\alpha} FE$ where $\alpha \ne 0$. So all states $\ket{\psi}$ in
  the code satisfy
  \begin{eqnarray*}
    \ket{\psi} & = & EF \ket{\psi} = \omega_p^{\alpha} FE \ket{\psi}\\
    & = & \omega_p^{\alpha} \ket{\psi}
  \end{eqnarray*}
  So it has eigenvalue +1 and $\omega_p^{\alpha}$ at the same time and
  therefore contradiction.
\end{proof}

Since we have seen that it suffices to look at a set of generators, we can
represent a stabilizer code in an easier way:

\begin{defn}
  A {\bf generator matrix} $\G$ of a stabilizer code is an $l\times 2n$-matrix
  \begin{eqnarray*}
    \G & = & \left( X | Z \right)
  \end{eqnarray*}
  where the first $n$ components represent the $X$ errors, the second $n$
  components the $Z$ errors. This matrix defines a $\qecc{n}{k}{d}$ quantum
  error correcting code with $k = n-l$.
\end{defn}

\begin{lemma}
  Elements $x = (x_1, \ldots, x_{2n}), \; y = (y_1, \ldots, y_{2n})$ of the
  Pauli group commute in the vector representation, i.e. $(x_1, \ldots,
  x_{2n})$ means $X_{x_1} Z_{x_{n+1}} \tensor \cdots \tensor X_{x_n}
  Z_{x_{2n}}$, if

  \begin{displaymath}
    \sip{x}{y} = \sum_{i=1}^n x_i y_{n+i} - x_{n+i} y_i = 0.
  \end{displaymath}
   We call $\sip{x}{y}$ the {\bf standard symplectic inner product}.
\end{lemma}

\begin{proof}
  A vector $(x_1,\ldots,x_{2n})$ in this representation denotes the operator
  \begin{displaymath}
    X_{x_1}Z_{x_{n+1}} \tensor \cdots \tensor X_{x_n}Z_{x_{2n}}.
  \end{displaymath}
  Therefore we get
    \begin{eqnarray*}
      &&(X_{y_1}Z_{y_{n+1}} \tensor \cdots \tensor
      X_{y_n}Z_{y_{2n}}) (X_{x_1}Z_{x_{n+1}} \tensor \cdots \tensor
      X_{x_n}Z_{x_{2n}})\\
      &&=  \omega_p^{\trace(\sum_{i=1}^n x_i y_{n+i} - x_{n+i} y_i)}
      (X_{x_1}Z_{x_{n+1}} \tensor \cdots 
      )(X_{y_1}Z_{y_{n+1}} \tensor \cdots \tensor
      X_{y_n}Z_{y_{2n}})\\
      &&=  \omega_p^{\trace(0)}
      (X_{x_1}Z_{x_{n+1}} \tensor \cdots \tensor
      X_{x_n}Z_{x_{2n}})(X_{y_1}Z_{y_{n+1}} \tensor \cdots \tensor
      X_{y_n}Z_{y_{2n}})\\
      &&=  (X_{x_1}Z_{x_{n+1}} \tensor \cdots \tensor
      X_{x_n}Z_{x_{2n}})(X_{y_1}Z_{y_{n+1}} \tensor \cdots \tensor
      X_{y_n}Z_{y_{2n}})
    \end{eqnarray*}    
  and therefore the operators commute.
\end{proof}

\begin{example}
  If $S$ is generated by $XYI = X \tensor Y \tensor \identity$ and $ZXX = Z
  \tensor X \tensor X$ over $\F_2$, the generator matrix is given by
  \begin{eqnarray*}
    \G & = & \left( 
    \begin{array}{ccc|ccc}
    1 & 1 & 0 & 0 & 1 & 0\\  
    0 & 1 & 1 & 1 & 0 & 0
    \end{array}
    \right)
  \end{eqnarray*}
  Let $x = \left( 
  \begin{array}{ccc|ccc}
    1 & 1 & 0 & 0 & 1 & 0
  \end{array}
  \right)$
  and $y = \left( 
  \begin{array}{ccc|ccc}
    0 & 1 & 1 & 1 & 0 & 0
  \end{array}
  \right)$, 
  then
  \begin{eqnarray*}
    \sip{x}{y} 
	& = & (1 \cdot 1 + 1 \cdot 0 + 0 \cdot 0) - (0 \cdot 0 + 1 \cdot 1 +1
	\cdot 0)\\
	& = & 1 - 1 = 0
  \end{eqnarray*}
\end{example}

We already introduced the 9-qubit code in the previous section. In the
following we give a description in terms of stabilizers.

\begin{example}
  {\bf 9-qubit code}
  
  The 9-qubit code encodes one qubit into 9. Therefore our generator matrix
  will have size $8 \times 18$.

  {\bf Claim}
  \begin{eqnarray*}
    \G & = &
    \left(
    \begin{array}{ccccccccc}
      Z & Z & 0 & 0 & 0 & 0 & 0 & 0 & 0 \\
      Z & 0 & Z & 0 & 0 & 0 & 0 & 0 & 0 \\
      0 & 0 & 0 & Z & Z & 0 & 0 & 0 & 0\\
      0 & 0 & 0 & Z & 0 & Z & 0 & 0 & 0\\
      0 & 0 & 0 & 0 & 0 & 0 & Z & Z & 0\\
      0 & 0 & 0 & 0 & 0 & 0 & Z & 0 & Z\\
      X & X & X & X & X & X & 0 & 0 & 0 \\
      X & X & X & 0 & 0 & 0 & X & X & X
    \end{array}
    \right)\\
    & = & 
    \left(
    \begin{array}{ccccccccc|ccccccccc}
      0 & 0 & 0 & 0 & 0 & 0 & 0 & 0 & 0 &
      1 & 1 & 0 & 0 & 0 & 0 & 0 & 0 & 0 \\
      0 & 0 & 0 & 0 & 0 & 0 & 0 & 0 & 0 &
      1 & 0 & 1 & 0 & 0 & 0 & 0 & 0 & 0 \\
      0 & 0 & 0 & 0 & 0 & 0 & 0 & 0 & 0 &
      0 & 0 & 0 & 1 & 1 & 0 & 0 & 0 & 0\\
      0 & 0 & 0 & 0 & 0 & 0 & 0 & 0 & 0 &
      0 & 0 & 0 & 1 & 0 & 1 & 0 & 0 & 0\\
      0 & 0 & 0 & 0 & 0 & 0 & 0 & 0 & 0 &
      0 & 0 & 0 & 0 & 0 & 0 & 1 & 1 & 0\\
      0 & 0 & 0 & 0 & 0 & 0 & 0 & 0 & 0 &
      0 & 0 & 0 & 0 & 0 & 0 & 1 & 0 & 1\\
      1 & 1 & 1 & 1 & 1 & 1 & 0 & 0 & 0 &
      0 & 0 & 0 & 0 & 0 & 0 & 0 & 0 & 0 \\
      1 & 1 & 1 & 0 & 0 & 0 & 1 & 1 & 1 &
      0 & 0 & 0 & 0 & 0 & 0 & 0 & 0 & 0 \\
    \end{array}
    \right)
  \end{eqnarray*}
  is a generator matrix for the stabilizer code of the 9-qubit code.
    
  \begin{proof}
    It is easy to check that the codewords are invariant under $\G$ and as $\G$
    has rang 8, it has to be a generator matrix for the code.
  \end{proof}
\end{example}

\subsection{Error Detection and Correction}

Like in classical coding theory we can detect and correct errors by syndrome
measurements, but in the quantum case we are not allowed to measure the state,
because this in general will destroy the superposition (see
\ref{appendix}).

\begin{defn}
  An {\bf ancilla} qudit is an extra qudit that is used during computations,
  but is not part of the input or output, and is prepared in some fixed state,
  usually $\ket{0}$.
\end{defn}

Before we can introduce error correction and the amount of errors we can
correct and detect, we have to define some more terms like weight and distance
for quantum codes.

\begin{defn}
  The {\bf weight} $wt$ of an operator $U_1 \tensor \cdots \tensor U_n$ is the
  number of elements $U_i$ that are not equal to the identity, e.g.
  \begin{displaymath}
    X \tensor \identity \tensor \identity \tensor Z \tensor Z
  \end{displaymath}
  has weight 3.
\end{defn}

In classical coding theory the distance of a stabilizer code (which is a
linear code) would be the minimal weight over all codewords. In the quantum
case this does not hold like that, because we have to detect different types
of errors.

\begin{defn}
  The {\bf normalizer} $N(S)$ of a stabilizer $S$ is the set of Pauli
  operators that commute with all stabilizer elements.
  \begin{eqnarray*}
    N(S) & = & \set{F \in \pauli_n \; | \; EF=FE \quad \forall E \in S}.
  \end{eqnarray*}
\end{defn}

This definition allows us to distinguish between three different types of
errors.

\begin{lemma}
  Let $F \in \pauli_n$ be an error operator. Then the following three cases
  are possible:
  \begin{enumerate}
  \item If $F \in S$ nothing happened to the code.
  \item If $F \in \pauli_n \backslash N(S)$ the error is detectable and can be
    corrected with the most likely error principle (see \ref{mostlikelyerror}
    and \ref{cosetleader}).
  \item If $F \in N(S)\backslash S$ the error cannot be detected and therefore
  is not correctable.
  \end{enumerate}
\end{lemma}

\begin{proof} Let $\ket{\psi}$ be any element of the code. Then $E\ket{\psi} =
  \ket{\psi}$ for all $E \in S$.
  \begin{enumerate}
  \item If $F \in S$, we get for all codewords
    \begin{eqnarray*}
      F \ket{\psi} & = & \ket{\psi}
    \end{eqnarray*}
    and nothing happens to the codeword.

  \item If $F \in \pauli_n \backslash N(S)$ for all $E \in S$
    \begin{eqnarray*}
      E(F\ket{\psi}) & = & (EF) \ket{\psi} = (\omega_p^{\alpha}FE) \ket{\psi}\\
      & = & \omega_p^{\alpha} F \ket{\psi}
    \end{eqnarray*}
    Therefore the faulty codeword is in the $\omega_p^{\alpha}$  eigenspace
    and detectable by syndrome measurements. With the
    method introduced in Chapter \ref{cec} it is correctable up to half the
    distance.
    
  \item If $F \in N(S)\backslash S$ for all $E \in S$
    \begin{eqnarray*}
      E(F\ket{\psi}) & = & (EF) \ket{\psi} = (FE) \ket{\psi}\\
      & = &  F \ket{\psi}
    \end{eqnarray*}
    and so we cannot detect the error by syndrome measurement, because $F
    \ket{\psi}$ is also in the $+1$ eigenspace.
  \end{enumerate}
\end{proof}

This lemma allows us to introduce the distance of a stabilizer code.

\begin{defn}
  The {\bf distance} $d$ of a quantum stabilizer code $C$ is the minimum
  weight of all normalizer elements that are not in the stabilizer.
  \begin{eqnarray*}
    d & = & \min \set{wt(x) \; | \; x \in N(S) \backslash S}.
  \end{eqnarray*}
  In other words, if we denote the normalizer as the set of all operators that
  commute with all elements in the stabilizer, then the normalizer is equal to
  the dual code $\Cperps$ with respect to the symplectic inner product
  $\sip{\,}{\,}$. We obtain that
  \begin{eqnarray*}
    d & = & \min \set{wt(x) \; | \; x \in \Cperps \backslash C}.
  \end{eqnarray*}
\end{defn}

To be able to correct errors smaller than $\floor{\frac{d-1}{2}}$ as in
classical coding theory, we can use syndrome measurements.

\begin{algorithm}\label{correction}
  $ $
  \begin{enumerate}
  \item Let $x_1,\ldots,x_l$ be the generators of the stabilizer code,
    i.e. the rows of the generator matrix.
  \item Prepare $l$ ancilla dits in the $\ket{+} = \frac{1}{\sqrt{2}} (\ket{0}
    + \ket{1})$ state. Every ancilla bit corresponds to one stabilizer
    generator.
  \item Perform controlled operations for every stabilizer generator as in
    Figure \ref{stabilizersyndrome}, i.e. if 
    \begin{eqnarray*}
      U^{(i)} & = & U^{(i)}_1 \tensor \cdots \tensor U^{(i)}_n,
    \end{eqnarray*}
    perform controlled $U^{(i)}_j$ operations on the $j$th qubit.
    \begin{figure}
      \centering
      \includegraphics[scale=0.8]{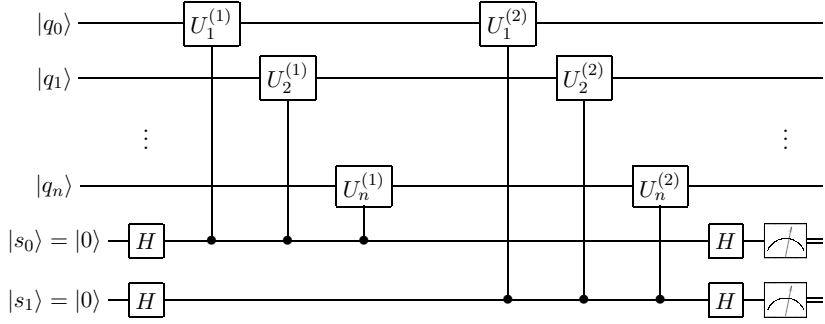}
      \caption{Example of a syndrome measurement in case where the stabilizer
	is given by the two generators $U^{(1)}_1 \tensor \cdots \tensor
	U^{(1)}_n$ and $U^{(2)}_1 \tensor \cdots \tensor U^{(2)}_n$}
      \label{stabilizersyndrome}
    \end{figure}
  \item Measure the ancilla dits. The results are our syndromes.
  \item Use classical theory (see Chapter \ref{cec}) to determine the errors.
  \item Apply the necessary gates to correct the errors.
  \end{enumerate}
\end{algorithm}

Note that this algorithm uses measurements and error detection and correction
are assumed to be perfect. Algorithm \ref{correction} is not fault-tolerant,
however, it has been shown how to make syndrome measurements fault-tolerant
for any stabilizer code \cite{gottesmanft}. One example how to omit
measurements is given by Example \ref{exrepnomeas} below.

\begin{example}
  \label{exrepnomeas}
  In Example \ref{exbitflip} we saw how to correct bit flip errors. In order
  to be able to locate errors, we had to measure the syndrome bits. An
  automatic error correction can be achieved by the circuit given in Figure
  \ref{bitflipwithoutmeas}. There we use a Toffoli gate to correct errors.

  A correction without measurement is in general possible if we use fresh
  ancilla qubits every time error correction is performed (see Box 10.1 in
  \cite{nielsen}).

  \begin{figure}
    \centering
    \includegraphics[scale=1]{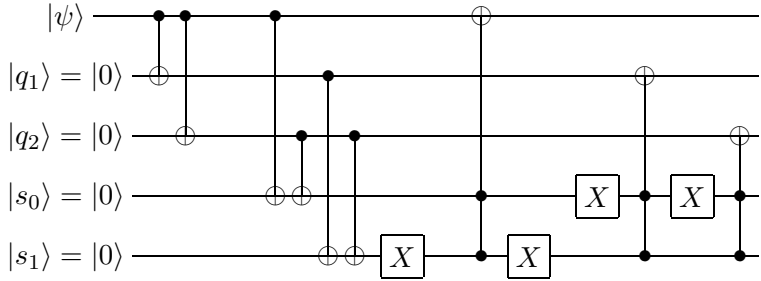}
    \caption{Example of an error correction without measurement}
    \label{bitflipwithoutmeas}
  \end{figure}

\end{example}

\subsection{Weighted Symplectic Inner Product}


In the following chapters we will see that our construction does not always
give codes that are orthogonal with respect to the standard symplectic inner
product. However, we can construct codes $C$ which are orthogonal with respect
to
\begin{eqnarray*}
  \sipa{x}{y} & := & \sum_{i=0}^c4n a_i (x_i\, y_{n+i} - x_{n+i} \, y_i) = 0
\end{eqnarray*}
for all $x,y \in C$ and all $a_i \ne 0$. We will call it {\bf weighted
symplectic inner product}.

The following lemma shows that we can construct a stabilizer code with respect
to the standard symplectic inner product with the same properties as the old
one with respect to the special symplectic inner product.

\begin{lemma}
  \label{codemod}
  Let $C$ be a linear code over $\F_{p^m}$ which is self-orthogonal with
  respect to the weighted symplectic inner product $\sipa{\;}{\,}$ and has a
  corresponding quantum code with parameters $\qecc{n}{k}{d}$ and generator
  matrix
  \begin{displaymath}
    \G = 
    \left(
    \begin{array}{ccc|ccc}
      c_{1,1} & \cdots & {c_{1,n}} &  {c_{1,n+1}} &
      \cdots & {c_{1,2n}}\\ 
      \vdots & \ddots & \vdots & \vdots & \ddots & \vdots\\
	     {c_{l,1}} & \cdots & {c_{l,n}} &  {c_{l,n+1}}
	     & \cdots & {c_{l,2n}}
    \end{array}
    \right).
  \end{displaymath}

  Then the code $C^{\prime}$ with generator matrix
  \begin{displaymath}
    \G^{\prime} = 
    \left(
    \begin{array}{ccc|ccc}
      {a_1 \cdot c_{1,1}} & \cdots & {a_n \cdot c_{1,n}} &  {c_{1,n+1}} &
      \cdots & {c_{1,2n}}\\ 
      \vdots & \ddots & \vdots & \vdots & \ddots & \vdots\\
	     {a_1 \cdot c_{l,1}} & \cdots & {a_n \cdot c_{l,n}} &  {c_{l,n+1}}
	     & \cdots & {c_{l,2n}} 
    \end{array}
  \right)
\end{displaymath}
where the $c_{i,j}$ are the elements of $\G$, defines a stabilizer code with
respect to the standard symplectic inner product $\sip{\;}{\,}$ with the same
parameters $\qecc{n}{k}{d}$.
\end{lemma}

\begin{proof}
  It is easy to show that $\G^{\prime}$ defines a stabilizer code with respect
  to the standard symplectic inner product:

  Let $x = (a_1 x_1,\ldots,a_n x_n, x_{n+1},\ldots, x_{2n}), y = (a_1
  y_1,\ldots,a_n y_n, y_{n+1},\ldots, y_{2n}) \in C^{\prime}$, then
  \begin{eqnarray*}
    \sip{x}{y} 
    & = & \sum_{i=1}^n (a_i x_i)\cdot y_{n+i} - x_{n+i}\cdot(a_i y_i)\\
    & = & \sum_{i=1}^n a_i \; (x_i \,y_{n+i} - x_{n+i}\, y_i)\\
    & = & \sipa{(x_1,\ldots, x_{2n})}{(y_1,\ldots, y_{2n})}\\
    & = & 0
  \end{eqnarray*}
  because $(x_1,\ldots, x_{2n}),(y_1,\ldots, y_{2n}) \in C$ and $C$ is a
  stabilizer code with respect to $\sipa{\;}{\;}$. Therefore $C^{\prime}$ is a
  stabilizer code.

  The reason why the parameters $\qecc{n}{k}{d}$ do not change when going to
  $\G^{\prime}$ is the following:
  \begin{itemize}
  \item The code length $n$ stays the same, because all codewords still have
    the length $2n$.
  \item The number of encoded qudits also does not change, because
    $\G^{\prime} = \G \cdot D$ where $D = \diag(a_1, \ldots, a_n, 1, \ldots,
    1)$. Since $D \in \GL(2n,\F_{p^m})$, we have that $\G$ and $\G^{\prime}$
    have the same rank.

   \item The distance $d$ of the code could only change, if the weights of the
   normalizer elements change. This is not possible, because all coefficients
   $a_i \ne 0$ and $\F_{p^m}$ is a field that has no zero divisors. Therefore
   the weights stay the same and the distance of the code, too, because
   \begin{eqnarray*}
     d & = & \min \set{wt(x) \; | \; x \in N(S) \backslash S}.
   \end{eqnarray*}
  \end{itemize}

\end{proof}

The following example illustrates the construction.

\begin{example}
  Let $\G$ be a generator matrix for a stabilizer code over $\F_5$ with
  \begin{eqnarray*}
    \G & = & \left(
    \begin{array}{cccc|cccc}      
      1&0&1&4&0&0&2&4\\
      4&1&1&0&2&3&3&0
    \end{array}
    \right)
  \end{eqnarray*}
  This code is self-orthogonal with respect to the symplectic inner
  product
  \begin{eqnarray*}
    \sipa{x}{y} & = & 2 (x_1 y_5 - x_5 y_1) + 1 (x_2 y_6 - x_6 y_2)\\
    & & + 1 (x_3 y_7 - x_7 y_3) + 4 (x_4 y_8 - x_8 y_4).
  \end{eqnarray*}
  Indeed, we obtain that
  \begin{eqnarray*}
    \sipa{(1,0,1,4,0,0,2,4)}{(4,1,1,0,2,3,3,0)} 
    & = & 2 (1 \cdot 2 - 0 \cdot 4) + 1 (0 \cdot 3 - 0 \cdot 1)\\
    & & + 1 (1 \cdot 3 - 2 \cdot 1) + 4 (4 \cdot 0 - 4 \cdot 0)\\
    & = & 2 \cdot 2 + 1\\
    & = & 0.
  \end{eqnarray*}
  We can now apply Lemma \ref{codemod} to get the new generator matrix
  \begin{eqnarray*}
    \G^{\prime} & = & \left(
    \begin{array}{cccc|cccc}      
      2&0&1&1&0&0&2&4\\
      3&1&1&0&2&3&3&0
    \end{array}
    \right).
  \end{eqnarray*}
  With the standard symplectic inner product we get
  \begin{eqnarray*}
    \sip{(2,0,1,4,0,0,2,4)}{(3,1,1,0,2,3,3,0)} 
    & = &  (2 \cdot 2 - 0 \cdot 3) + (0 \cdot 3 - 0 \cdot 1)\\
    & & + (1 \cdot 3 - 2 \cdot 1) + 4 (1 \cdot 0 - 4 \cdot 0)\\
    & = & 2 \cdot 2 + 1\\
    & = & 0
  \end{eqnarray*}
  and therefore orthogonal vectors.
\end{example}

\section{CSS Construction}

One idea of how to use classical codes for quantum codes was
introduced by Calderbank, Shor and Steane
$\cite{caldershor,steanecss}$. Therefore the construction is known as the CSS
construction.

\begin{theorem}
  Let $C_1=\ecc{n}{k_1}{d_1}_q$ and $C_2 = \ecc{n}{k_2}{d_2}_q$ be classical
  error correcting codes over $\F_q$ for which $C_1 \subseteq C_2^{\perp}$
  holds. Let $\G_1$ respectively $\G_2$ be their generator matrices.  Then $C$
  defined by the generator matrix
  \begin{eqnarray*}
    \G & = & \left(
    \begin{array}{c|c}
      \G_1 & 0\\ 0 & \G_2
    \end{array}
    \right)
  \end{eqnarray*}
  defines an $\qecc{n}{n-(k_1+k_2)}{\ge \min(d_1,d_2)}_q$ quantum
  error-correcting code $C$ over $\F_q$.
\end{theorem}

\begin{proof}

  The idea of this construction is that the condition $C_1 \subseteq
  C_2^{\perp}$ suffices to make the quantum code self-orthogonal with respect
  to the standard symplectic inner product. This condition is proved in the
  following.

  For $x,y \in C$, we have to check the different cases:
  \begin{eqnarray*}
    \sip{(x_1,\ldots,x_n,0,\ldots)}{(y_1,\ldots,y_n,0,\ldots)} & = &
    \sum_{i=1}^n x_i \cdot 0 - 0 \cdot y_i = 0\\
    \sip{(0,\ldots,x_{n+1},\ldots,
      x_{2n})}{(0,\ldots,y_{n+1},\ldots,y_{2n})} & = & \sum_{i=1}^n 0
    \cdot y_{n+i} - x_{n+i} \cdot 0  = 0\\
    \sip{(x_1,\ldots,x_n,0,\ldots)}{(0,\ldots,y_{n+1},\ldots,y_{2n})} 
    & = & \sum_{i=1}^n x_i y_{n+i} - 0 \cdot 0\\
    & = & \sum_{i=1}^n x_i y_{n+i} \stackrel{C_1 \subseteq \, C_2^{\perp}}{=} 0
  \end{eqnarray*}
  Therefore all codewords satisfy the symplectic inner product and $C$ is a
  stabilizer code.

  The properties of the quantum code are the following:
  \begin{itemize}
  \item The codeword length is $n$, because the generator matrix has length
  $2n$.
  \item $C_1$ has dimension $k_1$, $C_2$ has dimension $k_2$. Therefore $\G$
    has $k_1 + k_2$ rows. The dimension of the stabilizer code is $n$ minus
    the number of rows of the generator matrix and so $k = n - (k_1 + k_2)$
  \item The distance of the code cannot become smaller than the smallest one
    of the classical codes and therefore $d \ge \min(d_1,d_2)$.
  \end{itemize}
\end{proof}

\begin{example}
  Let $C_1$ be a classical $\ecc{3}{1}{3}_7$ code generated by
  \begin{eqnarray*}
    \G_1 & = & \left(
    \begin{array}{ccc}
      3&3&4
    \end{array}
    \right)
  \end{eqnarray*}
  and $C_2$ a classical $\ecc{3}{1}{3}_7$ code defined by
  \begin{eqnarray*}
    \G_2 & = & \left(
    \begin{array}{ccc}
      5&3&1
    \end{array}
    \right)
  \end{eqnarray*}
  over $\F_7$. Then $\Cperp_2$ is generated by
  \begin{displaymath}
    \left(
    \begin{array}{ccc}
      1&2&3\\2&1&1
    \end{array}
    \right),
  \end{displaymath}
  because 
  \begin{eqnarray*}
    \G_2 \cdot
    \left(
    \begin{array}{cc}
      1&2\\2&1\\3&1
    \end{array}
    \right)
    & = &
    \left(
    \begin{array}{ccc}
      5&3&1
    \end{array}
    \right) \cdot
    \left(
    \begin{array}{cc}
      1&2\\2&1\\3&1
    \end{array}
    \right)
     =  \left(
    \begin{array}{cc}
      0&0
    \end{array}
    \right).
  \end{eqnarray*}
  Since
  \begin{eqnarray*}
    (3,3,4) & = & (1,2,3) + (2,1,1) \in \Cperp_2,
  \end{eqnarray*}
  we have that $C_1 \subseteq \Cperp_2$ and we can apply the CSS
  construction. Hence we get a $\qecc{3}{1}{2}$ quantum code $C$ generated by
  the matrix
  \begin{eqnarray*}
    \G & = & \left(
    \begin{array}{ccc|ccc}
      3&3&4 & 0&0&0\\
      0&0&0 & 5&3&1
    \end{array}
    \right).
  \end{eqnarray*}
\end{example}

For later constructions we need a theorem of \cite{ashknill} for a special CSS
construction.  Since Matsumoto works in \cite{matsu} with $C \supseteq
\Cperps$ and not $C \subseteq \Cperps$, like usually done, the following
theorem is stated the other way round. In general, it does not matter, which
way around we construct it, because $(\Cperps)^{\perp_s} = C$. Then we just
have to use the dual code for constructions.

\begin{theorem}[\cite{ashknill}]
  \label{kdqecc}
  If there is an $(n+k)$-dimensional subspace $C \subseteq \F_{p^m}^{2n}$ such
  that $C \supseteq \Cperps$, then we can construct a $\floor{(d(C\backslash
  \Cperps)-1)/2}$-error-correcting quantum code $Q \subseteq \hilbert^{\tensor
  n}$ of dimension $(p^m)^k$, where
  \begin{eqnarray*}
    d(C\backslash \Cperps) & = & \min \set{wt(x) \; | \; x \in C\backslash
    \Cperps}.
  \end{eqnarray*}
\end{theorem}


\chapter{Algebraic Geometry}
\label{ag}

This chapter gives an overview over the basics of algebraic geometry which we
need for the later constructions of quantum error correcting codes. It is
rather technical but lays the foundation for a general method to generate
symplectic self-orthogonal codes over finite fields.

\section{Introduction to Algebraic Geometry}

The theory of algebraic geometry (AG) which includes algebraic curves and
algebraic function fields, is a very complex and technical, but powerful
theory. In this section we will give a brief introduction containing all main
results necessary for quantum AG codes. We use notations, definitions, and
theorems mainly from \cite{stichtenoth}, but also from \cite{s-selfdual},
\cite{gs-tower}, and \cite{sugiyama}.

\subsection{Plane Curves}

\begin{defn}
  A {\bf plane curve} over a field $K$ is the set
  \begin{displaymath}
    \Gamma(K) = \set{(x,y) \in K^2 \; | \; f(x,y) = 0} 
  \end{displaymath}
  where $f$ is a polynomial with coefficients in K. The elements of $\Gamma(K)$
  are called {\bf $K$-rational points} of $\Gamma$. A {\bf projective plane
  curve} $\hat{\Gamma}$ is defined by
  \begin{displaymath}
    \hat{\Gamma} = \hat{\Gamma}_f = \{ (X_0: Y_0: Z_0) \in
    \places^2(\complex) \; | \; f(X_0, Y_0, Z_0) = 0 \}
  \end{displaymath}
  where $f(X, Y, Z) \in \complex[X,Y,Z]$ is a homogeneous irreducible
  polynomial. (''Irreducible'' means that $f$ cannot be written as a product
  of two polynomials where each has degree $\ge 1$. $\complex[X,Y,Z]$ is the
  polynomial ring over the complex numbers.)
\end{defn}

\begin{example}
  Take $y^2 = x(x-1)(x+1)$, then $\Gamma_f = \{(x,y)\in \complex^2 \; | \;
  f(x,y) = x(x-1)(x+1)-y^2 = 0\}$. We cannot plot the graph of this curve, but
  if we take $\{(x,y)\in \mathbb{R}^2 \; | \; f(x,y) = 0\}$ instead of
  $\Gamma_f(\complex)$, we get the graph drawn in Figure \ref{ellcurve}.
  
  In the compactification the curve can be identified with a torus. The given
  curve is an example of an elliptic curve.

  \begin{figure}
    \centering
    \includegraphics[scale=0.6]{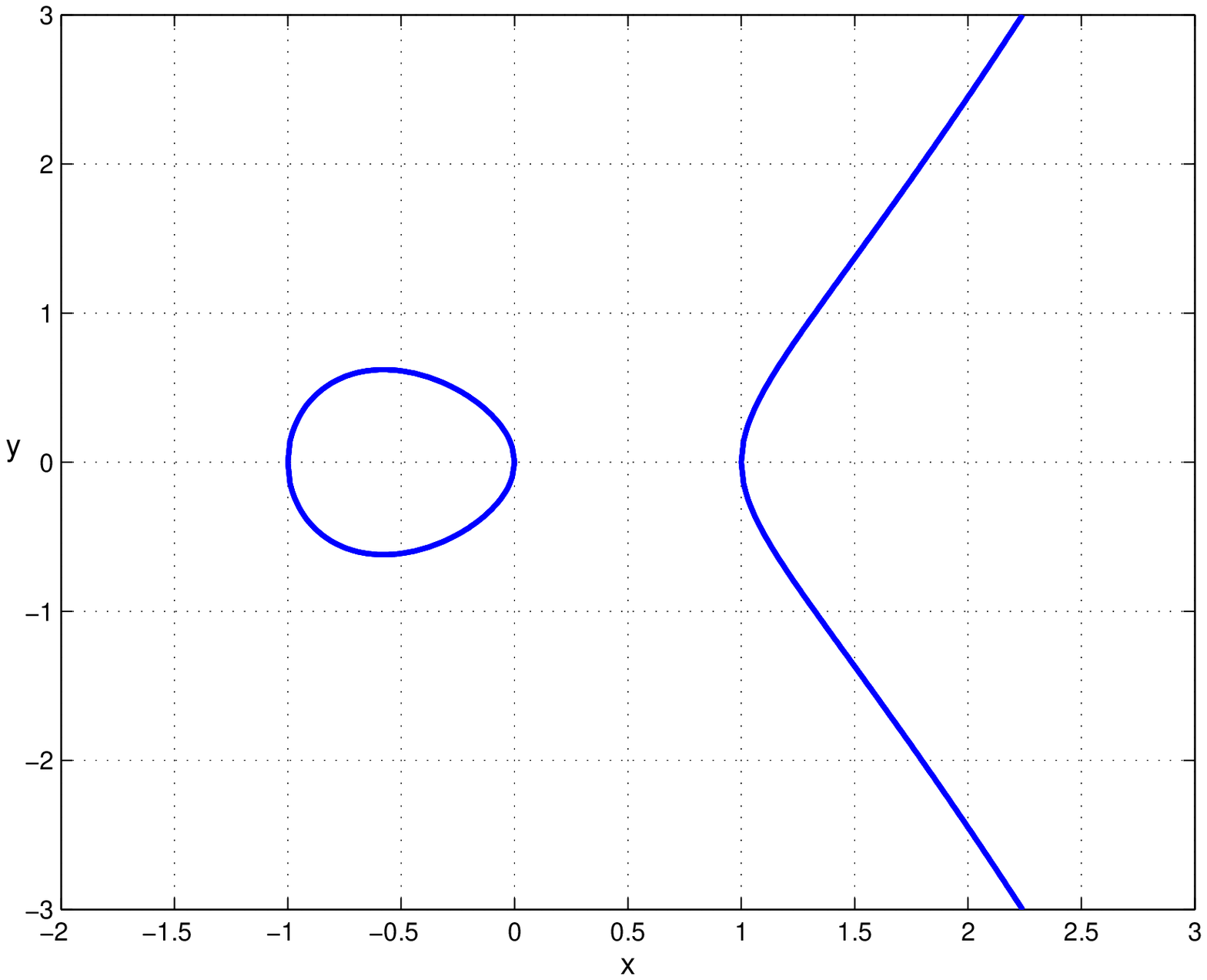}
    \caption{Example of the elliptic curve $y^2 = x(x-1)(x+1)$ over $\reals$.}
    \label{ellcurve}
  \end{figure}
\end{example}

\begin{note}
  In this example we see that it might be difficult to look at a curve over an
  arbitrary field, because we do not have a connected graph like in $\reals$
  or $\complex$. However, we can carry the essential properties of curves over
  to arbitrary fields. To this end we first establish an equivalence between
  algebraic curves and algebraic function fields in one variable. These fields
  are much easier to handle than the curves. Because of that, in most cases,
  the construction of codes will be over algebraic function fields, however we
  start the constructions with an algebraic curve.
\end{note}

\subsection{Coordinate Rings and Function Fields}

\begin{defn}\label{singular}
  A point on a projective curve $\hat{\Gamma} = \set{ (X,Y,Z) \; | \; f(X,Y,Z)
    = 0}$ is called {\bf singular}, if all partial derivations of $f$ are
    equal to zero. Otherwise it is called {\bf non-singular}.
\end{defn}

\begin{defn}
  Take an irreducible homogeneous polynomial
  \begin{displaymath}
    f(X,Y,Z) \in \K[X,Y,Z].
  \end{displaymath}
{\sloppy
  Two polynomials $g(X,Y,Z), h(X,Y,Z) \in \K[X,Y,Z]$ are called {\bf
    congruent modulo $f(X,Y,Z)$}, if there exists $q(X,Y,Z) \in
    \K[X,Y,Z]$ such that
  \begin{displaymath}
    g(X,Y,Z) - h(X,Y,Z) = q(X,Y,Z)f(X,Y,Z). 
  \end{displaymath}
  We denote this by $g(X,Y,Z) \equiv h(X,Y,Z) (\mod \, f(X,Y,Z))$.}
\end{defn}

\begin{defn}
  Let $\hat{\Gamma} = \{ (X:Y:Z) \in \places^2(\K) \; | \; f(X,Y,Z)
  = 0 \}$, where $f(X,Y,Z) \in \K[X,Y,Z]$ is an irreducible
  homogeneous polynomial. Then we call
  \begin{eqnarray*}
    R(\hat{\Gamma}) & = & \K[X,Y,Z]/ (f(X,Y,Z))
\end{eqnarray*}
the {\bf coordinate ring} of $\hat{\Gamma}$.  It can be split up into
``$d$-homogeneous'' parts $R_d$, i.e. $R(\hat{\Gamma})$ becomes a graded ring
\cite[I.1.5]{eisenbud} and can be written as
\begin{displaymath}
  R(\hat{\Gamma}) = \bigoplus_{d=0}^{\infty} R_d.
\end{displaymath}
Homogeneous polynomials of the same degree belong to the same $R_d$, e.g. $X$,
$Y$, and $X+Y \in R_1$.  If we take $\tilde{X}, \tilde{Y}, \tilde{Z}$ as new
variables of the congruence classes modulo $f(X,Y,Z)$, then we can denote
$R(\hat{\Gamma})$ by $\K[\tilde{X},\tilde{Y}, \tilde{Z}]$. Since
$R(\hat{\Gamma})$ is a graded ring, we can take the quotient field and define:
\end{defn}

\begin{defn}
  The {\bf function field}
  $\K(\hat{\Gamma})$ of a projective curve $\hat{\Gamma}$ is defined by
  \begin{eqnarray*}
    \K(\hat{\Gamma}) 
    & = & \Quot f(R(\hat{\Gamma}))\\
    & = & \set{
    \frac{g(\tilde{X},\tilde{Y},\tilde{Z})}{h(\tilde{X},\tilde{Y},\tilde{Z})}
    \; | \; g(\tilde{X},\tilde{Y},\tilde{Z}), h(\tilde{X},\tilde{Y},\tilde{Z})
    \in R_d, \; d \in \N, h \ne 0}.
  \end{eqnarray*}
\end{defn}

\begin{example}
  Let $f(X,Y,Z) = X^2 + Y^2 - Z^2$ be a homogeneous polynomial over the
  complex numbers. Then $f(X,Y,Z)$ defines a sphere in $\mathbb{P}^2$.

  The polynomials $g(X,Y,Z) =  X^2 + Y^2$ and $h(X,Y,Z) =  Z^2$ are congruent
  modulo $f(X,Y,Z)$, because
  \begin{eqnarray*}
    g(X,Y,Z) - h(X,Y,Z) 
    & = & X^2 + Y^2 - Z^2\\
    & = & 1 \cdot f(X,Y,Z).
  \end{eqnarray*}
  Hence $R(\hat{\Gamma}) = \complex[X,Y,Z]/(f(X,Y,Z))$.

\end{example}

\subsection{Valuation Rings}

The idea of valuations is to capture the multiplicity of zeros and poles of
functions. We will first look at an example which illustrates the meaning of
the term ``valuation''.

\begin{example}
  Let $\hat{\Gamma}$ be a projective plane curve. Take a point $P
  \in \hat{\Gamma}$ and define
  \begin{center}
    ${\cal{O}}_P(\hat{\Gamma}) = \{ \varphi \in \K(\hat{\Gamma}) \; | \;
    \varphi $ is defined at $ P \}$.
  \end{center} 
  If $P$ is non-singular (see Definition \ref{singular}) we have that $\K \;
  ^{\subset}_{\neq} \; {\cal{O}}_P(\hat{\Gamma}) \; ^{\subset}_{\neq} \;
  \K(\hat{\Gamma})$. The latter chain of inclusions is strict because
    \begin{itemize}
      \item The constants are defined at $P$, but there are functions which
	are not constant, e.g. those with zero in $P$, and are as well defined
	at $P$.
      \item But on the other hand, not all functions are defined at $P$
	because there exist functions that have a pole in $P$.
    \end{itemize}
    Furthermore, if $\varphi \in \K(\hat{\Gamma})$, then $ \varphi \in
    {\cal{O}}_P(\hat{\Gamma})$ or $\varphi^{-1} \in
    {\cal{O}}_P(\hat{\Gamma})$. Indeed, if a function is not defined at $P$,
    then it has a pole there. If a function has a pole at a certain point $P$,
    then its inverse has a zero at that $P$ and therefore its inverse is
    defined at $P$.
\end{example}

The previous example shows how to think of a valuation ring in terms of
algebraic curves. In the following we define a valuation ring in terms of
algebraic curves.

\begin{defn}
  A {\bf valuation ring} is a subring $ {\cal{O}} \subset
  \K(\hat{\Gamma})$ with the properties:
  \begin{enumerate}
  \item $\K \; ^{\subset}_{\neq} \; {\cal{O}} \; ^{\subset}_{\neq} \;
    \K(\hat{\Gamma})$
  \item If $\varphi \in \K(\hat{\Gamma})$, then $ \varphi \in {\cal{O}}$
    or $\varphi^{-1} \in {\cal{O}}$.
  \end{enumerate}
\end{defn}

We cite the following theorem from \cite[B.10]{stichtenoth}.
\begin{theorem}
  Let $\hat{\Gamma}$ be a non-singular projective plane curve. Then there
  exists a one-to-one correspondence between the points of $\hat{\Gamma}$ and
  the valuation rings of $\K(\hat{\Gamma})$ over an algebraically closed field
  $\K$:
  \begin{center}$
    \begin{array}{ccc}
      \{$points of $\hat{\Gamma} \} &
      \stackrel{1-1}{\longleftrightarrow} & \{$valuation rings
      $\K(\hat{\Gamma}) \}\\
      P & \longleftrightarrow  & {\cal{O}}_P(\hat{\Gamma})
    \end{array}$
  \end{center}
\end{theorem}

This theorem establishes a connection between the points on plane projective
curves and the valuation rings in function fields \cite[Thm. 1.1]{moreno}.

Using this correspondence of points on a curve and its relation to valuation
rings, we can look at a more abstract theory which leads to
divisors and the theorem of Riemann-Roch:

\begin{defn}
  \label{algff}
  An {\bf algebraic function field} of one variable $F/\K$ is a field extension
  of $\K$ such that $F$ is an algebraic extension of $\K(x)$ where $x \in F$ is
  transcendental over $K$. ( Recall that algebraic means that every $z \in F$
  is zero of a polynomial over $\K(x)$ \cite[Chapter V.1]{lang}. An element $x$
  is transcendental over $K$ if there exists no polynomial $f$ over
  $K$ such that $f(x) = 0$ \cite[Chapter II.3]{lang}.)
\end{defn}

\begin{defn}
  A {\bf valuation ring} of the function field $F/\K$ is a ring $\cal{O}$ with
  the following properties:
  \begin{enumerate}
  \item $K \; ^{\subset}_{\neq} \; {\cal{O}} \; ^{\subset}_{\neq} \; F$
  \item for any $z \in F$ we have that $z \in {\cal{O}}$ or $z^{-1} \in
  {\cal{O}}$.
  \end{enumerate}
\end{defn}

\begin{defn}
  A {\bf place} $P$ of the function field $F/K$ is the (unique) maximal ideal
  of a valuation ring ${\cal{O}}$ of $F/\K$. We define:
  \begin{center}
    $\places_F = \{ P \; | \; P$ is a place of $F/\K \}$
  \end{center}
\end{defn}

\begin{defn}
  Let $P$ be a place in $\places_F$. Let $t \in {\cal{O}}$ be such that $t
  \cdot {\cal{O}} = P$. For every $P \in \places_F$ we define a {\bf
  valuation} $v_P$ by
  \begin{displaymath}
    \begin{array}{rccl}
      v_P: & F & \rightarrow & \mathbb{Z} \cup \{ \infty \}\\
      & z & \mapsto & 
      \left\{
      \begin{array}{ll}
	n & $if $ z \neq 0, z = t^n u, u $ a unit$\\
	\infty & $if $ z = 0.
      \end{array}
      \right.
    \end{array}
  \end{displaymath}
\end{defn}

\begin{defn}
  $F_P := {\cal{O}}_P / P$ is called the {\bf residue class field} of $P$.
  The following map is called the {\bf residue map} with respect to $P$:
    \begin{displaymath}
      \begin{array}{ccl}
	F & \rightarrow & F_P \cup \{ \infty \}\\
	z & \mapsto & 
	\left\{
	\begin{array}{ll}
	  z(P) & $if $ z \in {\cal{O}}_P\\
	  z(P) = \infty & $otherwise.$ 
	\end{array}
	\right.
      \end{array}
    \end{displaymath}
\end{defn}

\begin{defn}
  \label{rationalplaces}
  The index $[F_P : \K]$ of the field extension $ F_P / \K$, $\K$ regarded as a
  subfield of $F_P$, is called the {\bf degree} of $P$. In symbols $\deg P =
  [F_P : \K]$.

  If $P$ is rational, i.e. $\deg P = 1$, then $[F_P:\K]=1$ and therefore $F_P =
  \K$, what is important for Goppa codes because then $F_P$ is not a base field
  extension and we get codewords over the field we started with.
\end{defn}

\begin{example}
  Let $\K = \F_2$, let $\K(x)$ be the rational function field, and observe that
  the polynomial $x+1$ has a zero at 1. Hence $\K[x]$ is a valuation ring if
  $x$ is not infinity. Hence $\K[x]/(x+1) \cong \F_2$
  because $x+1 = 0$ and therefore $x = -1 = 1$.

  If we have a polynomial
  \begin{eqnarray*}
    h(x) & = & (x+1)^k g(x)
  \end{eqnarray*}
  with $(x+1) \; \not| \; g(x)$, then
  \begin{eqnarray*}
    v_{(x+1)}(h(x)) & = & k.
  \end{eqnarray*}
\end{example}

\subsection{Divisors}

\begin{defn}
  A {\bf divisor} $D$ is a formal sum $ D = \sum_{P \in \places_F} n_P P$ with
  $n_P \in \mathbb{Z}$, where almost all\footnote{Recall that almost all means
  all but a finite number \cite[Chapter I.1]{lang}.} $n_P = 0$, $P \in
  \places_F$. A divisor of the form $D = P$ with $P \in \places_F$ is called a
  {\bf prime divisor}.  We add two divisors $ D = \sum_{P \in \places_F} n_P
  P$ and $ D^{\prime} = \sum_{P \in \places_F} n_P^{\prime} P$ as follows:
  \begin{eqnarray*}
    D + D^{\prime} & := & \sum_{P \in \places_F} (n_P + n_P^{\prime}) P
  \end{eqnarray*}
  The divisors form a group denoted by ${\cal{D}}_F = \{ A \; | \; A$ is
  divisor over $ F / K \}$. Furthermore, we define a partial order on the
  divisor group $\divgroup$ by
  \begin{eqnarray*}
    D \le D^{\prime} & :\Longleftrightarrow & v_P(D) \le v_P(D^{\prime})
    \qquad \forall P \in \places_F.
  \end{eqnarray*}
  The {\bf degree} of a divisor $D = \sum_{P \in \places_F} n_P P$ is
  given by the following formula:
  \begin{eqnarray*}
    \deg D & = & \sum_{P \in \places_F} n_P \deg P.
  \end{eqnarray*}
\end{defn}

\subsection{Theorem of Riemann-Roch}

The theorem of Riemann-Roch is an important result, since it establishes a
relation between the dimension (see Def. \ref{divisordimension}) and the
degree of a divisor. It will help us in the following chapters to estimate the
properties of our constructed codes.

\begin{defn}
  Let $ x \in F\backslash \{ 0 \}$. We define the {\bf principal divisor}
  $(x)$ of $x$ by
  \begin{displaymath}
    (x) := \sum_{P \in \places_F} v_P(x) P
  \end{displaymath}
  It is known that principal divisors have degree 0 \cite[Corollary
  I.4.11]{stichtenoth}. We define the {\bf pole divisor} $(x)_{\infty}$ by
  \begin{eqnarray*}
    (x)_{\infty} & := & - \sum_{P \in \places_F, \; v_P(x)<0} v_P(x) P
  \end{eqnarray*}
\end{defn}

\begin{example}
  Let
  \begin{eqnarray*}
    f(x) & = & \frac{(x+1)(x-2)}{(x^2+2)^2}
  \end{eqnarray*}
  be an element of $\reals(x)$. This function has roots in -1, 2 and a double
  root in infinity because it has a double pole at infinity in the enumerator
  and a fourfold pole in the denominator. The function has a double pole in
  $(x^2+2)$ because this polynomial is irreducible over the reals. Hence the
  principal divisor corresponding to $f(x)$ is given by
  \begin{eqnarray*}
    (f(x)) & = & P_{(x+1)} + P_{(x-2)} - 2 P_{\infty} - 2 P_{(x^2+2)} + 4
    P_{\infty}\\
    & = & P_{(x+1)} + P_{(x-2)} - 2 P_{(x^2+2)} + 2 P_{\infty}\\
    & = & P_{(x+1)} + P_{(x-2)} - 2 P_{(x^2+2)} + 2
    P_{\left(\frac{1}{x}\right)}
  \end{eqnarray*}
  where $(x+1)$, $(x-2)$, $(x^2+2)$, and $\left(\frac{1}{x}\right)$ are the
  irreducible factors over $\reals$ and $P_{(\cdot)}$ the corresponding
  places.

  Furthermore
  \begin{eqnarray*}
    \deg (f(x)) 
    & = & \deg(P_{(x+1)}) + \deg(P_{(x-2)}) - 2 \cdot \deg((x^2+2)) + 2 \cdot
    \deg(P_{\infty})\\
    & = & 1 + 1 - 2\cdot2 + 2 \cdot 1 \; = \; 0.
  \end{eqnarray*}
\end{example}

\begin{defn}
  For any $ A \in {\cal{D}}_F$ set
  \begin{eqnarray*}
    \codeL(A) & := & \set{ x \in F \; | \; (x) \ge - A} \cup \set{0},
  \end{eqnarray*}
  then $\codeL(A)$ is a vector space over $\K$ \cite[Lemma I.4.6]{stichtenoth}.
\end{defn}

\begin{example}
  Let $P_1, \, P_2, \, P_3, \, P_4$ be places of a function field and
  \begin{eqnarray*}
    A & = & 2 P_1 + P_2 - P_3
  \end{eqnarray*}
  a divisor, then the following $(x)$ are examples of principal divisors in
  $\codeL(A)$:
  \begin{eqnarray*}
    (x) & = & - 2 P_1 + 2 P_3,\\
    (x) & = & - P_1 - P_2 + P_3 + P_4,\\
    (x) & = & 0.
  \end{eqnarray*}
\end{example}

\begin{defn}
  \label{divisordimension}
  For $A \in {\cal D}_F$ we define $\dim A = \dim \codeL(A)$ and call it the
  {\bf dimension} of $A$, where $\dim \codeL(A)$ is the dimension of the
  vector space $\codeL(A)$ over $K$. The {\bf genus} $g$ of $F / \K$ is
  defined as:
  \begin{eqnarray*}
    g & := & \max \set{ \deg A - \dim A + 1 \; | \; A \in {\cal D}_F}.
  \end{eqnarray*}
  For a non-singular projective plane curve defined by the irreducible
  homogeneous polynomial $f(X,Y,Z)$ of degree $d$, we can calculate the genus
  by the following formula \cite[I.7.2]{hartshorne}
  \begin{eqnarray*}
    g & = & \frac{(d-1)(d-2)}{2}.
  \end{eqnarray*}
\end{defn}

We cite the following lemma from \cite[Cor. I.4.12]{stichtenoth}:
\begin{lemma} \label{degdim}
  Let $A$ be a divisor, then holds: If $\deg A < 0$ then $\dim A = 0$.
\end{lemma}

\begin{theorem}[Riemann-Roch {\cite[Thm. I.5.15]{stichtenoth}}]\label{rieroch}
  Let $F / \K$ be an algebraic function field in one variable.
  \begin{enumerate}
  \item For any $A \in {\cal{D}}_F$, $\dim A$ is finite.
  \item For any $A \in {\cal{D}}_F$ we have
    \begin{eqnarray*}
      \dim A & = & \deg A + 1 - g + \dim(W - A)
    \end{eqnarray*}
    for $g$ the genus of $F/\K$  and $W$ any divisor of degree $2g-2$  called
    a canonical divisor.
  \item If $A \in {\cal{D}}_F$ is of degree $\ge 2g-1$, then
    \begin{eqnarray*}
      \dim A & = & \deg A + 1 - g.
    \end{eqnarray*}
  \end{enumerate}
\end{theorem}

An important theorem for valuations and approximations is the following that
we cite from \cite[Theorem I.6.4]{stichtenoth}:

\begin{theorem}[Strong Approximation Theorem] 
  \label{strongapproxthm}
  Let $F/ \K$ be an algebraic function field in one variable. Let $S \;
  ^{\subset}_{\neq} \; \places_F$ be a proper subset of $\places_F$ and $ P_1,
  \ldots , P_r \in S$. Suppose we are given $x_1, \ldots , x_r \in F$ and $
  n_1, \ldots , n_r \in \mathbb{Z}$. Then there exists an element $x \in F$
  such that
  \begin{eqnarray*}
    v_{P_i} (x - x_i) & = & n_i \quad (i = 1, \ldots , r),\; \mbox{and}\\
    v_P(x) & \ge & 0 \quad \mbox{for all} \; P \in S \backslash \set{P_1,
    \ldots , P_r}.
  \end{eqnarray*}
\end{theorem}

\subsection{Differential Forms}

\begin{defn}
  The {\bf adele space} $\adelespace$ of $F/K$ is defined by 
  \begin{center}
    $\adelespace = \{ \alpha = (\alpha_P)_{P \in \places_F} \;| \; \alpha_P
      \in F $ and $ v_P (\alpha) := v_P (\alpha_P) \ge 0$ for almost all $P
      \in \places_F\}$.
  \end{center}

  Define also $ \adelespace (A) = \{ \alpha \in \adelespace \; | \; v_P
  (\alpha) \ge - v_P (A)$ for all $P \in \places_F\}$.
\end{defn}

\begin{defn}\label{weil}
  A {\bf Weil differential} $\eta$ of $F/ \K$ is a $K$-linear map $\eta:
  \adelespace \rightarrow K$ that vanishes on $\adelespace (A) + F$ for some
  divisor $A \in \divgroup$.  Define $\Omega_F$ to be the set of all Weil
  differentials and
  \begin{eqnarray*}
    \Omega_F (A) & := & \set{ \eta \in \Omega_F \; | \; \eta \hbox{ vanishes
	  on } \adelespace (A) + F}.
  \end{eqnarray*}

  For each $0 \neq \eta \in \Omega_F$ we can find a unique divisor $ W =
  (\eta) \in \divgroup$ called the {\bf canonical divisor} with the
  following properties \cite[Def. I.5.11]{stichtenoth}:
  \begin{enumerate}
  \item $\eta$ vanishes on $\adelespace(W)+F$.
  \item If $\eta$ vanishes on $\adelespace((A))+F$, then $ A \le (\eta)$.
  \end{enumerate}
  Canonical divisors have degree $2g-2$ where $g$ is the genus of $F/ \K$
  \cite[Corollary I.5.16]{stichtenoth}.
\end{defn}

\begin{defn}
  To define the {\bf residue} of a Weil differential $\eta \in \Omega_F$ we
  need the following: For a given place $P \in \places_F$, we take an element
  $t$ which is $P$-prime. Then for $z \in F$ , we can find $ a_i \in K$ and $n
  \in \mathbb{Z}$ such that
  \begin{eqnarray*}
    z & = & \sum\limits_{i=n}^{\infty} a_i t^i.
  \end{eqnarray*}
  This representation is unique and is called the {\bf P-adic power series
  expansion} and can be compared to a Laurent expansion in complex
  analysis. We define the {\bf residue} of $z$ with respect to $P$ and $t$ as
  \begin{eqnarray*}
    \res{{P,t}}{z} & := & a_{-1}.
  \end{eqnarray*}
  Note that $\Omega_F$ is a one-dimensional vector space over $F$ and every
  differential $\eta$ can be written in the form $\eta = z \, dt$
  \cite[Prop. I.5.9]{stichtenoth}. The {\bf residue} of a differential $\eta
  \in \Omega_F$ with $\eta = z \, dt$ is defined as
  \begin{eqnarray*}
    \res{P}{\eta} & := & \res{{P,t}}{z}.
  \end{eqnarray*}
\end{defn}

\begin{example}
  The theory of residues and differentials is equivalent to the one known from
  complex analysis if the field $\K = \complex$ \cite[Chapter
  IV]{stichtenoth}. To give an example of a residue, let
  \begin{eqnarray*}
    \eta & = & \frac{1}{x} \; dx
  \end{eqnarray*}
  be a differential over $\reals(x)$. Then
  \begin{eqnarray*}
    \res{P=1}{\eta} & = & \res{P=1,x}{\frac{1}{x}} = 1.
  \end{eqnarray*}
\end{example}

\subsection{Algebraic Field Extensions}
\label{algextensions}

To get good codes in the proceeding chapters, we need to modify the algebraic
function field and the places. This will be done by field extensions. In the
following we give an overview of extension fields and how to create a whole
sequence of extensions, called a tower of function fields, that can be used
for the construction of asymptotically good codes.

\begin{defn}
  An algebraic function field $F^{\prime} / K^{\prime}$ is called an {\bf
  algebraic extension} of an algebraic function field in one variable $F/K$,
  if $F^{\prime} \supseteq F$ is an algebraic field extension and $K^{\prime}
  \supseteq K$. It is said to be {\bf Galois} if for any $z \in F^{\prime}$
  the minimal polynomial $f(X) \in F[X]$ splits completely into different
  linear factors over $F^{\prime}$. For a Galois extension we define the {\bf
  Galois group} $\Gal(F^{\prime} / F)$ to be all automorphism on $F^{\prime}$
  that do not move elements of $F$. The {\bf fixed field} $F^{\sigma}$ of an
  automorphism $\sigma \in \Gal(F^{\prime} / F)$ is given by
  \begin{eqnarray*}
    F^{\sigma} & := & \set{z \in F^{\prime} | \sigma(z) = z}.
  \end{eqnarray*}
\end{defn}

\begin{defn}
  An {\bf Artin-Schreier Extension} is a field extension $F(\gamma) / F$,
  $\chara \, F = p$, where
  \begin{displaymath}
    \gamma^{p^m} - \gamma = c \in F, \quad \mathrm{and} \; c \ne \alpha^{p^m}
    - \alpha \quad \forall \, \alpha \in F.
  \end{displaymath}
\end{defn}

\begin{defn}
  Let $F^{\prime}/K^{\prime}$ be an algebraic extension of $F/K$. A place
  $P^{\prime} \in \places_{F^{\prime}}$ is {\bf lying over} $P \in \places_F$
  if $P = P^{\prime} \cap F$. In this case we write $P^{\prime} | P$. It is
  known that there exists an integer $e(P^{\prime} | P)$ such that
  \begin{eqnarray*}
    v_{P^{\prime}}(x) & = & e(P^{\prime} | P) \cdot v_P(x)
  \end{eqnarray*}
  for all $ x \in F$ \cite[Prop. III.1.4]{stichtenoth}. This number
  $e(P^{\prime} | P) \ge 1$ is called the {\bf ramification index} of $P$. The
  pair $P^{\prime} | P$ is said to be {\bf ramified} if $e(P^{\prime} | P) >
  1$, it is said to be {\bf unramified} if $e(P^{\prime} | P) = 1$. A place
  $P$ is {\bf totally ramified} if there is only one extension $P^{\prime}$ of
  $P$ with $e(P^{\prime} | P) = [F^{\prime}:F]$.
\end{defn}

We cite the following proposition from \cite[Prop. III.1.7]{stichtenoth}:
\begin{prop}
  Let $F^{\prime} / \K^{\prime}$ be an algebraic extension of the algebraic
  function field in one variable $F/ \K$. Any place $P \in \places_F$ has at
  least one, but only finitely many, extensions $P^{\prime} \in
  \places_{F^{\prime}}$
\end{prop}

\begin{theorem}[{\cite[Thm. III.1.11]{stichtenoth}}]
  Let $F^{\prime}/K^{\prime}$ be a finite\footnote{Finite extension means an
  algebraic extension of finite degree, i.e. $[F^{\prime} : F] < \infty$
  \cite[Def. III.1.1]{stichtenoth}.} extension of $F/K$, $P$ a place of $F/K$
  and $P_1,\ldots,P_m$ all the places of $F^{\prime}/K^{\prime}$ lying over
  $P$ with $f(P_i|P) = [F_{P_i}^{\prime}:F_P]$. Then
  \begin{eqnarray*}
    \sum_{i=1}^m e(P_i|P)\cdot f(P_i|P) & = & [F^{\prime}:F].
  \end{eqnarray*}
\end{theorem}

\begin{example}
  \label{exramification}
  This example shows how we can use the theorem above to calculate the
  possible number of places lying over a place $P$, if we know the degree of
  the field extension $F^{\prime}/F$.

  If $[F^{\prime}:F] = 2$, we have three different kinds of places $P$ in $F$:
  \begin{enumerate}
  \item $P$ is ramified, hence $e(P_i|P) > 1$ for all $P_i | P$. Therefore
    there exists only one place $P^{\prime}|P$ with
    \begin{eqnarray*}
      e(P^{\prime}|P) & = & 2\\
      f(P^{\prime}|P) & = & 1.
    \end{eqnarray*}
  \item $P$ is unramified, hence $e(P_i|P) = 1$ for all $P_i | P$. Therefore
    there exist two possibilities how to split 2:
    \begin{enumerate}
    \item There exists $P^{\prime}|P$ with $f(P^{\prime}|P) = 2$, then there
      is only one place lying over $P$ and
      \begin{eqnarray*}
	e(P^{\prime}|P) & = & 1\\
	f(P^{\prime}|P) & = & 2.
      \end{eqnarray*}
    \item For all places $P_i|P$ holds $f(P_i|P) = 1$, then there are
    two places $P_1$ and $P_2$ lying over $P$ with
      \begin{eqnarray*}
	e(P_i|P) & = & 1\\
	f(P_i|P) & = & 1
      \end{eqnarray*}
      and
      \begin{eqnarray*}
	e(P_1|P) \cdot f(P_1|P) + e(P_2|P) \cdot f(P_2|P) & = & 2 \quad = \quad
	[F^{\prime}:F].
      \end{eqnarray*}
    \end{enumerate}
  \end{enumerate}
\end{example}

\begin{theorem}\label{deddifferent}
  We define the {\bf different exponent} $d(P^{\prime}|P)$ of a place
  $P^{\prime}$ lying over $P$ according to {\it Dedekind's Different theorem}
  \cite[Thm. III.5.1]{stichtenoth}:
  \begin{enumerate}
  \item $ d(P^{\prime}|P) = e(P^{\prime}|P) - 1 \qquad \mathrm{iff} \; \chara
    \, K \not| \; e(P^{\prime}|P)$
  \item $ d(P^{\prime}|P) \ge e(P^{\prime}|P) - 1 \qquad \mathrm{otherwise}$
  \end{enumerate}
  For a more precise definition see
  \cite[Def. III.4.3]{stichtenoth}. Furthermore we define the {\bf different}
  \begin{eqnarray*}
    \Diff(F^{\prime} / F) & := & \sum_{P \in \places_F} \sum_{P^{\prime} | P}
    d(P^{\prime}|P) \cdot P^{\prime}.
  \end{eqnarray*}
\end{theorem}

\begin{example}
  If we continue with Example \ref{exramification} and assume that $\chara \;
  \K \ne 2$, we can calculate the different exponent explicitly with
  the help of Dedekind's different formula:
  \begin{eqnarray*}
    d(P^{\prime}|P) & = & 
    \left\{
    \begin{array}{ll}
      1 & {\rm if}\; P \;{\rm ramified}\\
      0 & {\rm otherwise}
    \end{array}
    \right.
  \end{eqnarray*}
  and the different of the field extension is given by
  \begin{eqnarray*}
    \Diff(F^{\prime} / F) & = & \sum_{P^{\prime} \; {\rm with} \;
    e(P^{\prime}|P)=1} 1 \cdot P^{\prime}
  \end{eqnarray*}
\end{example}

\subsection{Towers of Artin-Schreier Extensions}
\label{as-extensions}

This section presents some results of \cite{gs-tower} on Artin-Schreier
extensions.

\begin{defn}
  A {\bf tower} of Artin-Schreier extensions is a set of function fields
  \begin{displaymath}
    \F_{q^2} \subseteq F_1 \subseteq F_2 \subseteq F_3 \subseteq
    \cdots
  \end{displaymath}
  where $F_1 := \F_{q^2}(x_1)$ is the rational function field over
  $\F_{q^2}$, $q = p^m$, $p$ a prime, $m \ge 1$, and 
  \begin{eqnarray*}
    F_{n+1} & := & F_n(z_{n+1}).
  \end{eqnarray*}
  Here $z_{n+1}$ is defined by the equation
  \begin{eqnarray*}
    z_{n+1}^q + z_{n+1} & = & x_n^{q+1}
  \end{eqnarray*}
  and
  \begin{eqnarray*}
    x_n & := & \frac{z_n}{x_{n-1}} \in F_n, \quad n \ge 2.
  \end{eqnarray*}
\end{defn}

The goal of this section is to determine the ramification index of all the
places in every extension. To this end we have to define the following
hierarchy of sets.

\begin{defn}
  Let $Q_n \in \places_{F_n}$ be the unique place which is a common zero of
  the functions $x_1,\, z_2, \, \ldots, \, z_n$ \cite[Lemma 2.3]{gs-tower}.
  \begin{enumerate}
  \item For $n \ge 2$, let
    \begin{eqnarray*}
      S_0^{(n)} & := & \set{P \in \places_{F_n} | P \cap F_{n-1} = Q_{n-1}
      \quad \mathrm{and} \quad P \ne Q_n}
    \end{eqnarray*}
  \item For $1 \le i \le \lfloor \frac{n-3}{2} \rfloor$, let
    \begin{eqnarray*}
      S_i^{(n)} & := & \set{ P \in \places_{F_n} | P \cap F_{n-1} \in
      S_{i-1}^{(n-1)}}.
    \end{eqnarray*}
  \item Let $P_{\infty} \in \places_{F_1}$ denote the pole of $x_1$ in $F_1$,
    \begin{eqnarray*}
      S^{(1)} & := & \set{ P_{\infty} \} \; \mathrm{and} \; S^{(2)} := \{ P \in
      \places_{F_2} | P \in S_0^{(2)} \; \mathrm{or} \; P \cap F_1 \in
      S^{(1)}}
    \end{eqnarray*}
  \item For $n \ge 3$ and $n$ odd, define
    \begin{eqnarray*}
      S^{(n)} & := & \set{ P \in \places_{F_n} \; | \; P \cap F_{n-1} \in
      S^{(n-1)}},
    \end{eqnarray*}
    and for $n \ge 4$ and $n$ even,
    \begin{eqnarray*}
      S^{(n)} & := & \set{ P \in \places_{F_n} \; | \; P \cap F_{n-1} \in
      S^{(n-1)} \cup S_{\frac{n-4}{2}}^{(n-1)}}.
    \end{eqnarray*}
  \end{enumerate}
\end{defn}

\begin{figure}
  \centering
  \includegraphics[scale=0.5]{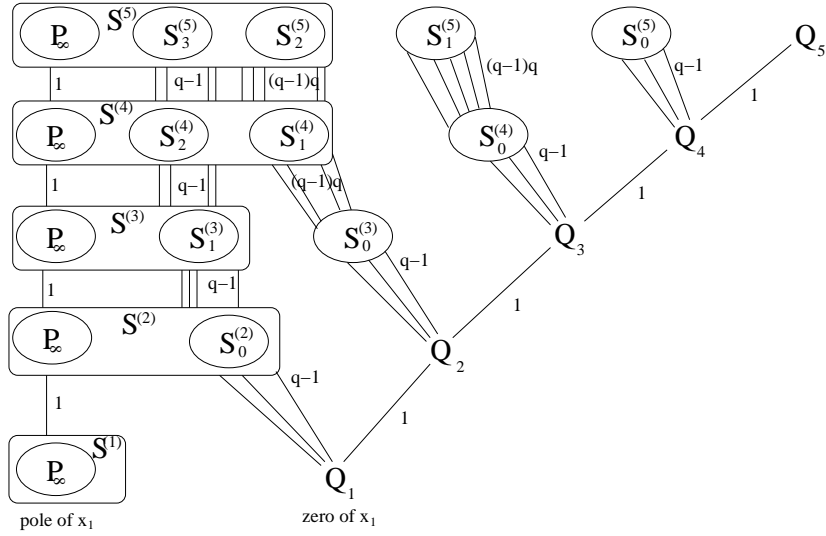}
  \caption{An example for a Garcia-Stichtenoth tower \cite{gs-tower}. The
    places lying over the zero and pole divisor of $x_1$ are shown. The pole
    divisor $P_{\infty}$ and all places in the boxes are totally ramified. The
    places $Q_i$ lying over the zero divisor are unramified and split
    completely over an algebraically closed field.}
  \label{as-figure}
\end{figure}

Figure \ref{as-figure} shows the structure of these sets and the number of
places lying over every place. This tower has the following properties.

\begin{lemma}[{\cite[Lemma 2.8]{gs-tower}}] \label{gstowerval}
  For $P \in S^{(n)}$, we have $v_P(x_n) = -1$.
\end{lemma}

\begin{lemma}[{\cite[after Lemma 2.9]{gs-tower}}] \label{gstowerdiff}
  The different exponent of a place $P^{\prime} \in \places_{F_{n+1}}$ lying
  over $P \in S^{(n)}$ is given by
  \begin{eqnarray*}
    d(P^{\prime}) & = & d(P^{\prime}|P) = (q-1)(q+2).
  \end{eqnarray*}
\end{lemma}

\begin{theorem}[{\cite[Thm. 2.10]{gs-tower}}] \label{gstowergen}
  The genus $g_n$ of $F_n$ is given by the following formula:
  \begin{eqnarray*}
    g_n = \left\{
    \begin{array}{lllll}
      q^n + q^{n-1} - q^{\frac{n+1}{2}} - 2 q^{\frac{n-1}{2}} + 1 &
      \mathrm{if} \; n & \equiv & 1 & \mod \; 2,\\
      q^n + q^{n-1} - \frac{1}{2} q^{\frac{n}{2}+1} - \frac{3}{2}
      q^{\frac{n}{2}}-  q^{\frac{n}{2}-1} + 1 &
      \mathrm{if} \; n & \equiv & 0 & \mod \; 2.\\
    \end{array}
    \right.
  \end{eqnarray*}
\end{theorem}

\begin{lemma}[{\cite[Lemma 3.A]{gs-tower}}] \label{gstowernpl}
  Let $P \in \places_{F_1}$ be the zero of $x_1 - \alpha$, with $0 \ne \alpha
  \in \F_{q^2}$. Then, the place $P$ splits completely in $F_n / F_1$:
  i.e., there are exactly $q^{n-1}$ places above $P$ in $\places_{F_n}$, all
  of them having degree one. Altogether, there are $(q^2 - 1)q^{n-1}$ places
  of this type in $\places_{F_n}$.
\end{lemma}

\subsection{Some more Galois Theory}
\label{theorems}
This section cites some important theorems from Galois theory which we will
need for the construction of quantum AG codes.

\begin{theorem}[{\cite[Thm. III.7.1]{stichtenoth}}] \label{galoistrans}
  Let $F^{\prime} / K^{\prime}$ be a Galois extension of $F/K$ and $P_1, \,
  P_2 \in \places_{F^{\prime}}$ be extensions of $P \in \places_F$. Then $P_2
  = \sigma(P_1)$ for some $\sigma \in \Gal(F^{\prime} / F)$. In other words,
  the Galois group acts transitively on the set of extensions of $P$.
\end{theorem}

\begin{prop}[{\cite[Prop. 1.1]{gs-tower}}] \label{galoisgroup1}
  Suppose that $F/K$ is an algebraic function field over $K = \F_{q^2}, \; q =
  p^n$ ( $K$ is algebraically closed in $F$). Let $w \in F$ and assume there
  exists a place $P \in \places_F$ such that
  \begin{eqnarray*}
    v_P (w) & = & -m, \; \mathrm{where} \; m > 0 \; \mathrm{and} \; \gcd(m,q)
    = 1,
  \end{eqnarray*}
  then the polynomial $T^q + T - w \in F[T]$ is absolutely irreducible (this
  follows e.g. from {\it Eisenstein's Criterion}).
\end{prop}

\begin{prop}[{\cite[Prop. III.7.10]{stichtenoth}}] \label{galoisgroup2}
  Consider an algebraic function field $F/ \K$ of characteristic $p > 0$. Let
  $w \in F$ and assume there exists a place $P \in \places_F$ such that
  \begin{eqnarray*}
    v_P (w) & = & -m, \; \mathrm{where} \; m > 0 \; \mathrm{and} \; \gcd(m,q)
    = 1.
  \end{eqnarray*}
  Let $F^{\prime} = F(z)$ with
  \begin{eqnarray*}
    z^q + z & = & w.
  \end{eqnarray*}
  Then $F^{\prime} / F$ is a Galois extension of degree $[F^{\prime} : F] =
  q$, and for the Galois group of $F^{\prime} / F$ we have $\Gal(F^{\prime} /
  F) \cong (\mathbb{Z} / p \mathbb{Z})^n$.
\end{prop}

\begin{lemma}[{\cite[Remark IV.3.7]{stichtenoth}}] \label{candivisor}
  For a canonical divisor $(z\,dx)$, the following formula holds:
  \begin{eqnarray*}
    (z \, dx) = (z) + (dx) & = & (z) - 2 (x)_{\infty} + \Diff(F/K(x))
  \end{eqnarray*}
  where $(x)_{\infty}$ denotes the pole divisor of $x$.
\end{lemma}

\section{Hyperelliptic Curves}


The main ideas in Chapter \ref{qechyper} use the properties of hyperelliptic
curves. Therefore we will spend one section to introduce this special kind of
algebraic varieties and Kummer extensions, i.e. algebraic curves. 

Throughout this section let $\K$ denote the finite field $\F_{p^m}$.

\begin{defn}
  A {\bf hyperelliptic curve} over a field $\K$ is the point at infinity and
  the set of solutions of the equation
  \begin{eqnarray*}
    y^2 & = & f(x)
  \end{eqnarray*}
  where $f(x)$ is a square-free polynomial of degree $\ge 5$. Similarly a {\bf
  projective hyperelliptic curve} over a field $\K$ is the set of solutions
  (including the point at infinity) of the equation
  \begin{eqnarray*}
    Y^2 \cdot Z^{d-2} & = & F(X,Z)
  \end{eqnarray*}
  where $F(X,Z)$ is a homogenous square-free polynomial of degree $d \ge 5$.
\end{defn}

\begin{defn}
  A {\bf hyperelliptic function field} over a field $K$ is an algebraic
  function field $F/K$ of genus $g \ge 2$ which contains a rational subfield
  $K(x) \subseteq F$ with $[F:K(x)] = 2$.
\end{defn}

The context of the following proposition \cite[Prop. VI.2.3]{stichtenoth} is
to show that these two definitions are equivalent and can be used alternatively.
\begin{prop}
  \label{hypercurvefield}
  {\rm
  Assume that $\chara K \ne 2$. Then the following statements hold:
  \begin{enumerate}
  \item Let $F/K$ be a hyperelliptic function field of genus $g$. Then there
    exist $x,y \in F$ such that $F = K(x,y)$ and
    \begin{eqnarray*}
      y^2 & = & f(x) \in K[x]
    \end{eqnarray*}
    with a square-free polynomial $f(x)$ of degree $2g+1$ or $2g+2$.
    
  \item Conversely, if $F = K(x,y)$ and $y^2 = f(x) \in K[x]$ with a
    square-free polynomial $f(x)$ of degree $m \ge 5$, then $F/K$ is
    hyperelliptic of genus
    \begin{eqnarray*}
      g & = &
      \left\{
      \begin{array}{ll}
	(m-1)/2 & {\rm if} \; m \equiv 1 \;\mod \; 2,\\
	(m-2)/2 & {\rm if} \; m \equiv 0 \;\mod \; 2.
      \end{array}
      \right.
    \end{eqnarray*}

  \item Let $F = K(x,y)$ with $y^2 = f(x)$ as above. Then the places $P \in
  \places_{K(x)}$ which ramify in $F/K(x)$ are the following:
  \begin{center}
    \begin{tabular}{l}
      all zeros of $f(x)$ if $\deg f(x) \equiv 0 \;\mod \; 2$,\\
      all zeros of $f(x)$ and the pole of $x$ if $\deg f(x) \equiv 1 \;\mod
      \;2$.
    \end{tabular}
  \end{center}
  In particular if $f(x)$ decomposes into linear factors, then exactly $2g+2$
  places of $K(x)$ are ramified in $F/K(x)$.
  \end{enumerate}
  }
\end{prop}

\begin{note}$ $
  \begin{itemize}
    
  \item Example \ref{exramification} can be applied to hyperelliptic curves,
    i.e. a hyperelliptic function field $F$ is a field extension of degree
    2. All places have ramification index 1 or 2. The places $P$ with
    $e(P^{\prime} | P) = 2$ are described in Proposition
    \ref{hypercurvefield}. As we work with $\chara F = p > 2$, it does not
    divide the ramification index and we can use Dedekind's different theorem
    and get the values $d(P^{\prime} | P) = e(P^{\prime} | P) - 1$ for the
    different index.

  \item Observe that hyperelliptic curves are symmetric: If $(\alpha,\beta)$
    is a point on the curve, then also $(\alpha,- \beta)$ is a point on the
    curve. This follows from the fact that there are a positive and a negative
    square root of an element.
  \end{itemize}
  
\end{note}

\begin{example}
  An illustration of a hyperelliptic curve over the reals is given in Figure
  \ref{hyperellcurve}. It shows a curve of genus $g = 3$ over the complex
  numbers, but plotted only over the reals. Topologically, a hyperelliptic
  curve over the complex numbers respectively over the algebraic closure is a
  torus with $g$ holes.

  \begin{figure}
    \centering
    \includegraphics[scale=0.6]{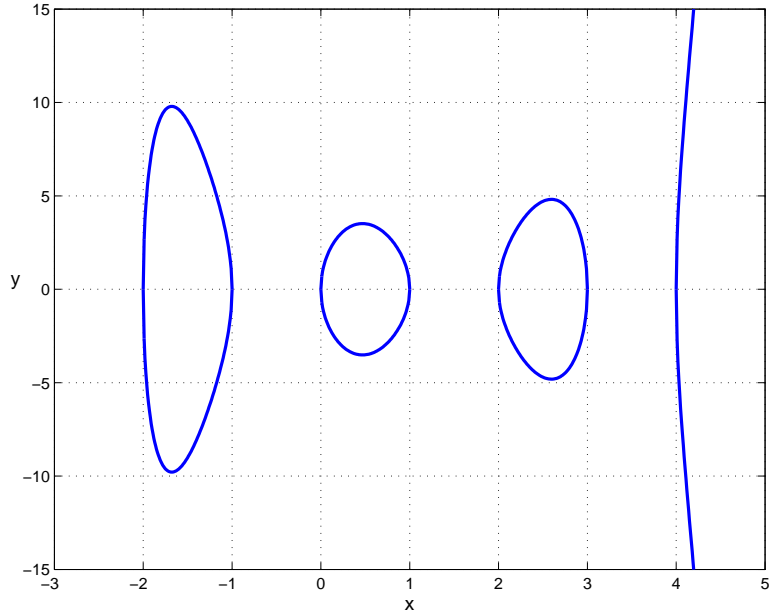}
    \caption{The hyperelliptic curve $y^2=x(x+1)(x-1)(x+2)(x-2)(x-3)(x-4)$}
    over $\reals$.
    \label{hyperellcurve}
  \end{figure}

\end{example}

The following lemma characterizes the splitting behaviour of the places of the
rational function field, i.e., if they split into two, are ramified or give a
constant field extension. This lemma makes use of Proposition
\ref{hypercurvefield}, Part 3. and expresses the same idea from a function
field point of view. Furthermore, it distinguishes between the cases
$f(P_i|P)=1$ and $f(P_i|P)=2$, presented in Example \ref{exramification}.

\begin{lemma}[{\cite[Lemma 1]{xing}}]
  \label{splitcases}
  If $P$ is a finite prime divisor and $p(x)$ the associated monic irreducible
  polynomial in $K[x]$, then exactly one of the following three cases holds:
  \begin{enumerate}
  \item If $p(x)|f(x)$, then $P$ is ramified.
  \item If \hspace{1mm} $p(x)\not|f(x)$ \hspace{1mm} and \hspace{1mm}
  $(f(x)/p(x))=1$, where $(\cdot/p(x))$ denotes the Jacobi symbol
  \cite{neukirch}, then $P$ splits.
  \item If $p(x)\not|f(x)$ and $(f(x)/p(x))=-1$, then $P$ is inert, i.e. a
  constant field extension.
  \end{enumerate}
\end{lemma}

\begin{lemma}
  \label{hyperconj}
  Every hyperelliptic curve respectively hyperelliptic function field has an
  bijective map $\sigma$ of order two called {\bf conjugation}, flipping the
  two places lying over one place $P$ of $K(x)$, if $P$ splits, otherwise the
  one place lying over $P$ is mapped to itself.
\end{lemma}

\begin{proof}
  If there are two places $P_1,\,P_2 \in \places_F$ lying over $P \in
  \places_{K(x)}$, Theorem \ref{galoistrans} states that there exists an
  isomorphism $\sigma$ with $\sigma(P_1) = P_2$ respectively $\sigma(P_2) =
  P_1$, because there are only two places lying over $P$. Therefore
  $\sigma(\sigma(P_1)) = \sigma(P_2) = P_1$ and $\sigma$ is of order 2. If
  there is just one place $P^{\prime}|P$, we have $\sigma(P^{\prime}) =
  P^{\prime}$, because elements of $K(x)$ are invariant under $\sigma$.
\end{proof}

\begin{figure}
  \centering
  \includegraphics[scale=0.6]{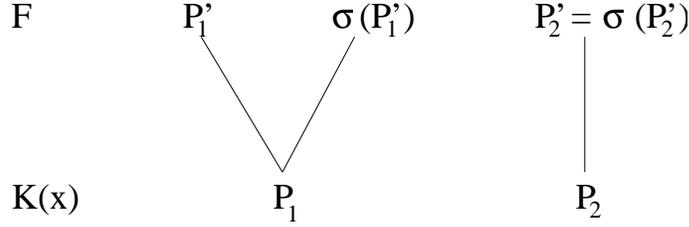}
  \caption{Splitting behaviour of places of the hyperelliptic function field
  $\ff$ lying over the rational function field $K(x)$.}
  \label{split}
\end{figure}

Figure \ref{hyperellcurve} shows that hyperelliptic curves are
symmetric. Pictorially, we see that the map $\sigma$ is given by
$\sigma(\alpha,\beta) = (\alpha,-\beta)$. Hence we have that $\sigma
(\sigma(\alpha,\beta)) = (\alpha,\beta)$ and therefore $\sigma$ is of order
two. Figure \ref{split} illustrates this.

\begin{example}
  \label{exsplit}
  The following Magma \cite{magma} calculations show how the rational places
  split or are ramified. We work over $\K = \F_{19}$ and use the equation
  \begin{eqnarray*}
    y^2 & = & (x-1)(x-2)(x-3)(x-4)(x-5) \in \F_{19}[x,y].
  \end{eqnarray*}
  We observe that the second component of each vector is either 0 or there are
  two vectors with the same first component and the sum of the second
  components is $0$.
  
  \begin{verbatim}
    F<w> := GF(19);
    P2<x,y,z> := ProjectiveSpace(F,2);
    //Galois field + projective space
    f := 18*y^2*z^3 + (x-z)*(x-2*z)*(x-3*z)*(x-4*z)*(x-5*z);
    
    X := Curve(P2,f);
    g := Genus(X);
    //construction of a curve X corresponding to f

    > Places(X,1); //rational places of X;
    [
    Place at (0 : 1 : 0),
    Place at (2 : 0 : 1),
    Place at (4 : 0 : 1),
    Place at (16 : 14 : 1),
    Place at (16 : 5 : 1),
    Place at (7 : 13 : 1),
    Place at (7 : 6 : 1),
    Place at (17 : 11 : 1),
    Place at (17 : 8 : 1),
    Place at (15 : 17 : 1),
    Place at (15 : 2 : 1),
    Place at (11 : 12 : 1),
    Place at (11 : 7 : 1),
    Place at (3 : 0 : 1),
    Place at (6 : 14 : 1),
    Place at (6 : 5 : 1),
    Place at (12 : 13 : 1),
    Place at (12 : 6 : 1),
    Place at (5 : 0 : 1),
    Place at (1 : 0 : 1)
    ]
  \end{verbatim}
  In this list Place at $(x:y:z)$ means the point in ${\mathbb P}^2$ in
  homogeneous coordinates. Hence the point $(0:1:0)$ defines the place at
  infinity.
\end{example}

\section{Goppa Codes}
\label{goppa}

\subsection{Standard Goppa Codes} 

This section gives the basic ideas of Goppa codes. The definitions and
theorems are taken from \cite{popp}, \cite{stichtenoth}, and
\cite{s-selfdual}. In the following, let $\F_q$ be a finite field with
$q=p^m$, let $p$ be a prime number, and let $F/\F_q$ be an algebraic function
field.

\begin{defn}
  Let $F/\F_q$ be an algebraic function field in one variable. Let $P_1,
  \ldots, P_n$ be places of degree one and let $D = P_1 + \cdots +
  P_n$. Furthermore let $G$ be a divisor with $\supp(G) \cap \supp(D) =
  \emptyset$. Then the {\bf Goppa code} (respectively {\bf AG code}) $C_\codeL
  \subseteq \F_q^n$ is defined by
  \begin{eqnarray*}
    C_\codeL(D,G) & = & \set{ (f(P_1), \ldots , f(P_n)) \; | \; f \in
    \codeL(G) } \subseteq \F_q^n
  \end{eqnarray*} 
  Define the following linear {\bf evaluation map}
  \begin{eqnarray*}
    \varphi: & &
    \left\{
    \begin{array}{crl}
      \codeL(G) & \rightarrow & \F_q^n\\
      f & \mapsto & (f(P_1), \ldots, f(P_n)).
    \end{array}
    \right.
  \end{eqnarray*}
  Then the Goppa Code is given by $C_\codeL(D,G) = \varphi (\codeL(G))$.
\end{defn}

\begin{theorem}[{\cite[Thm. II.2.2]{stichtenoth}}]
  \label{goppaparam}
  The code $C_\codeL(D,G)$ is a linear $\ecc{n}{k}{d}$ code with parameters
  \begin{eqnarray*}
    k & = & \dim G - \dim (G-D)\\
    d & \ge & n - \deg G =: d_{des}.
  \end{eqnarray*}
  The parameter $d_{des}$ is called the {\bf designed distance} of the Goppa
  code.
\end{theorem}

\begin{corollary}[{\cite[Thm. II.2.3]{stichtenoth}}]
  Assume $\deg G < n$ and let $g$ be the genus of $F/\F_q$. Then we have:
  \begin{enumerate}
  \item $ \varphi: \codeL(G) \rightarrow C_\codeL(D,G)$ is injective and
    $C_\codeL(D,G)$ is an $[n,k,d]$ code with
    \begin{eqnarray*}
	k & = & \dim G \ge \deg G + 1 - g\\
	d & \ge & n - \deg G.
    \end{eqnarray*}
  \item If in addition $2g-2 < \deg G < n$, then
    \begin{eqnarray*}
      k  & = & \deg G + 1 - g.
    \end{eqnarray*}
  \item If $(f_1, \ldots, f_k)$ is a basis of $\codeL(G)$, then
    \begin{eqnarray*}
      M & = &
      \left(
      \begin{array}{ccc}
	f_1(P_1) & \cdots & f_1(P_n)\\
	\vdots   &        & \vdots\\
	f_k(P_1) & \cdots & f_k(P_n)\\	
      \end{array}
      \right)
    \end{eqnarray*}
    is a {\it generator matrix} for $C_{\cal L}(D,G)$.
  \end{enumerate}
\end{corollary}

\begin{example}
  \label{exgoppa}
  This example was created with Magma \cite{magma} and uses the curve
  constructed in Example \ref{exsplit}. We use only the pairs of rational
  points. Therefore we will be able to use this example for a quantum code
  construction.

  In the following we give an example of a Goppa code in Magma:
  \begin{verbatim}
    //Function to construct a finite field and a projective space
    constr_field := function(q)
    local F, P2;
      F<w> := GF(q);
      P2<x,y,z> := ProjectiveSpace(F,2);
    return F, P2;    
    end function;
    
    //Function to construct a curve
    constr_curve := function(P2, f)
    local X, g;
      X := Curve(P2,f);
      g := Genus(X);
    return X,g;
    end function;



    //Example of the construction of a Goppa code
    //Construction of the finite field with 19 elements
    //and the corresponding projective space
    F<w>,P2<x,y,z> := constr_field(19);
    //Definition of a polynomial f
    f := 18*y^2*z^3 + (x-z)*(x-2*z)*(x-3*z)*(x-4*z)*(x-5*z);
    
    //Construction of the curve X corresponding to f
    X,g := constr_curve(P2,f);
    //Divisor group to X
    DG := DivisorGroup(X);
    //Rational places of X
    place1 := Places(X,1);
    //Function field corresponding to X
    F<a,b> := FunctionField(X);
    
    //Exclusion of the place at infinity 
    //and the non splitting place
    D := Exclude(place1,place1[1]); Exclude(~D,place1[14]);
    //Generation of the divisor of rational places for the code
    D3 := &+[DG!D[i]: i in [3..16]];
    
    //Divisor G for the generation of codewords is an 
    //evaluation at the place at infinity
    G := 7* DG!place1[1];
    C := AlgebraicGeometricCode(D,G);
  \end{verbatim}
  This construction leads to the following code:
  \begin{verbatim}
    > C;
    [14, 6] Linear Code over GF(19)
    Generator matrix:
    [ 1  0  0  0  0 13  0 14 15 11 17 15  4 16]
    [ 0  1  0  0  0  6  0  5  6 10 13 15 13  1]
    [ 0  0  1  0  0 10  0  7 12 11  2  6  6 14]
    [ 0  0  0  1  0  9  0 12  4  5 16 12  2 13]
    [ 0  0  0  0  1  1  0  0  4  4  5  5  3  3]
    [ 0  0  0  0  0  0  1  1 17 17  5  5 11 11]
  \end{verbatim}

\end{example}

\begin{note}
  To characterize the dual code of a Goppa code we need to look at the
  original definitions of Goppa by means of differential forms and its
  relations to the code defined above.
\end{note}

\begin{defn}
  Let $D = P_1 + \cdots + P_n$ be a divisor, where the $P_i$'s are places of
  degree one of an algebraic function field $F/\F_q$, Furthermore let $G$ be a
  divisor with $supp(G) \cap supp(D) = \emptyset$. Then we define the code
  $C_{\Omega}(D,G)$ by
  \begin{eqnarray*}
  C_{\Omega}(D,G) & := & \set{ (\res{P_1}{\omega}, \ldots, \res{P_n}{\omega})
  \; | \; \omega \in \Omega_F(G-D)} \subseteq \F_q^n.
  \end{eqnarray*}
\end{defn}

\begin{prop}[{\cite[Thm. 2.4 and 2.5]{s-selfdual}}] \label{dualcode}
  The code $C_{\Omega}(D,G)$, where $D$ and $G$ are as above has the following
  properties:
  \begin{enumerate}
  \item $C_\codeL(D,G)^{\perp} = C_{\Omega}(D,G)$.
  \item $C_{\Omega}(D,G) = a \cdot C_\codeL(D,H)$ with $H = D - G + (\eta)$
    where $\eta$ is a differential, $v_{P_i}(\eta) = -1$ for $i = 1,
    \ldots, n$, and $a = ( \res{P_1}{\eta}, \ldots, \res{P_n}{\eta})$.
  \item $C_{\cal L}(D,G)^{\perp} = a \cdot C_\codeL(D,H)$.
  \end{enumerate}
\end{prop}

The following proposition is cited from \cite[Prop. VII.1.2]{stichtenoth}. It
allows to construct differentials with special properties that help to
construct a self-orthogonal code.

\begin{prop} \label{constrdiff}
  Let $x$ and $y$ be elements of $F$ such that $v_{P_i}(y) = 1$, $v_{P_i}(x) =
  0$ and $x(P_i) = 1$ for $i = 1, \ldots, n$. Then the differential $\eta := x
  \cdot \frac{dy}{y}$ satisfies $v_{P_i}(\eta) = -1$ and $\res{P_i}{\eta} = 1$
  for $i = 1, \ldots, n$.
\end{prop}

\subsection{Weighted Self-Orthogonal Goppa Codes}


Chapter \ref{qecc} shows that quantum codes have to satisfy certain
self-orthogonality properties. In order to find codes satisfying them, we will
have to work with some codes which are what we call ``weighted
self-orthogonal''.

\begin{defn}
  We call a Goppa code {\bf weighted self-orthogonal} if it satisfies
  \begin{eqnarray*}
    C & \subseteq & C^{\perp^a}
  \end{eqnarray*}
  with respect to the inner product
  \begin{eqnarray*}
    \ipa{x}{y} & := & \sum_{i=1}^n a_i \; x_i \, y_i,
  \end{eqnarray*}
  where $a = (a_i)^n \in \F_q^n$.
\end{defn}

\begin{corollary}
  $\ipa{x}{y} := \sum_{i=1}^n a_i \; x_i \, y_i$ defines an inner product over
  $\F_q^n$.
\end{corollary}

\begin{proof}
We have to show bilinearity and symmetry to proof this corollary. Bilinearity
is shown in the following.
\begin{eqnarray*}
  \ipa{\lambda x+z}{y} 
  & = & \sum_{i=1}^n a_i \; (\lambda x_i+z_i) \, y_i\\
  & = & \lambda \sum_{i=1}^n a_i \; x_i \, y_i + \sum_{i=1}^n a_i \; z_i
  \, y_i\\
  & = & \lambda \ipa{x}{y} + \ipa{z}{y},
\end{eqnarray*}
for $\lambda \in \F_q$ and $x,y,z \in \F_q^n$.
Symmetry is given by
    \begin{eqnarray*}
      \ipa{x}{y}
      & = & \sum_{i=1}^n a_i \; x_i \, y_i\\
      & = & \sum_{i=1}^n a_i \; y_i \, x_i\\
      & = & \ipa{y}{x}.
    \end{eqnarray*}

\end{proof}

\begin{corollary}
  Similarly, $\sipa{x}{y} = \sum_{i=1}^n a_i \; (x_i y_{n+i} - x_{n+i} y_i)$
  defines a symplectic inner product over $\F_q^{2n}$.
\end{corollary}

\begin{proof}
We have to show that $\sipa{x}{y}$ is bilinear, anti-symmetric and that every
vector is self-orthogonal. First we show bilinearity in the following.
\begin{eqnarray*}
  \sipa{\lambda x+z}{y} 
  & = & \sum_{i=1}^n a_i \; ((\lambda x_i+z_i) y_{n+i} -  (\lambda
  x_{n+i}+z_{n+i})\, y_i)\\
  & = & \lambda \sum_{i=1}^n a_i \; (x_i y_{n+i} - x_{n+i} y_i) +
  \sum_{i=1}^n a_i \;(z_i y_{n+i} - z_{n+i} y_i)\\
  & = & \lambda \sipa{x}{y} + \sipa{z}{y}
\end{eqnarray*}
for $\lambda \in \F_q$ and $x,y,z \in \F_q^{2n}$. The form is anti-symmetric:
\begin{eqnarray*}
  \sipa{x}{y}
  & = & \sum_{i=1}^n a_i \; (x_i y_{n+i} - x_{n+i} y_i)\\
  & = & - \sum_{i=1}^n a_i \; (y_i x_{n+i} - y_{n+i} x_i)\\
  & = & - \sipa{y}{x}.
\end{eqnarray*}
And every vector has length zero:
\begin{eqnarray*}
  \sipa{x}{x}
  & = & \sum_{i=1}^n a_i \; (x_i x_{n+i} - x_{n+i} x_i)\\
  & = & \sum_{i=1}^n a_i \; (x_i x_{n+i} - x_i x_{n+i})\\
  & = & 0.
\end{eqnarray*}
\end{proof}

\vspace{0.5cm}

This concludes the chapter about basics in algebraic geometry in which we have
established a
sufficient background to construct quantum codes over algebraic curves.

\chapter{Good Binary Quantum Goppa Codes}
\label{matsumoto}

In this chapter we give a detailed account of {\it R. Matsumoto's} paper
{\it ``Algebraic geometric construction of a quantum
stabilizer code''} \cite{matsu} and an explicit description of
his code construction.

Throughout this chapter let $\F_q$ be a finite field with $q=p^m$ for some
prime number $p$. In the explicit construction of a code we use a binary
field, i.e. $q=2^m$, but most of the theorems also hold for non-binary fields
and can be used in Chapter \ref{qechyper}. Therefore we state them in the more
general case.

\section{Existence and Decoding of Quantum Codes}

First, we show under which circumstances a quantum stabilizer code can be
obtained from an algebraic geometric construction. Subsequently we present a
method how decoding and error correction can be implemented for these codes.

\subsection{Existence and Encoding}

\begin{prop} \label{matsuI}
  Let $F/\F_q$ be an algebraic function field of one variable, $\sigma$ an
  automorphism of order 2 of $F$ which leaves $\F_q$ invariant, and $P_1,
  \ldots ,P_n$ pairwise distinct places of degree one such that $\sigma P_i
  \neq P_j, \; \forall \, i,j = 1, \dots , n$. Let $\eta$ be a differential
  with the following properties:
  \begin{equation}\label{matsuIeq}
    \left\{ \begin{array}{l} 
      v_{P_i}(\eta) = v_{\sigma P_i}(\eta) = -1,\\
      res_{P_i}(\eta) = 1,\\
      res_{\sigma P_i}(\eta) = -1.
    \end{array}
    \right.
  \end{equation}
  The existence of such $\eta$ is guaranteed by the strong approximation
  theorem of discrete valuations (see Theorem \ref{strongapproxthm}). Further
  assume that we have a divisor $G$ such that $\sigma G = G$, $ v_{P_i}(G) =
  v_{\sigma P_i}(G) = 0$ for all $i$. Define
  \begin{eqnarray*}
    C(G) & = & \set{ (f(P_1), \ldots , f(P_n), f(\sigma P_1), \ldots ,
      f(\sigma P_n)) \; | \; f \in {\cal{L}}(G)} \subseteq \F_q^{2n}.
  \end{eqnarray*}
  Let
  \begin{eqnarray*}
     H & = & (P_1 + \cdots + P_n + \sigma P_1 + \cdots + \sigma P_n) - G +
    (\eta),
  \end{eqnarray*}
  where $\eta$ is as in Equation (\ref{matsuIeq}).  Then we have
  $C(G)^{\perp_s} = C(H)$
\end{prop}

\begin{proof}
  Let $x = (x_1, \ldots, x_{2n}) \in \F_q^{2n}$ and
  $y = (y_1, \ldots, y_{2n}) \in \F_q^{2n}$.

  \renewcommand{\labelenumi}{(\roman{enumi})}
  \begin{enumerate}
  \item Note that $\sigma(\sigma(f)) = f$, because $\sigma$ is of order 2.
    
  \item Let $\codeL^{\sigma}(G) := \set{\sigma(x) \in F \; | \; (x) \ge -G}
    \cup \set{0}$. As $\sigma G = G$, we get for $f \in \codeL(G)$:
    \begin{displaymath}
      \sigma(f) \in \codeL^{\sigma}(G)
    \end{displaymath}
    so
    \begin{eqnarray*}
      \sigma(f) \in \codeL(\sigma G) & = & \codeL(G).
    \end{eqnarray*}
    
  \item This gives us for $(x_1, \ldots, x_{2n}) \in C(G)$
    \begin{displaymath}
      \begin{array}{cl}
	& (x_1, \ldots, x_{2n}) \in C(G)\\ 
	\Longleftrightarrow & \exists \; f
	\in \codeL(G) \hbox{ with }\\
	& (x_1, \ldots, x_{2n}) = (f(P_1), \ldots , f(P_n), f(\sigma P_1),
	\ldots , f(\sigma P_n)) \in C(G).
      \end{array}
    \end{displaymath}
    Remark (ii) gives the equivalence to
    \begin{eqnarray*}
      (\sigma(f)(P_1), \ldots , \sigma(f)(P_n), \sigma(f)(\sigma P_1), \ldots,
	\sigma(f)(\sigma P_n)) \in C(G).
    \end{eqnarray*}
    With \cite[\rm Prop. VII.3.3]{stichtenoth} we can rewrite this expression
    as
    \begin{eqnarray*}
      (f(\sigma P_1), \ldots , f(\sigma P_n), f(\sigma \sigma P_1), \ldots,
	f(\sigma \sigma P_n)) \in C(G).
    \end{eqnarray*}
    Finally Remark (i) gives equivalence to 
    \begin{eqnarray*}
      &&(f(\sigma P_1), \ldots , f(\sigma P_n), f(P_1), \ldots, f(P_n)) \in
	C(G)\\
	&\Longleftrightarrow & (x_{n+1}, \ldots, x_{2n}, x_1, \ldots,
	x_n) \in C(G).
    \end{eqnarray*}

  \item With help of this observation, we can show that $C(G)^{\perp_s} =
    C(H)$:
    
    Let $x \in C(H)$. Then with Proposition \ref{dualcode} we get for all $y
    \in C(G)$:
    \begin{eqnarray*}
	  0
	  & = & \sum\limits_{i=1}^{n} res_{P_i}(\eta) \; x_i y_i +
	  \sum\limits_{i=n+1}^{2n} res_{\sigma P_i}(\eta) \; x_i y_i\\
	  &  = & \sum\limits_{i=1}^n 1 \cdot x_i y_i + \sum\limits_{i=n+1}^{2n}
	  (-1) \cdot x_i y_i\\
	  & = & \sum\limits_{i=1}^n x_i y_i - \sum\limits_{i=n+1}^{2n} x_i y_i.
      \end{eqnarray*}
    With (iii) this is equivalent to that for all $y \in C(G)$
    \begin{eqnarray*}
      \sum\limits_{i=1}^n x_i y_{n+i} - \sum\limits_{i=1}^{n} x_{n+i} y_i & =
      & 0, 
    \end{eqnarray*}
    and therefore $x \in C(G)^{\perp_s}$.  
  \end{enumerate}

  Hence by choosing an automorphism of order 2 and suitable residues, we can
  render the identity $C(G)^{\perp} = a \cdot \, C(H)$, which we know from
  classical coding theory into the identity $C(G)^{\perp_s} = C(H)$.
\end{proof}

This proposition tells us that the orthogonal code of $C(G)$ is generated by
$H$. The following corollary is a result of the combination of
Theorem \ref{goppaparam} and Proposition \ref{matsuI} above and gives us the
wanted quantum code:

\begin{corollary} \label{matsuII}
  We use the same notations as in Proposition \ref{matsuI}. Furthermore, we
  assume that $G \ge H$. Then we can construct an $[[n,k,d]]$ quantum code
  $Q$, where
  \begin{eqnarray*}
    k & = & \dim G - \dim (G - P_1 - \cdots - P_n - \sigma P_1 - \cdots -
    \sigma P_n) - n.
  \end{eqnarray*}
  For the minimum distance $d$ of $Q$, we have
  \begin{eqnarray*}
    d & \ge & n - \left\lfloor \frac{\deg G}{2} \right\rfloor.
  \end{eqnarray*}
\end{corollary}

\begin{proof}
  
  Theorem \ref{kdqecc} and Proposition \ref{matsuI} show that we can construct
  a quantum stabilizer code from $C(D,G)$, because
  \begin{eqnarray*}
    && G \ge H\\
    &\Leftrightarrow & \codeL(G) \supseteq \codeL(H)\\
    & \Leftrightarrow & C(D,G)  \supseteq C(D,H) = C(D,G)^{\perp_s}.
  \end{eqnarray*}
  By Theorem \ref{kdqecc} we get $k = \dim C(D,G) - n$.
  Then Theorem \ref{goppaparam} implies
  \begin{eqnarray*}
    \dim C(D,G) & = & \dim G - \dim (G - P_1 - \cdots - P_n - \sigma P_1 -
    \cdots - \sigma P_n)
  \end{eqnarray*}
  and the statement for the dimension of the code $k$ is proven.

  The next step is to prove the stated bound on the minimum distance
  $d$. Suppose that
  \begin{eqnarray*}
    \weight((f(P_1),\ldots, f(\sigma P_n))) & = & \delta \ne 0
  \end{eqnarray*}
  for $f \in \codeL(G)$. Then there exists a set $\set{i_1,\ldots,
  i_{n-\delta}}$ such that
  \begin{eqnarray*}
    f(P_{i_1}) = f(\sigma P_{i_1}) = \cdots =
  f(P_{i_{n-\delta}}) = f(\sigma P_{i_{n-\delta}}) = 0,
  \end{eqnarray*}
  which implies that $f \in \codeL(G-\sum_{j=1}^{n-\delta} (P_{i_j} + \sigma
  P_{i_j}))$. Since $f \ne 0$, we have
  \begin{eqnarray*}
    && \dim (G - \sum_{j=1}^{n-\delta} (P_{i_j} + \sigma P_{i_j})) > 0 \\
    & \stackrel{{\rm Cor. \; \ref{degdim}}}{\Longrightarrow} & \deg (G -
    \sum_{j=1}^{n-\delta} (P_{i_j} + \sigma P_{i_j})) \ge 0\\
    & \Longleftrightarrow & \deg G - 2(n-\delta) \ge 0\\
    & \Longleftrightarrow & 2 \delta \ge 2n - \deg G\\
    & \Longleftrightarrow & \delta \ge n - \floor{\frac{\deg G}{2}}.
  \end{eqnarray*}
\end{proof}

As in classical AG codes, this construction provides good codes only when $q$
is large. Matsumoto \cite{matsu} uses the following theorem to construct
$q$-ary quantum codes from $q^m$ ones, if $q$ is small:

\begin{theorem}[Ashikhmin and Knill \cite{ashknill}] \label{ashknill}
  Let $m$ be a positive integer and $\{ \alpha_1, \ldots , \alpha_m \}$ an
  $\F_q$-basis of $\F_{q^m}$. Define $\F_q$-linear maps $\alpha : \F_q^m
  \rightarrow \F_{q^m}$ sending $(x_1, \ldots , x_m)$ to $x_1 \alpha_1 +
  \cdots + x_m \alpha_m$, and $\beta: \F_q^m \rightarrow \F_{q^m}$ sending
  $(x_1, \ldots , x_m)$ to
  \begin{displaymath}
    (\alpha_1, \ldots, \alpha_m) M 
    \left(  
    \begin{array}{c}
      x_1 \\ \vdots \\ x_m
    \end{array}
    \right)
    \in \F_{q^m}
  \end{displaymath}
  where $M$ is an $m \times m$ matrix defined by $M_{ij} =
  \trace_q^{q^m}(\alpha_i,\alpha_j)$, where $\trace$ denotes the trace map from
  $\F_{q^m}$ onto $\F_q$. For $C \subseteq \F_{q^m}^{2n}$, let
  \begin{displaymath}
    \gamma(C) = \set{ (\alpha^{-1}(x_1), \ldots, \alpha^{-1}(x_n),
      \beta^{-1}(x_{n+1}), \ldots, \beta^{-1}(x_{2n})) | (x_1, \ldots, x_{2n})
      \in C} 
  \end{displaymath}
  $ \subseteq \F_{q}^{2mn}$. If $\Cperps \subseteq C$ for $C \subseteq
  \F_{q^m}^{2n}$, then $(\gamma(C))^{\perp_s} \subseteq \gamma(C)$. We also
  have $d(\gamma(C) \backslash (\gamma(C))^{\perp_s}) \ge d(C \backslash
  \Cperps)$.
\end{theorem}

Now, as we have seen the existence of a quantum code from curves, we still
need to know how to decode it and correct errors. This is described in the
following paragraph:

\subsection{Decoding and Error Correction}

To decode this code, we can use the Algorithm of Farr\'an
\cite{farran}. Matsumoto reduced his code to Farr\'an's algorithm as follows:

\begin{red}
  If there exists a vector $e \in \F_q^{2n}$ such that
  $\langle e, b_i \rangle_s = s_i$ for $i = 1, \ldots , n-k$
  and that
  \begin{equation}
    \label{weighterror}
    2\weight(e) + 1 \le n - \left\lfloor \frac{\deg G}{2} \right\rfloor,
  \end{equation}
  where $ \{ b_1, \ldots, b_{n-k} \} $ is a basis of $C(G)^{\perp_s} = C(H)$
  constructed in Corollary \ref{matsuII}, then we can efficiently find $e$
  from $s_1, \ldots, s_{n-k}$ as follows.  The algorithm of Farr\'an
  \cite{farran} efficiently finds the unique vector $x$ having the minimum
  Hamming weight $\weight_H(x)$ in the set $\set{y \in \F_q^{2n} |
  \sip{y}{b_i} = s_i \;{\rm for} \; i = 1, \ldots, n-k}$ from given $s_1,
  \ldots, s_{n-k}$, provided that $2 \weight_H(x) + 1 \le 2n - \deg G$, where
  $\ip{x}{b_i}$ is the standard inner product of $x$ and $b_i$ and $\set{ b_1,
  \ldots, b_{n-k}}$ is a basis of $C(H)$ in $O(n^{2.81})$. Let $e = (e_1,
  \ldots, e_{2n})$ and $e^{\prime} = (-e_{n+1}, \ldots, -e_{2n}, e_1, \ldots,
  e_n)$. Then $s_i = \ip{e^{\prime}}{b_i} = \sip{e}{b_i}$. Since
  $\weight_H(e^{\prime}) \le 2 \weight(e)$, Equation (\ref{weighterror})
  implies
  \begin{eqnarray*}
    2 \weight_H(e^{\prime}) + 1 & \le & 2n - \deg G,
  \end{eqnarray*}
  and the algorithm of Farr\'an finds $e^{\prime}$ from $s_1, \ldots, s_{n-k}$
  correctly. We can easily find $e$ from $e^{\prime}$ with the map
  \begin{eqnarray*}
    (e_1, \ldots, e_n, e_{n+1}, \ldots, e_{2n}) & \mapsto & (e_{n+1}, \ldots,
    e_{2n}, - e_1, \ldots, - e_n).
  \end{eqnarray*}
\end{red}

\section{Construction and Bounds}

This section gives the explicit construction of a quantum stabilizer code over
$\F_{2^m}$, its properties, and shows some bounds on the rate $k/n$ and the
relative minimum distance $d/n$.

\begin{prop}
  For an integer $m \ge 2$ there exists a sequence of binary quantum
  stabilizer codes with parameters $\qecc{n_i}{k_i}{d_i}$ such that

  \begin{eqnarray*}
    \lim \limits_{i \rightarrow \infty} n_i & = & \infty,\\
    \liminf \limits_{i \rightarrow \infty} \frac{k_i}{n_i} & \ge &
    R_m^{(1)}(\delta),\\
    \liminf \limits_{i \rightarrow \infty} \frac{d_i}{n_i} & \ge &
    \delta,\\
  \end{eqnarray*}
  where
  \begin{eqnarray*}
    R_m^{(1)}(\delta) & = & 1 - \frac{2}{2^m - 1} - 4m\delta.
  \end{eqnarray*}
\end{prop}

\begin{proof}
  The proof is divided into two parts: First we will explicitly construct a
  tower of algebraic function fields and Goppa codes derived from these
  fields. Later we will prove that these codes have the claimed properties.
  \paragraph{Construction}

  \begin{enumerate}
    
  \item We use the Garcia-Stichtenoth function field \cite{gs-tower}. Its
    properties and how to construct it is explained in Section
    \ref{as-extensions}. Here we use it in the following way: Let $q=2^m$, let
    $F_i = \F_{q^2}(x_1, z_2, \ldots, z_n)$, $i \ge 2$ and
    \begin{eqnarray*}
      z_i^q + z_i - x_{i-1}^{q+1} & = & 0,\\
      x_i & = & \frac{z_i}{x_{i-1}}.
    \end{eqnarray*}

  \item {\bf Claim 1 } The Galois group of $F_i / F_{i-1}$ is isomorphic to
    the additive group of $\F_2^m$ and there exists a $\sigma \in \Gal(F_i /
    F_{i-1})$ of order 2.

    {\it Proof.} Apply Proposition \ref{galoisgroup1} and Proposition
    \ref{galoisgroup2} of Section \ref{theorems}. Here $[F_i : F_{i-1}] = q =
    2^m$, so we get that $\Gal(F_i / F_{i-1})$ is isomorphic to $(\mathbb{Z}/2
    \mathbb{Z})^m$ which is equal to the additive group of $\F_2^m$. The
    existence of $\sigma$ follows from the fact that in $\F_2^m$ all elements
    are self-inverse.
    
  \item Let $n_i := \frac{(q^2 - 1)q^{i-1}}{2}$ and $y = x_1^{q^2-1}-1$.\\
    {\bf Claim 2 } The zero divisors of $y$ consist of $2n_i$ places of degree
    1.

    {\it Proof.} As our field extension satisfies the conditions of Lemma
    \ref{gstowernpl} of Section \ref{as-extensions}, we can apply it. So we get
    that there are exactly $q^2 -1$ places $P$ of degree one in $F_1$ and
    $q^{i-1}$ over every $P$, so all together $(q^2 -1) \cdot q^{i-1} = 2 n_i$
    places of degree one in $F_i$. By construction, all zeros of $y$ are of
    degree one.
    
  \item Let $F_i^{\sigma}$ be the fixed field of $\sigma$. Let $Q$ be a zero
    of $y$.

    {\bf Claim 3 } There exists a zero $Q^{\prime}$ of $y$ such that
    $Q^{\prime} \ne Q$ and $Q \cap F_i^{\sigma} = Q^{\prime} \cap
    F_i^{\sigma}$.

    {\it Proof.} The isomorphism $\sigma$ is an element of the Galois group of
    order 2. This means that every zero of $y$ is mapped to another one and
    applying $\sigma$ twice gives the identity. Therefore the zeros come in
    pairs as shown in the picture below. Hence $\sigma Q = Q^{\prime}$ and
    $\sigma Q^{\prime} = Q$. For all $x \in Q \cap F_i^{\sigma}$ we have
    $\sigma(x) = x$, whence $x \in Q^{\prime}$ and $x \in F_i^{\sigma}$,
    i.e. $x \in Q^{\prime} \cap F_i^{\sigma}$. The other inclusion can be
    shown similarly.
    \begin{center}
      \includegraphics[scale=0.35]{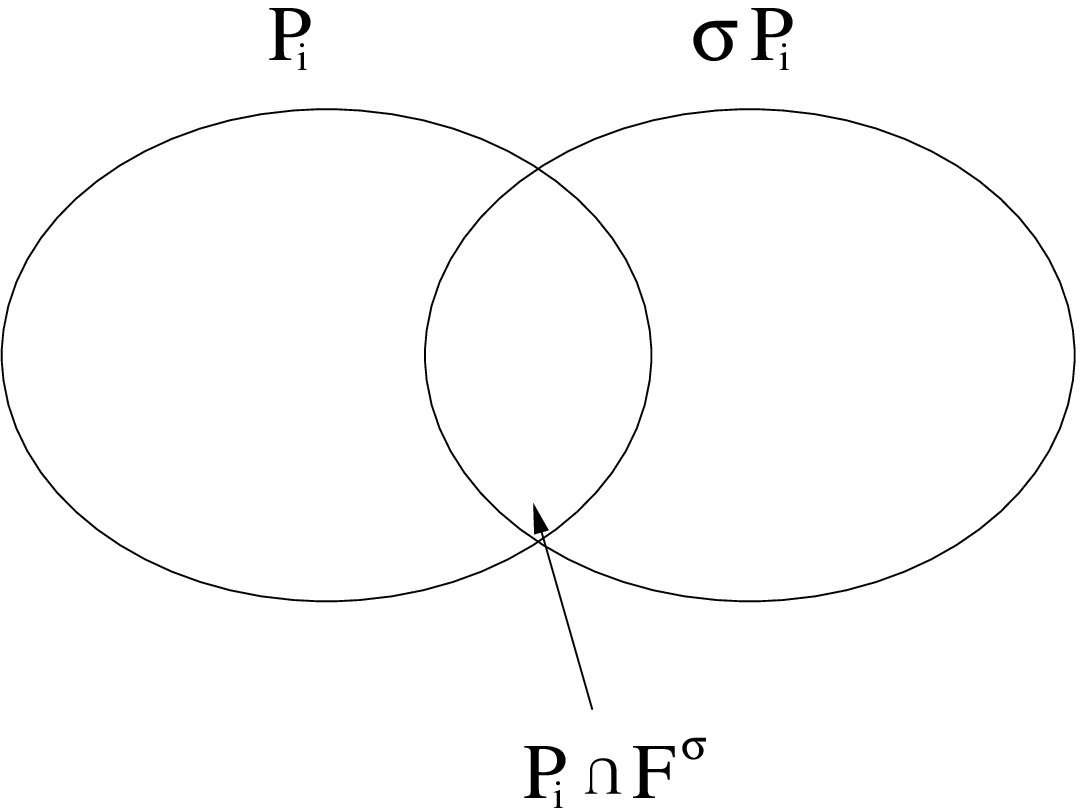}
    \end{center}

  \item {\bf Claim 4 } Since $F_i / F_i^{\sigma}$ is Galois, we have that
    $\sigma Q = Q^{\prime}$. Therefore we can write the zero divisor of
    $y$ as
    \begin{displaymath}
      P_1 + \sigma P_1 + \cdots + P_{n_i} + \sigma P_{n_i} 
    \end{displaymath}
    such that $\sigma P_j \ne P_\ell$ for all $1 \le j,\; \ell \le n_i$.

    {\it Proof.} The first part is clear by the proof of Claim 3 and Theorem
    \ref{galoistrans}. With the property that all zeros are conjugated by
    $\sigma$, it is clear that one can write the zero divisor of $y$ as
    claimed.

  \item Let 
    \begin{eqnarray*}
      \eta & = & \frac{dy}{y} \; = \; \frac{dy}{dx_1} \; \frac{dx_1}{y} \; =
      \; \frac{d(x_1^{q^2-1})}{dx_1} \; \frac{dx_1}{y}\\ & = & (q^2-1)
      x_1^{q^2-2} \; \frac{dx_1}{y}.
    \end{eqnarray*}
    As we work over a binary field, we have $q^2-1 = 1$, therefore
    \begin{eqnarray*}
      \eta & = & x_1^{q^2-2} \; \frac{dx_1}{y}.
    \end{eqnarray*}

    {\bf Claim 5 } $\eta$ satisfies the conditions of Equation (\ref{matsuIeq})
    in Proposition \ref{matsuI} and can be used for our construction.

    {\it Proof.}
    \begin{enumerate}
    \item $ v_{P_i}(1) = v_{\sigma P_i}(1) = 0$
    \item $ v_{P_i}(y) = v_{\sigma P_i}(y) = 1$, because the elements $P_i$
      and $\sigma P_i$ are simple poles of $y = x_1^{q^2-1} - 1$.
    \item $1(P_i) = 1(\sigma P_i) = 1$.
    \item Now we can apply Proposition \ref{constrdiff}, which leads to
      \begin{eqnarray*}
	v_{P_i}(\eta) & = & v_{\sigma P_i}(\eta) = -1,\\
	\res{P_i}{\eta} & = & 1,\\
	\res{\sigma P_i}{\eta} & = & 1 = -1 \; \mathrm{ since } \;
	\chara(\F_{q^2}) = 2.\\
      \end{eqnarray*}
    \end{enumerate}
    Hence all conditions are satisfied and we can use $\eta$ to construct
    codes.
    
  \item Let $G_0^{\prime} := (\eta) + P_1 + \sigma P_1 + \cdots + P_{n_i} +
    \sigma P_{n_i}$, and $P_{\infty}$ the unique pole of $x_1$ in $F_i$.  (For
    existence see Section \ref{as-extensions}.) Then we can apply Lemma
    \ref{candivisor} and get
    \begin{eqnarray*}
      (\eta) & = & (x_1^{q^2-2} \; \frac{dx_1}{y})\\ & = & (x_1^{q^2-2})+
      (dx_1) - (y)\\ & = & (q^2 - 2)(x_1) + (dx_1)\\ & & - (P_1 + \sigma P_1 +
      \cdots + P_{n_i} + \sigma P_{n_i} +
      \underbrace{v_{P_{\infty}}(x_1^{q^2-1})P_{\infty}}_{\mathrm{pole} \;
      \mathrm{divisor} \; \mathrm{of}\; y})\\ & = & (q^2 - 2)(x_1) + (dx_1)\\
      & & - (P_1 + \sigma P_1 + \cdots + P_{n_i} + \sigma P_{n_i} +
      (q^2-1)v_{P_{\infty}}(x_1)P_{\infty}).
    \end{eqnarray*}
    From this follows that we can write $G_0^{\prime}$ as
    \begin{eqnarray*}
      G_0^{\prime}
      & = & (q^2 - 2)(x_1) + (dx_1)\\
      && - (P_1 + \sigma P_1 + \cdots + P_{n_i} +
      \sigma P_{n_i} + (q^2-1)v_{P_{\infty}}(x_1)P_{\infty})\\
      && + P_1 + \sigma P_1 + \cdots + P_{n_i} + \sigma P_{n_i}\\
      & = & (q^2 - 2)(x_1) - (q^2-1)v_{P_{\infty}}(x_1)P_{\infty} + (dx_1)
    \end{eqnarray*}
 
  \item {\bf Claim 6 } The valuation of the divisor $G_0^{\prime}$ is an even
    integer at every place of $F_i$.\\
    {\it Proof.} To show this, we will have a closer look at the individual
    summands of $G_0^{\prime}$.
    \begin{enumerate}
    \item $v_{P_{\infty}}(x_1) = - q^{i-1}$\\ By Lemma \ref{gstowerval} in
      Section \ref{as-extensions}, we know that $v_{P_{\infty}(F_1)}(x_1) =
      -1$ and the unique pole is totally ramified in all extensions. Denote
      $P_{\infty}$ in $F_j$ by $P_{\infty}(F_j)$.We get that
      $e(P_{\infty}(F_j)) = e(P_{\infty}(F_j) \, | \, P_{\infty}(F_{j-1})) =
      q$ for all $j \ge 2$ and therefore
      \begin{eqnarray*}
	v_{P_{\infty}(F_i)}(x_1) 
	& = & e(P_{\infty}(F_i)) \cdot v_{P_{\infty}(F_{i-1})}(x_1)\\ & = &
	\cdots \, = \, q^{i-1} v_{P_{\infty}(F_1)}(x_1) \, = \, q^{i-1} \cdot
	(-1)\\
	& = & - \,q^{i-1}
      \end{eqnarray*}
    \item The discrete valuation of $(dx_1)$ is given by
      \begin{displaymath}
	(dx_1) = (1) - 2(x_1)_{\infty} + \Diff(F_i / F_1)
      \end{displaymath}
      This follows immediately from Lemma \ref{candivisor} in Section
      \ref{theorems}.
    \item The discrete valuation of $(dx_1)$ is even at every place of $F_i$.\\
      For all $P_j \in \{ P_1, \sigma P_1, \ldots, P_{n_i}, \sigma P_{n_i}\}$
      \begin{eqnarray*}
	v_{P_j}((dx_1))
	& = & v_{P_j}((1)) + v_{P_j}(-2(x_1)_{\infty}) + v_{P_j}(\Diff(F_i /
	F_1))\\
	& = & 0 - 2\cdot v_{P_j}(P_{\infty}) + \sum\limits_{P \in
	  \mathbb{P}_{F_1}}\sum\limits_{P^{\prime}|P} d(P^{\prime}|P) \cdot
	v_{P_j}(P^{\prime})\\
	& = & \underbrace{- 2\cdot v_{P_j}(P_{\infty})}_{\mathrm{even}} +
	\underbrace{\sum\limits_{P \in
	\mathbb{P}_{F_1}}\sum\limits_{P^{\prime}|P}
	\underbrace{d(P^{\prime}|P)}_{\mathrm{ even}} \cdot
	v_{P_j}(P^{\prime})}_{\mathrm{even}}\\
      \end{eqnarray*}
      The last equation follows because the elements $d(P^{\prime}|P)$ are all
      even: By Lemma \ref{gstowerdiff} in Section \ref{algextensions}, all
      places of degree one in $F_i$ are either totally ramified over $F_{i-1}$
      with different exponent $d(P^{\prime}) = (q-1)(q+2)$, or unramified, so
      $e(P^{\prime}) = 1$ and $char(\F_{q^2}) = 2 \, \not| \; 1$. Therefore we
      can apply Dedekind's Different Theorem (see Theorem \ref{deddifferent})
      and get $d(P^{\prime}) = e(P^{\prime}) - 1 = 0$. As $2 \, |
      \,(q-1)(q+2)$ and $2 \, | \, 0$, $d(P^{\prime})$ is also even over $F_1$
      for all $P^{\prime}$ with $P^{\prime} | P \in \mathbb{P}_{F_1}$.
    \end{enumerate}
    Hence we conclude
    \begin{eqnarray*}
      G_0^{\prime}
      & = & (q^2 - 2)(x_1) - (q^2-1)v_{P_{\infty}}(x_1)P_{\infty} + (dx_1)\\
      & = & (q^2 - 2)(x_1) -
      (q^2-1)\underbrace{(-q^{i-1})}_{\mathrm{(a)}}P_{\infty}\\
      & & + \underbrace{(1) - 2(x)_{\infty} + \Diff(F_i /
	F_1)}_{\mathrm{(b)}}\\
      & = & \underbrace{(q^2 - 2)(x_1)}_{\mathrm{even}} +
      \underbrace{(q^2-1)
	\underbrace{q^{i-1}}_{\mathrm{even}}}_{\mathrm{even}}P_{\infty} +
      \underbrace{(1) - 2(x)_{\infty} + \Diff(F_i / F_1)}_{\mathrm{(c)} \;
	\Rightarrow \; \mathrm{ even}}\\
    \end{eqnarray*}
    and the valuation of $G_0^{\prime}$ is even.

  \item With the result of Claim 6, we can define $G_0 := \frac{1}{2}
    G_0^{\prime}$ and get
    \begin{eqnarray*}
      \deg G_0 
      & = & \frac{1}{2} \deg G_0^{\prime}\\
      & = & \frac{1}{2} ((q^2-2) \underbrace{\deg (x_1)}_{= \, 0 \;
	\mathrm{(a)}} + 
      \underbrace{(q^2-1)q^{i-1}}_{= \, 2n_i} \underbrace{\deg P_{\infty}}_{=
	\, 1} + \deg (dx_1))\\
      & = & \frac{1}{2} (2n_i + \deg (dx_1))\\
      & = & \frac{1}{2} (2n_i + \underbrace{2g_i - 2}_{\mathrm{(b)}})\\
      & = & n_i + g_i - 1
    \end{eqnarray*}
    where $g_i$ is the genus of $ F_i / \F_{q^2}$.
    \begin{enumerate}
    \item $(x_1)$ is a principal divisor and therefore it has degree zero.
    \item $(dx_1)$ is a canonical divisor and therefore it has degree $2g_i -
    2$.
    \end{enumerate}

  \item Let $j$ be a nonnegative integer.\\
    {\bf Claim 7 } $G_0 + j P_{\infty}$ satisfies the conditions on $G$ in
    Proposition \ref{matsuI}.
    
    {\it Proof.} We have to show that $\sigma (G_0 + j P_{\infty}) = G_0 + j
    P_{\infty}$. Denote that poles are mapped to poles and zeros are mapped
    to zeros under $\sigma$ and $P_{\infty}$ is unique.
    \begin{eqnarray}
      \sigma (G_0 + j P_{\infty}) 
      & = & \sigma G_0 + j \sigma P_{\infty} \; = \;
      \sigma G_0 + j P_{\infty} \nonumber\\
      & = & \sigma \left( \frac{1}{2} \left((q^2-2)(x_1) - (q^2-1)q^{i-1}
      P_{\infty} + (dx_1)\right)\right) \nonumber\\
      & & + j P_{\infty} \nonumber\\
      & = & \frac{q^2-2}{2}\; \sigma \left(\sum_{P} v_P(x_1) P \right) -
      \frac{(q^2-1)q^{i-1}}{2} \sigma P_{\infty} \nonumber\\
      & & + \sigma \left(\sum_{P} v_P(dx_1) P \right) + j P_{\infty}
      \nonumber\\
      & = & \frac{q^2-2}{2} \sigma \left(\sum_{P \ne
      P_{\infty}, \; \ne P_{x_1}} 0 \cdot P + 1 \cdot P_{x_1} +
      (-1)P_{\infty}\right) \nonumber\\
      & & - \frac{(q^2-1)q^{i-1}}{2} P_{\infty} \nonumber\\
      & & + \sigma \left(\sum_{P \ne P_{\infty}} v_P(1) P +
      v_{P_{\infty}}(dx_1) P_{\infty} \right) + j P_{\infty}
      \label{matsueqnsigma}\\
      & = & \frac{q^2-2}{2}(P_{x_1} - P_{\infty}) - \frac{(q^2-1)q^{i-1}}{2}
      P_{\infty} - 2 P_{\infty} + j P_{\infty} \nonumber\\
      & = & \frac{q^2-2}{2}\; \sum_{P} v_P(x_1) P -  \frac{(q^2-1)q^{i-1}}{2}
      P_{\infty} \nonumber\\
      & & + \sum_{P} v_P(dx_1) P + j P_{\infty} \nonumber\\
      & = &  \frac{1}{2} ((q^2-2)(x_1) - (q^2-1)q^{i-1} P_{\infty} +
      (dx_1)) + j P_{\infty} \nonumber\\
      & = & G_0 + j P_{\infty} \nonumber
    \end{eqnarray}
    Equation (\ref{matsueqnsigma}) follows because of the following: Let us
    have a look at the properties of $dx_1$. Denote that $x_1$ is a $P$-prime
    element for all $P$ except $P_{\infty}$. For $P_{\infty}$ we can use
    $\frac{1}{x_1}$ as $P$-prime element. Therefore for $P \ne P_{\infty}$
    \begin{eqnarray*}
      v_P(dx_1) & = & v_P(1 \cdot dx_1) \stackrel{\mathrm{Def.}}{=} v_P(1).
    \end{eqnarray*}
    For  $P = P_{\infty}$ we get
    \begin{eqnarray*}
      dx_1 
      & = & \frac{dx_1}{d\frac{1}{x_1}} \; d\frac{1}{x_1}
      = - \left(\frac{1}{x_1}\right)^{-2} \; d\frac{1}{x_1}
    \end{eqnarray*}
    which implies that
    \begin{displaymath}
      v_{P_{\infty}}(dx_1) = v_{P_{\infty}}\left(-
      \left(\frac{1}{x_1}\right)^{-2}
      d\frac{1}{x_1}\right) = v_{P_{\infty}}\left(-
      \left(\frac{1}{x_1}\right)^{-2}\right) = -2
    \end{displaymath}
    
  \item Let
    \begin{eqnarray*}
      H 
      & = & ( P_1 + \cdots + P_{n_i} + \sigma P_1 + \cdots + \sigma P_{n_i}) -
      (G_0 + j P_{\infty}) + (\eta)\\
      & = & ( P_1 + \cdots + P_{n_i} + \sigma P_1 + \cdots + \sigma P_{n_i})\\
      & & - \left(\frac{1}{2}( (\eta) + P_1 + \sigma P_1 + \cdots + P_{n_i} +
      \sigma
      P_{n_i}) + j P_{\infty}\right) + (\eta)\\
      & = & \frac{1}{2}\left( (\eta) + P_1 + \sigma P_1 + \cdots + P_{n_i} +
      \sigma P_{n_i}\right) - j P_{\infty}\\
      & = & G_0 - j P_{\infty}.
    \end{eqnarray*}
    Then $G_0 + j P_{\infty} \ge G_0 - j P_{\infty} = H$, and therefore by
    Proposition \ref{matsuI}
    \begin{eqnarray*}
      C(G_0 + j P_{\infty})^{\perp_s} & = & C(H) \; = \; C( G_0 - j
      P_{\infty})\\
      & \subseteq & C(G_0 + j P_{\infty}).
    \end{eqnarray*}
    By Corollary \ref{matsuII}, we can construct a quantum stabilizer code
    from $C$.
  \end{enumerate}

  \paragraph{Properties of the Constructed Code}
  
  \begin{enumerate}
  
  \item {\bf Claim 8 } By Corollary \ref{matsuII} for $i,j \in \N$ we can
    construct an $[[n_i,k_{ij},d_{ij}]]$ stabilizer code with
    \begin{displaymath}
      k_{ij} \ge j, \; d_{ij} \ge (n_i - g_i - j + 1)/2
    \end{displaymath}
    
	{\it Proof.} Corollary \ref{matsuII} implies that
	\begin{displaymath}
	    \renewcommand{\arraystretch}{1.5}
	  \begin{array}{cclcl}
	    d_{ij}
	    & \ge & n_i - \left\lfloor \frac{\deg G}{2} \right\rfloor
	    
	    & = & n_i - \overbrace{\left\lfloor \frac{\deg G_0 + j
	    P_{\infty}}{2}\right\rfloor}^{\ge 0}\\ 
	    & \stackrel{\mathrm{(9.)}\;\mathrm{constr.}}{=} & n_i -
	    \left\lfloor \frac{n_i + g_i - 1 + j}{2} \right\rfloor & \ge & n_i
	    - \frac{n_i + g_i - 1 + j}{2}\\
	    & = & \frac{n_i - g_i - j + 1}{2}.
	  \end{array}
	\end{displaymath}
	For $k_{ij}$ we have to assume $j \le n_i - g_i$ and obtain
	then
	\begin{eqnarray}
	  k_{ij}
	  & = & \dim G \nonumber\\
	  & & - \dim (G - P_1 - \cdots - P_{n_i} - \sigma P_1 - \cdots -
	  \sigma P_{n_i}) - n_i \nonumber\\
	  & = & \dim (G_0 + j P_{\infty}) - \dim (- G_0 + j P_{\infty} +
	  (\eta)) - n_i \nonumber\\
	  & \ge & \deg (G_0 + j
	  P_{\infty}) + 1 - g_i 
	  - \dim (- G_0 + j P_{\infty} + (\eta)) - n_i \label{kijthmrieroch}\\
	  & = & [ (n_i + g_i -1) +
	  j + (1 - g_i) - n_i ] \nonumber\\
	  & & - \dim (- G_0 + j P_{\infty} + (\eta)) \label{kijixconstr}\\
	  & = & j \label{kijlast}
	\end{eqnarray}
	In these calculations, Equation (\ref{kijthmrieroch}) follows from
	Theorem \ref{rieroch}, Equation (\ref{kijixconstr}) follows from 9. of
	the code construction and Equation(\ref{kijlast}) is a consequence of
	the following calculations:
	\begin{eqnarray*}
	  \deg (- G_0 + j P_{\infty} + (\eta))
	  & = & - (n_i + g_i - 1) + j + (2 g_i - 2)\\
	  & = & g_i - n_i - 1 + j\\
	  & \le & g_i - n_i - 1 + (n_i - g_i)\\
	  & = & - 1\\
	  & < & 0
	\end{eqnarray*}
	By Corollary \ref{degdim} it follows immediately that $\dim(- G_0 +
	j P_{\infty} + (\eta))= 0$.  
	
      \item Let $R$ be a real number such that $ 0 \le R \le 1$ and $\lfloor
	Rn_i \rfloor \le n_i - g_i$. Set $j := \lfloor Rn_i \rfloor$. By
	Theorem \ref{ashknill} we can construct a sequence of $[[ n_i, k_i,
	d_i]]$ binary quantum stabilizer codes.

	{\bf Claim 9 } This code has the following properties
	\begin{eqnarray*}
	  \liminf \limits_{i \rightarrow \infty} \frac{k_i}{n_i} & \ge &
	  R_m^{(1)}(\delta), \qquad \liminf \limits_{i \rightarrow \infty}
	  \frac{d_i}{n_i} \quad \ge \quad \delta,\\
	\end{eqnarray*}
	where
	\begin{eqnarray*}
	  R_m^{(1)}(\delta) & = & 1 - \frac{2}{2^m - 1} - 4m\delta.
	\end{eqnarray*}
	{\it Proof.} 
	\begin{eqnarray*}
	  \liminf \limits_{i \rightarrow \infty} \frac{k_i}{n_i} & \ge &
	  \liminf \limits_{i \rightarrow \infty} \frac{j_i}{n_i} \quad = \quad
	  \liminf \limits_{i \rightarrow \infty} \frac{\lfloor Rn_i
	  \rfloor}{n_i} \quad = \quad R
	\end{eqnarray*}
	The binary code obtained from Theorem \ref{ashknill} has rate $
	\frac{2m \cdot k_i}{2m \cdot n_i} = \frac{k_i}{n_i}$ and therefore the
	same rate as the higher dimensional case. Moreover,
	\begin{eqnarray}
	  \liminf \limits_{i \rightarrow \infty} \frac{d_i}{n_i} 
	  & \ge & \liminf \limits_{i \rightarrow \infty} \frac{(n_i - g_i -
	  j_i + 1)/2}{n_i} \nonumber\\
	  & = & \frac{1}{2} \liminf \limits_{i \rightarrow \infty} \frac{n_i -
	    g_i - \lfloor R n_i \rfloor + 1}{n_i}\nonumber\\
	  & = & \frac{1 - R - 2/(2^m - 1)}{2} \label{dini}\\
	  & \ge  & \frac{1 - R - 2/(2^m - 1)}{4m} =: \delta \nonumber
	\end{eqnarray}
	And for the binary case
	\begin{eqnarray*}
	  \liminf \limits_{i \rightarrow \infty} \frac{d_i}{2m \cdot n_i} 
	  & \ge & \liminf \limits_{i \rightarrow \infty} \frac{(n_i - g_i -
	  j_i + 1)/2}{2m \cdot n_i}\\
	  & =  & \frac{1 - R - 2/(2^m - 1)}{4m} =: \delta.
	\end{eqnarray*}

	Equation (\ref{dini}) follows if we use the genus formula of Theorem
	\ref{gstowergen} and calculate
	\begin{itemize}
	\item For $i = 1 \; \mod \; 2$:
	  \begin{eqnarray*}
	    \lim\limits_{i \rightarrow \infty} \frac{n_i}{g_i}
	    & = &  \lim\limits_{i \rightarrow \infty}
	    \frac{(2^{2m}-1)(2^m)^{i-1}/2}{(2^m)^i + (2^m)^{i-1} -
	      (2^m)^{\frac{i-1}{2}} + 1}\\
	    & = & \frac{1}{2} \frac{2^{2m}-1}{2^m + 1}\\
	    & = & \frac{1}{2} \frac{(2^m + 1)(2^m - 1)}{2^m + 1}\\
	    & = & \frac{2^m - 1}{2}.
	  \end{eqnarray*}
	\item and for $i = 0 \; \mod \; 2$:
	  \begin{displaymath}
	    \begin{array}{l}
	      \lim\limits_{i \rightarrow \infty} \frac{n_i}{g_i}\\
	      \begin{array}{ll}
		= & \lim\limits_{i \rightarrow \infty}
		\frac{(2^{2m}-1)(2^m)^{i-1}/2}{(2^m)^i + (2^m)^{i-1} -
		  \frac{1}{2} (2^m)^{\frac{i}{2} + 1} - \frac{3}{2}
		  (2^m)^{\frac{i}{2}} - (2^m)^{\frac{i}{2} - 1} + 1}\\
		= & \frac{1}{2} \frac{2^{2m}-1}{2^m + 1}\\
		= & \frac{1}{2} \frac{(2^m + 1)(2^m - 1)}{2^m + 1}\\
		= & \frac{2^m - 1}{2}.
	      \end{array}
	    \end{array}
	  \end{displaymath}
	\end{itemize}
	
	We now can calculate $R_m^{(1)}(\delta) = R$:
	\begin{displaymath}
	  \begin{array}{cccc}
	    & \delta & = & \frac{1 - R - 2/(2^m - 1)}{4m}\\
	    \Longleftrightarrow & 4m \delta & = & 1 - R - 2/(2^m - 1)\\
	    \Longleftrightarrow & R & = & 1 - \frac{2}{2^m - 1} - 4m \delta,
	  \end{array}
	\end{displaymath}
	
  \end{enumerate}
  which establishes the claimed result.
  
\end{proof}

\section{A Small Example for the Construction}

Next, we give an example for a code we get from Matsumoto's construction. To
do so, we consider the shortest possible length. Let $m = 2$ and $i = 2$, so
$q = 2^2 = 4$ and therefore we use the field $\F_{4^2} = \F_{16}$. Then $F_1 =
\F_{16}(x_1)$ is the rational function field and $F_2 = F_1(z_2) =
\F_{16}(x_1,z_2)$ with
\begin{eqnarray*}
  z_2^4 + z_2 - x_1^5 & = & 0.
\end{eqnarray*}
Using {\it Magma} \cite{magma} we can calculate the genus of the
extension:

\begin{verbatim}  
Magma V2.11-5
Type ? for help.  Type <Ctrl>-D to quit.
> F<w> := GF(16);
> P2<x,y,z> := ProjectiveSpace(F,2);
> f := y^4*z + y*z^4 - x^5;
> X := Curve(P2, f);
> g := Genus (X);
> g;
6
\end{verbatim}

Now we are able to estimate $k$ and $d$:
\begin{eqnarray*}
  k_{2,j} \ge j, \qquad d_{2,j} \ge \frac{n_2 - g_2 - j + 1}{2} 
  & = & \frac{30 - 6 - j + 1}{2} = 12 - \frac{j-1}{2}
\end{eqnarray*}
This gives us the possibility to choose a suitable value for $R$. We will have
a look at two examples.
\begin{enumerate}
\item Set $R = \frac{1}{2}$, then $j = \frac{1}{2} \cdot 30 = 15$. We get
  \begin{displaymath}
    k_{2,15} \ge 15, \qquad d_{2,15} \ge 12 - \frac{15-1}{2} = 5.
  \end{displaymath}
  Using Theorem \ref{ashknill}, we get for the parameters of the binary code
  \begin{displaymath}
    n = 4 \cdot n_2 = 120, \qquad k = 4 \cdot k_{2,15} \ge 60, \qquad d =
    d_{2,15} \ge 5.
  \end{displaymath}
  Therefore we can calculate the ratios
  \begin{displaymath}
    \frac{k}{n} \ge \frac{60}{120} = \frac{1}{2}, \qquad \frac{d}{n} \ge
    \frac{5}{120} = \frac{1}{24}
  \end{displaymath}
  and the limits are bounded by
  \begin{displaymath}
    \liminf \limits_{i \rightarrow \infty} \frac{k_i}{n_i} \ge \frac{1}{2},
    \qquad 
    \liminf \limits_{i \rightarrow \infty} \frac{d_i}{n_i} \ge \frac{1 -
    \frac{1}{2} - \frac{2}{4-1}}{4 \cdot 2} = - \frac{1}{48}.
  \end{displaymath}
  This gives an estimate $< 0$ and therefore is not a valid parameter.

\item Set $R = \frac{1}{3}$, then $j = \frac{1}{3} \cdot 30 = 10$. This gives
  us
  \begin{displaymath}
    k_{2,10} \ge 10, \qquad d_{2,10} \ge 12 - \frac{10-1}{2} = 7.5, \;
    \mathrm{so} \; d_{2,10} \ge 8.
  \end{displaymath}
  From Theorem \ref{ashknill}, we get for the binary code
  \begin{displaymath}
    n = 4 \cdot n_2 = 120, \qquad k = 4 \cdot k_{2,10} \ge 40, \qquad d =
    d_{2,10} \ge 8.
  \end{displaymath}
  Therefore we can calculate the ratios
  \begin{displaymath}
    \frac{k}{n} \ge \frac{40}{120} = \frac{1}{3}, \qquad \frac{d}{n} \ge
    \frac{8}{120} = \frac{1}{15}
  \end{displaymath}
  and the limits are bounded by
  \begin{displaymath}
    \liminf \limits_{i \rightarrow \infty} \frac{k_i}{n_i} \ge \frac{1}{3},
    \qquad 
    \liminf \limits_{i \rightarrow \infty} \frac{d_i}{n_i} \ge \frac{1 -
    \frac{1}{3} - \frac{2}{4-1}}{4 \cdot 2} = 0.
  \end{displaymath}
  Hence $R = \frac{1}{3}$ is the threshold in order to get $\frac{d}{n} \ge
  0$. Therefore we should use $R \le \frac{1}{3}$. This gives an
  $\qecc{120}{\ge 40}{\ge 8}$ binary quantum error correcting code.

\end{enumerate}

\chapter{Codes over Hyperelliptic Curves}
\label{qechyper}
The following chapter uses the machinery introduced in the preceding chapters
to construct quantum error correcting codes from Goppa codes. We will work
over fields of odd characteristic. If we would work in characteristic $2$ we
would get the first step of the hierarchy used in Matsumoto's construction
explained in the preceding chapter.

In this chapter we will first look at the CSS construction that can be applied
to all self-orthogonal AG codes. We will generalise this construction to
weighted self-orthogonal codes. Afterwards we will directly construct quantum
Goppa codes, comparable to Matsumoto's construction in $\cite{matsu}$. Finally
we shall see some examples that illustrate the construction.

\section{The CSS Construction Revisited}


This section shows how we can use hyperelliptic curves to
construct quantum AG codes. The construction is similar to the one in the next
section, but is easier to prove. 

\subsection{General Construction}
\label{constructioncss}
\begin{enumerate}
  
\item Let $\K := \F_{p^m}$ be a finite field of odd characteristic, and choose 
  a hyperelliptic curve 
  \begin{eqnarray*}
    y^2 & = & f(x),
  \end{eqnarray*}
  where $f(x)$ is a square-free polynomial of degree $\ge 5$. Let $F :=
  K(x,y)$ be the corresponding function field. (see Chapter \ref{ag})

\item Then $F$ has a set of rational places (see Definition
\ref{rationalplaces}). We choose any subset of pairs and denote it by $SP =
\set{P_1,\ldots, P_n}$.

Set $D = P_1 + \cdots + P_n$ and $G = (\floor{\frac{n}{2}}+g-1-r)P_{\infty}$,
where $r$ can be chosen with $0 \le r \le n-g$ and $P_{\infty}$ the place at
infinity.  We also set $(\eta) = W = -D + (n + 2g - 2)P_{\infty}$.  Then $W$
is a canonical divisor to the differential
  \begin{eqnarray*}
    \eta & = & \frac{1}{y \; \prod_{P_i
	\,\rm{corr.}\,\alpha_i}(x-\alpha_i)} {\rm d} x.
  \end{eqnarray*}
  \begin{proof}
    The proof is similar to the one given in Lemma \ref{etaequalsw}.
  \end{proof}

\item Denote the residues of $\eta$ at the places $P_1,\ldots,P_n$ by
  \begin{displaymath}
    a_i = \res{P_i}{\eta}.
  \end{displaymath}
  for $i = 1,\ldots,n$.

\item Now we can construct a classical Goppa code $C(D,G)$ (see Section
  \ref{goppa}) with
  \begin{displaymath}
    C(D,G)^{\perp} = C(D,H) \cdot
    \diag(a_1,\ldots,a_n)
  \end{displaymath}
  where $H = D - G + W$ is defined as usual (proof see Theorem 2.5 in
  \cite{s-selfdual}).  The code satisfies $C(D,G) \subseteq C(D,H)$ and
  therefore $C(D,G) \subseteq C(D,G)^{\perp^a}$ with respect to the inner
  product $\ipa{\;}{\,}$. This will be shown in Corollary
  \ref{clcquasiselforth}.

\item If not all coefficients $a_i$ are in the base field, we can transform
  our code to a code $C^{\prime}(D,G)$ that satisfies $C^{\prime}(D,G)
  \subseteq C^{\prime}(D,G)^{\perp^b}$ with respect to a new inner product
  $\ipb{x}{y} = \sum_{i=1}^n b_i x_i y_i$ where all $b_i$ are in $\F_p$, if
  $m$ is odd. This is proved in Proposition \ref{aitobasefield}.

\item Now we take two copies of $C^{\prime}(D,G)$ and multiply in the first
  copy every codeword component wise with the corresponding coefficient
  $b_i$. Then we can apply the CSS construction and get a generator matrix
  \begin{eqnarray*}
    \G & = & \left(
    \begin{array}{ccc|ccc}
      {b_1 \cdot c_{1,1}} & \cdots & {b_n \cdot c_{1,n}} & 0 & \cdots & 0\\ 
      \vdots & \ddots & \vdots & \vdots & \ddots & \vdots\\
      {b_1 \cdot c_{l,1}} & \cdots & {b_n \cdot c_{l,n}} & 0 & \cdots & 0\\ 
      0 & \cdots & 0 & {c_{1,1}} & \cdots & {c_{1,n}}\\
      \vdots & \ddots & \vdots & \vdots & \ddots & \vdots\\
      0 & \cdots & 0 & {c_{l,1}} & \cdots & {c_{l,n}}
    \end{array}
    \right),
  \end{eqnarray*}
  if the classical code $C^{\prime}(D,G)$ has generator matrix
  \begin{displaymath}
    \left(
    \begin{array}{ccc}
      c_{1,1} & \cdots & {c_{1,n}}\\ 
      \vdots & \ddots & \vdots\\
      {c_{l,1}} & \cdots & {c_{l,n}}\\
    \end{array}
    \right).
  \end{displaymath}
  A proof of this is given in Corollary \ref{cssforquasiselforth}.

\item If we want to project this code and used the modifications of the
  original code above, we can apply the special case where all coefficients
  $a_i$ are in the base field $\F_p$. Therefore this CSS construction has a
  nice down projection that allows us to write the new code as a symmetric one
  with weights $a_i$ in the $X$ component.

\end{enumerate}

\subsection{Correctness of the Construction}

To justify and clarify the above construction, in the following subsection
we prove all statements claimed in the last section.

\begin{corollary}
  \label{clcquasiselforth}

   Let $C(D,G)$ be the Goppa code constructed in Section
   \ref{constructioncss} with $G = (\floor{\frac{n}{2}}+g-1-r)P_{\infty}$
   and
   \begin{displaymath}
     a = (a_1,\ldots,a_n)= (\res{P_1}{\eta},\ldots,\res{P_n}{\eta}).
   \end{displaymath}
   Then the code satisfies $C(D,G)
   \subseteq C(D,H)$ with $H = D-G+(\eta)$ and therefore
   \begin{displaymath}
     C(D,G) \subseteq C(D,G)^{\perp^a}
   \end{displaymath}
   with respect to the inner product $\ipa{\;}{\,}$.
\end{corollary}

\begin{proof}
  First we have to calculate $H$:
  \begin{eqnarray*}
    H & = & D-G+(\eta)\\ 
    & = & D - \left(\floor{\frac{n}{2}} +g-1-r
    \right)P_{\infty} + (-D + (n + 2g - 2)P_{\infty})\\
    & = & \left(\ceil{\frac{n}{2}} + g-1+r\right)P_{\infty}\\ 
    &\ge& G.
  \end{eqnarray*}
  Therefore $C(D,G) \subseteq C(D,H)$ and by \cite[Thm. 2.5]{s-selfdual}
  \begin{eqnarray*}
    C(D,G)^{\perp} & = & C(D,H)\cdot \diag(a_1,\ldots,a_n).
  \end{eqnarray*}
  It follows that all
  codewords $x,y$ of $C(D,G)$ satisfy
  \begin{eqnarray*}
    \sum_{i=1}^n a_i \,x_i y_i & = & 0,
  \end{eqnarray*}
  and $C(D,G)$ is self orthogonal with respect to the inner product
  \begin{eqnarray*}
    \ipa{x}{y} & = & \sum_{i=1}^n a_i \,x_i y_i.
  \end{eqnarray*}
\end{proof}

\begin{prop}
  \label{aitobasefield}
  If not all weights $a_i$ are in the base field, we can transform
  our code to a code $C^{\prime}(D,G)$ that satisfies $C^{\prime}(D,G)
  \subseteq C^{\prime}(D,G)^{\perp^b}$ with respect to a new inner product
  $\ipb{x}{y} = \sum_{i=1}^n b_i x_i y_i$ where all $b_i$ are in $\F_p$, if
  $m$ is odd.
\end{prop}

\begin{proof}
 
  The idea is to use Stichtenoth's proof how to transform residue squares into
  residues with value one (Corollary 3.4 in $\cite{s-selfdual}$).
  
  \begin{enumerate}
  
  \item For all elements that satisfy $\res{P_i}{\eta} = b_i^2$ for some $b_i
  \in \F_{p^m}$, set $u(P_i) = b_i$. For all the others with $\res{P_i}{\eta}
  = d_i$ and $d_i$ is not a square, set $u(P_i) = 1$. The existence of u is
  given by the Strong Approximation Theorem (see Theorem
  \ref{strongapproxthm}). Set $G^{\prime} = G - (u)$ with new residues
  $\res{P_i}{u^{-2} \eta} =\res{P_i}{\eta^{\prime}} = 1$ for all elements that
  were squares. The residues of the other $P_i$ stay $d_i$.
  
\item The residues for our new code over $G^{\prime}$ are 1 if possible. All
  elements of $\F_{p^m}$ are of the form $\alpha^r$ with $r = 1,\ldots, p^m$
  and $\alpha$ a generator of $\F_{p^m}$. The idea is to transform all
  residues to elements of the base field. It turns out that this is only
  possible under the circumstances that $m$ is odd. We get this result from
  the following calculations: The elements of the base field are those which
  satisfy the equation
  \begin{displaymath}
    \beta^{p-1} = 1 = \beta^{p^m-1}
  \end{displaymath}
  We can use this property to find elements $b_i$ satisfying
  \begin{displaymath}
    (b_i^{-2} \alpha^{r_i})^{p-1} = 1 = \alpha^{p^m-1},
  \end{displaymath}
  where $\alpha^{r_i} = d_i$. Denote $b_i := \alpha^{k_i}$ for some $k_i$. Our
  goal is to find $k_i$ that satisfies the equation above. We have that
  \begin{eqnarray*}
    (b_i^{-2} \alpha^{r_i})^{p-1} & = & \alpha^{p^m-1},\\
    \alpha^{(-2k_i + r_i)(p-1)} & = & \alpha^{p^m-1},\\
    (r_i - 2k_i)(p-1) & = & c \cdot (p^m-1),
  \end{eqnarray*}
  for some $c$ (because of $\mod \; p^m-1$). We know that
  \begin{eqnarray*}
    s := (p^m-1)/(p-1) & = & p^{m-1} + \cdots + 1
  \end{eqnarray*}
  and $s$ is odd iff $m$ is odd. Our equation can be transformed to $2k_i = r_i
- c \cdot s$. As the elements $r_i$ are odd (otherwise they are squares which
would have been transformed to 1 in step 1), $c \cdot s$ has to be
odd. Therefore $s$ has to be odd and this yields that $m$ odd. Hence $ r_i - c
\cdot s$ is even and we can divide it by 2. Therefore we can find $k_i$ such
that our equation $2k_i = r_i - c \cdot s$ is satisfied and our residues can be
transformed by
  \begin{eqnarray*}
    k_i & = & \frac{ r_i - c \cdot s}{2}
  \end{eqnarray*}
  to elements of $\F_p\ \set{0}$. Therefore we get a code $C^{\prime}(D,G)$
  that is self-orthogonal with respect to an inner product with weights
  (residues) $b_i$ in $\F_p$. We can apply the trace
  operation on the code and the inner product over $\F_{p^m}$ and get the
  weights out of the trace:
  \begin{eqnarray*}
    \trace(\ipb{x}{y})
    & = & \sum_{i=1}^n \trace( b_i\;x_i y_i)\\
    & = & \sum_{i=1}^n \trace( b_i \sum_{j=1}^m
    \sum_{k=1}^m  x_i^{(j)} y_i^{(k)} \alpha_j \alpha_k)\\
    & = &  \sum_{i=1}^n \sum_{k=1}^m \sum_{l=1}^m b_i\;
    x_i^{(j)} y_i^{(k)} \underbrace{\trace(\alpha_j \alpha_k)}_{\delta_{jk}}
  \end{eqnarray*}
  where $\{ \alpha_1, \ldots, \alpha_m \}$ is a self-dual basis for
  $\F_{p^m}/\F_p$, i.e. for all basis elements holds $\trace(\alpha_i \alpha_j)
  = \delta_{ij}$ (for existence see \cite[Ch. 2, Notes 3.]{lidl}) and 
  \begin{eqnarray*}
    x_i & = & x_i^{(1)} \alpha_1 + \cdots + x_i^{(m)} \alpha_m
  \end{eqnarray*}
  for all codewords $x$. Therefore
  we get a code over $\F_p$ that stays self-orthogonal with respect to
  an inner product $\ipb{\;}{\;}$.
  \end{enumerate}

\end{proof}

\begin{corollary}
  \label{cssforquasiselforth}

  Take two copies of $C^{\prime}(D,G)$ defined above and multiply in the first
  copy every codeword component wise with the coefficients $b_i$. Then we can
  apply the CSS construction and get a generator matrix
  \begin{eqnarray*}
    \G & = & \left(
    \begin{array}{ccc|ccc}
      {b_1 \cdot c_{1,1}} & \cdots & {b_n \cdot c_{1,n}} & 0 & \cdots & 0\\ 
      \vdots & \ddots & \vdots & \vdots & \ddots & \vdots\\
      {b_1 \cdot c_{l,1}} & \cdots & {b_n \cdot c_{l,n}} & 0 & \cdots & 0\\ 
      0 & \cdots & 0 & {c_{1,1}} & \cdots & {c_{1,n}}\\
      \vdots & \ddots & \vdots & \vdots & \ddots & \vdots\\
      0 & \cdots & 0 & {c_{l,1}} & \cdots & {c_{l,n}}
    \end{array}
    \right)
  \end{eqnarray*}
  if the classical code $C^{\prime}(D,G)$ has generator matrix
  \begin{displaymath}
    \G_{cl} = \left(
    \begin{array}{ccc}
      c_{1,1} & \cdots & {c_{1,n}}\\ 
      \vdots & \ddots & \vdots\\
      {c_{l,1}} & \cdots & {c_{l,n}}\\
    \end{array}
    \right).
  \end{displaymath}
\end{corollary}

\begin{proof}
  The classical code $C^{\prime}(D,G)$ satisfies
  \begin{eqnarray*}
    \ipb{x}{y} & = & 0
  \end{eqnarray*}
  for all codewords $x,y$. So if $ \G_{cl}$ is a generator matrix of the
  classical code
  \begin{eqnarray*}
    \G & = & \left(
    \begin{array}{c|c}
      \G_{cl} & 0\\
      0 & \G_{cl}
    \end{array}
    \right)
  \end{eqnarray*}
  is a valid CSS quantum stabilizer code with respect to the symplectic inner
  product 
  \begin{eqnarray*}
    \sipb{x}{y} & = & \sum_{i=1}^n b_i \, (x_i y_{n+i} - x_{n+i} y_i),
  \end{eqnarray*}
  because
  \begin{eqnarray*}
    \sipb{(x_1,\ldots,x_n,0,\ldots)}{(y_1,\ldots,y_n,0,\ldots)} & = &
    \sum_{i=1}^n b_i (x_i \cdot 0 - 0 \cdot y_i) = 0,\\
    \sipb{(0,\ldots,x_{n+1},\ldots,
    x_{2n})}{(0,\ldots,y_{n+1},\ldots,y_{2n})} & = & \sum_{i=1}^n b_i (0
    \cdot y_{n+i} - x_{n+i} \cdot 0)\\
    & = & 0,\\
    \sipb{(x_1,\ldots,x_n,0,\ldots)}{(0,\ldots,y_{n+1},\ldots,y_{2n})} 
    & = & \sum_{i=1}^n b_i \, (x_i y_{n+i} - 0 \cdot 0)\\
    & = & \sum_{i=1}^n b_i \, x_i y_{n+i} = 0.
  \end{eqnarray*}
  Now we can apply Lemma \ref{codemod} and transform our CSS code to a quantum
  code with respect to the standard symplectic inner product.
\end{proof}

\section{Direct Construction and Code Properties}

\label{qagcode}

The goal of this section is to show how it is possible to use Goppa codes over
hyperelliptic curves to construct quantum stabilizer codes.

\subsection{Construction of Weighted Self-Orthogonal Codes}

In order to make the construction more transparent, we will divide it into
several smaller steps. The construction is similar to the one given in Section 
\ref{constructioncss}.
First we will present the construction; the corresponding proofs can be found
in the following subsections.

\begin{enumerate}
  
\item see 1. in Construction \ref{constructioncss}.

\item Then $F$ has a set of splitting rational places (see
  \ref{rationalplaces}). Choose a set of pairs
  \begin{eqnarray*}
    SP & = & \set{P_1,\ldots, P_n, \sigma P_1,\ldots,\sigma P_n},
  \end{eqnarray*}
  where $\sigma$ denotes the hyperelliptic conjugation and $P_i$ and $\sigma
  P_i$ are the two places lying over one rational place in $K(x)$.

  The following elements are similar to Construction \ref{constructioncss},
  but the values are a bit different because we use $2n$ rational places. Set
  \begin{eqnarray*}
    D & = & P_1 + \cdots + P_n + \sigma P_1 + \cdots + \sigma P_n
  \end{eqnarray*}
  and
  \begin{eqnarray*}
    G & = & (n+g-1-r)P_{\infty}
  \end{eqnarray*}
  where $0 \le r \le n-g$ can be chosen and $P_{\infty}$ is the place at
  infinity.
 
  We also set
  \begin{eqnarray*}
    (\eta) & = & W = -D + (2n + 2g - 2)P_{\infty}.
  \end{eqnarray*}
  Then $W$ is a canonical divisor.

  \begin{proof}
    $W = (\eta)$ for 
    \begin{eqnarray*}
      \eta & = & \frac{1}{y \; \prod_{P_i
	  \,\rm{cor.}\,\alpha_i}(x-\alpha_i)} {\rm d} x
    \end{eqnarray*}
    is proved in Lemma \ref{etaequalsw}. Therefore $W$ comes from a
    differential and is canonical.
  \end{proof}

\item The residues of $\eta$ at the places $P_1,\ldots,P_n,\sigma
  P_1,\ldots,\sigma P_n$ satisfy
  \begin{displaymath}
    a_i = \res{P_i}{\eta} = - \res{\sigma P_i}{\eta}.
  \end{displaymath}
  for $i = 1,\ldots,n$ (proof see Lemma \ref{respisigmapi}).

\item Now we can construct a Goppa code $C(D,G)$ (see Section \ref{goppa})
  with
  \begin{displaymath}
    C(D,G)^{\perp_s} = C(D,H) \cdot
    \diag(a_1,\ldots,a_n,1,\ldots,1)
  \end{displaymath}
  where $H = D - G + W$ is defined as usual (proof see Proposition
  \ref{dualequalsh}).  
  The code satisfies $C(D,G) \subseteq C(D,H)$ and therefore $C(D,G) \subseteq
  C(D,G)^{\perp_s^a}$ with respect to the symplectic inner product
  $\sipa{\;}{\,}$ what is proved in Corollary \ref{cquasiselforth}.

\item With the help of Corollary \ref{transformcdg} we transform $C(D,G)$ to a
  self-orthogonal code $C^{\prime}(D,G)$ with respect to the standard
  symplectic inner product by multiplying each component $x_i$ of every
  codeword by the corresponding $a_i$, for $1 \le i \le n$. The difference to
  Construction \ref{constructioncss} is that we have self-orthogonality with
  respect to a symplectic inner product and not with respect to an inner
  product.

\item $C^{\prime}(D,G)$ defines a stabilizer code with parameters
 $\qecc{n}{k}{d}$, where $n$ is the number of used splitting rational places of
 the rational function field, $k \ge r$ and $d \ge \frac{n-g+1-r}{2}$ (see
 Proposition \ref{qeccprops}).

\end{enumerate}

The following subsections include the proofs for the construction above.

\subsubsection{The Canonical Divisor and its Residue Properties}


In the following construction of a weighted self-orthogonal Goppa code we
always set
\begin{eqnarray*}
  W & = & - D + (2n + (2g-2))P_{\infty}
\end{eqnarray*}
where $W$ is the canonical divisor.

\begin{lemma}
  \label{etaequalsw}
  $\eta$ is the differential corresponding to $W$ with
  \begin{eqnarray*}
    \eta & = & \frac{1}{y \; \prod_{P_i
	\,\rm{cor.}\,\alpha_i}(x-\alpha_i)} {\rm d} x
  \end{eqnarray*}
  where the elements $\alpha_i$ are the $x$-coordinates of the points on the
  curve corresponding to the places $P_i$.
\end{lemma}

\begin{proof} 
  We will calculate $(\eta)$ explicitly.
  \begin{eqnarray}
    (\eta) 
    & = & \left( \frac{1}{y} \right) + \left( \frac{1}{\prod
      (x-\alpha_i)} \right) + \left( {\rm d} x \right)\nonumber\\
    & = & \left( \frac{1}{y} \right) + 2n P_{\infty} - \sum P_i - \sum
    \sigma P_i - 2 (x)_{\infty} + \Diff(F/K(x))\nonumber\\
    & = & - \sum_{{\rm irred. \, of} \, f(x)} Q_i +
    (2g+1)P_{\infty} + 2n P_{\infty} - \sum P_i - \sum
    \sigma P_i \nonumber\\
    && - 2 \cdot 2 P_{\infty} + \sum_P \sum_{P^{\prime}|P}
    (e(P)-1) P^{\prime} \label{differentialhyper}\\
    & = & - D + (2n + 2g - 3)P_{\infty} - \sum Q_i + \sum 1 \cdot Q_i + 1 \cdot
    P_{\infty}\nonumber\\
    & = & - D + (2n + (2g-2))P_{\infty}\nonumber\\
    & = & W
  \end{eqnarray}
  
  Equation (\ref{differentialhyper}) follows from that fact that 
  \begin{eqnarray*}
    g = \frac{\deg f(x) -1}{2} & \Longleftrightarrow & \deg f(x) = 2g+1
  \end{eqnarray*}
  for $(2g+1)P_{\infty}$, and that $e(P) = 2$ for the irreducible components of
  $f(x)$ and $P_{\infty}$, otherwise $e(P) = 1$.

\end{proof}

\begin{lemma}
  \label{respisigmapi}
  The differential
  \begin{eqnarray*}
    \eta & = & \frac{1}{y \; \prod_{P_i
	\,\rm{cor.}\,\alpha_i}(x-\alpha_i)} {\rm d} x
  \end{eqnarray*}
  satisfies
  \begin{eqnarray*}
    \res{P_i}{\eta} & = & - \res{\sigma P_i}{\eta}.
  \end{eqnarray*}
\end{lemma}

\begin{proof}
  Let $(\alpha_i,\beta_i)$ be the corresponding point to $P_i$. Then
  $(\alpha_i, -\beta_i)$ is the corresponding to $\sigma P_i$. We get
  \begin{eqnarray*}
    \res{P_i}{\eta} & = & \frac{1}{\beta_i \prod_{i \ne j} (\alpha_i -
    \alpha_j)}\\
    \res{\sigma P_i}{\eta} & = & \frac{1}{-\beta_i \prod_{i \ne j}
    (\alpha_i - \alpha_j)}\\
      & = & - \frac{1}{\beta_i \prod_{i \ne j} (\alpha_i -
	\alpha_j)}\\
      & = & - \res{P_i}{\eta}.
  \end{eqnarray*}
\end{proof}

\subsubsection{Weighted Self-Orthogonality and Quantum Code Construction}


The following proposition is similar to Proposition \ref{matsuI} except that
the residues at the conjugate places do not have to be $1$ and $-1$, but
negative to each other. Proposition \ref{matsuI} is stronger, because it shows
the existence of a special differential $\eta$, but the proof is not
constructive. Here we prove the weaker version, because the differential
constructed in Lemma \ref{etaequalsw} and Lemma \ref{respisigmapi} satisfies
these conditions.

\begin{prop}
  \label{dualequalsh}
  Let $F/\F_q$ be an algebraic function field, $\sigma$ an automorphism of
  order 2 of $F$ not moving elements in $\F_q$, and $P_1,\ldots,P_n$ pairwise
  distinct places of degree one such that $\sigma P_i \ne P_j$ for all
  $i,j=1,\ldots,n$, $D=P_1+\cdots+P_n+ \sigma P_1+\cdots+ \sigma P_n$. Let
  $\eta$ be a differential with the properties
  \begin{displaymath}
    \left\{ \begin{array}{rcl} 
      v_{P_i}(\eta) & = & v_{\sigma P_i}(\eta) = -1,\\
      \res{P_i}{\eta} & = & - \res{\sigma P_i}{\eta}.
    \end{array}
    \right.
  \end{displaymath}
  Further assume that we have a divisor $G$ such that $\sigma G = G$,
  $v_{P_i}(G) = v_{\sigma P_i}(G) = 0$. Define
  \begin{eqnarray*}
    C(D,G) & = & \set{(f(P_1),\ldots , f(P_n), f(\sigma P_1), \ldots , f(\sigma
      P_n)) \; | \; f \in \codeL(G)}  \subseteq \mathbb{F}_q^{2n}.
  \end{eqnarray*}
  Let $H = D - G + (\eta)$, then we have $C(D,G)^{\perp_s^a} = C(D,H)$ where
  $a = (a_1, \ldots, a_n)$ are the weights of the symplectic inner product.
\end{prop}

\begin{proof}
  The proof is similar to the one of Proposition \ref{matsuI}. The only
  difference is the more general assumption that $\res{P_i}{\eta} = -
  \res{\sigma P_i}{\eta}$, instead of $1$ and $-1$.
\end{proof}

\begin{corollary}
   \label{cquasiselforth}
   Let $C(D,G)$ be the Goppa code constructed above and 
   \begin{displaymath}
     a = (a_1,\ldots,a_n)= (\res{P_1}{\eta},\ldots,\res{P_n}{\eta}).
   \end{displaymath}
   Then the code satisfies $C(D,G)
   \subseteq C(D,H)$ with $H = D-G+(\eta)$ and therefore
   \begin{displaymath}
     C(D,G) \subseteq C(D,G)^{\perp_s^a}
   \end{displaymath}
   with respect to the symplectic inner product $\sipa{\;}{\,}$.
\end{corollary}

\begin{proof}
  Note that $G = (n+g-1-r)P_{\infty}$ and $(\eta) = -D + (2n + 2g -
  2)P_{\infty}$. Therefore we can calculate $H$ in the following
  \begin{eqnarray*}
    H
    & = & D - G + (\eta)\\
    & = & D - (n+g-1-r)P_{\infty} + (-D + (2n + 2g - 2)P_{\infty})\\
    & = & (n+g-1+r)P_{\infty}\\
    &\ge& G.
  \end{eqnarray*}
  So we get $\codeL(G) \subseteq \codeL(H)$ and thereby $C(D,G) \subseteq
    C(D,H)$.  Now we apply Proposition \ref{dualequalsh} and use that
    $C(D,G)^{\perp_s^a} = C(D,H)$. This leads to the following implications
    and concludes the proof:
  \begin{enumerate}
  \item Then for all $x \in C(D,G)$, $y \in C(D,H)$ we have that
    \begin{eqnarray*}
      \sipa{x}{y} &=& \sum_{i=1}^n a_i (x_i y_{n+i} - x_{n+i} y_i) = 0.
    \end{eqnarray*}
  \item In particular because $C(D,G) \subseteq C(D,H)$ this equation holds
    for all $x,y \in C(D,G)$.
  \item Therefore $C(D,G) \subseteq C(D,G)^{\perp_s^a}$ and the claim is
  proved.
  \end{enumerate}
\end{proof}

\begin{corollary}
  \label{transformcdg}
  For $G \le H$, $G$ and $H$ as above, we have $C(D,G) \subseteq C(D,H)$. If
  we multiply
  \begin{eqnarray*}
    C^{\prime}(D,G) & = & C(D,G) \cdot
    \diag(\res{P_1}{\eta},\ldots,\res{P_n}{\eta},1,\ldots,1)
  \end{eqnarray*}
then $C^{\prime}$ is a linear self-orthogonal code with respect to the
standard symplectic inner product. This code modification does not change the
code properties and defines a quantum code.
\end{corollary}

\begin{proof}
   The proof of Corollary \ref{cquasiselforth} is independent of the choice of
   $D$ and $G$. Therefore $C(D,G) \subseteq C(D,G)^{\perp_s^a}$ holds under
   the more general assumptions of this corollary. Hence $C(D,G)$ meets the
   conditions of a stabilizer code and Lemma \ref{codemod} proves the
   corollary.
\end{proof}

\subsubsection{Code Properties}


\begin{prop}
  \label{qeccprops}
  Let $C^{\prime}(D,G)$ be the Goppa code constructed by Corollary
  \ref{cquasiselforth} and Corollary \ref{transformcdg}. Then
  $C^{\prime}(D,G)$ is a stabilizer code with parameters $[[n,k,d]]$
  where $n$ is the number of used splitting rational places of the rational
  function field, $k \ge r$ and $d \ge \frac{n-g+1-r}{2}$.
\end{prop}

\begin{proof}$ $
  \begin{enumerate}
  \item We can construct a stabilizer code, because $C^{\prime}(D,G) \subseteq
    C^{\prime}(D,G)^{\perp_s}$ (see Theorem \ref{kdqecc}). Since our code
    transformation did not change the code properties we can calculate the
    properties of $C(D,G)$ in order to get those of $C^{\prime}(D,G)$.
  \item Now we can apply Corollary \ref{matsuII} by changing $G$ and $H$ which
  enables us to calculate $k$ and $d$.
  \item For $d$ we get
    \begin{eqnarray*}
      d
      & \ge & n - \floor{\frac{\deg H}{2}}\\
      & = & n - \floor{\frac{n+g-1+r}{2}}\\
      & \ge & \frac{n-g+1-r}{2}.
    \end{eqnarray*}
  \item For the bound on $k$ we need the Theorem of Riemann-Roch
    \ref{rieroch}.
    \begin{eqnarray}
      k
      & = & \dim H - \dim(H-D) -n\nonumber\\
      &\ge& \deg H + 1 - g - \dim(H-D) -n\nonumber\\
      & = & (n+g-1+r)+ 1 - g - \dim(H-D) -n\nonumber\\
      & = & r.\label{khyper}
    \end{eqnarray}

    Equation (\ref{khyper}) follows because $\dim(H-D) = 0$

    {\it Proof.} Note that $0 \le r \le n-g$. Therefore
    the following equation holds:
    \begin{eqnarray*}
      \deg (H-D)
      & = & \deg H - \deg D\\
      & = & (n+g-1+r) - 2n\\
      & = & r - (n-g) -1\\
      &\le& -1.
    \end{eqnarray*}
    Now we can apply Corollary \ref{degdim} and get $\dim (H-D) = 0$.
  \end{enumerate}
\end{proof}

\section{Projection onto the Prime Field}


If we have a code over a field $\F_{p^m}$, it may be desirable to project it
onto the base field $\F_p$. For this projection we can use the ideas of
Ashikhmin and Knill in $\cite{ashknill}$. Note that this projection is always
possible. In the following we will also see under which circumstances it is
possible to find a nicer projection that allows us to see the coefficients
$a_i$ respectively the new coefficients $a_{ij}$ as weights of the symplectic
inner product.

\subsection{General Case}

The general case uses almost completely Ashikhmin and Knill's ideas in
$\cite{ashknill}$. It gives us the following projection properties.

\begin{prop}
  Let $\set{\alpha_1, \ldots, \alpha_m}$ be a self-dual basis for 
  $\F_{p^m}$ over $\F_p$ , then the projection 
  \begin{eqnarray*}
    (a_1 x_1,\ldots, a_n x_n,x_{n+1},\ldots, x_{2n}) & \mapsto &
    ((a_1 x_1)^{(1)},(a_1 x_1)^{(2)},\ldots, x_{2n}^{(m)}),
  \end{eqnarray*}
  where each component $x_i = x_i^{(1)} \alpha_1 + \cdots + x_i^{(m)}
  \alpha_m$ is represented in the self-dual basis, of the code
  $C^{\prime}(D,G)$ onto the base field gives a stabilizer code $\code(D,G)$
  with respect to the symplectic inner product
  \begin{eqnarray*}
    \sipp{x}{y} & = & \sum_{i=1}^n \sum_{j=1}^m ((a_i \,x_i)^{(j)} \,
    y_{n+i}^{(j)} - x_{n+i}^{(j)} \,(a_i \, y_i)^{(j)}).
  \end{eqnarray*}

\end{prop}

\begin{proof}
  First observe that if a symplectic inner product of two vectors is zero,
  then so is the trace over it \cite{ashknill} where the trace of an element
  $\alpha \in \F_{p^m}$ over the base field is defined as
  \begin{eqnarray*}
    \trace(\alpha) & = & \sum_{\nu=1}^m \sigma_\nu \, \alpha,
  \end{eqnarray*}
  where the elements $\sigma_i$ are the Galois automorphisms of
  $\F_{p^m}/\F_p$ \cite[Chapter VI.5]{lang}. Next we will show that the new
  code is still self-orthogonal with respect to the new inner product. For
  $x,y \in C^{\prime}(D,G)$ we get that
  \begin{eqnarray*}
    0 = \trace(\sip{x}{y}) & = & \trace \left(\sum_{i=1}^n (a_i \,x_i \,
    y_{n+i} - x_{n+i} \,a_i \, y_i)\right)\\
    & = & \sum_{i=1}^n \sum_{j=1}^m \sum_{k=1}^m ((a_i \,x_i)^{(j)} \,
    y_{n+i}^{(k)} - x_{n+i}^{(j)} \,(a_i \, y_i)^{(k)})
    \underbrace{\trace(\alpha_j \alpha_k)}_{= \delta_{jk}}\\
    & = & \sum_{i=1}^n \sum_{j=1}^m ((a_i \,x_i)^{(j)} \, y_{n+i}^{(j)} -
    x_{n+i}^{(j)} \,(a_i \, y_i)^{(j)}),
  \end{eqnarray*}
  and therefore $\code(D,G) \subseteq \code(D,G)^{\perp_s^p}$.  Finally, we
  have to show that this code is again a linear code. Therefore let 
    \begin{eqnarray*}
      {\cal X} & = & ((a_1 x_1)^{(1)},(a_1 x_1)^{(2)},\ldots, x_{2n}^{(m)}) \;
    \mathrm{and}\\
    {\cal Y} & = & ((a_1 y_1)^{(1)},(a_1 y_1)^{(2)},\ldots, y_{2n}^{(m)}) \in
    \code(D,G),
    \end{eqnarray*}
    then there exist $x, y \in C^{\prime}(D,G)$ with
    \begin{eqnarray*}
      x & = & ((a_1 x_1)^{(1)} \alpha_1
      + \cdots + (a_1 x_1)^{(m)} \alpha_m, \ldots, x_{2n}^{(1)}\alpha_1 +
      x_{2n}^{(m)} \alpha_m)\\
      y & = & ((a_1 y_1)^{(1)} \alpha_1 + \cdots + (a_1 y_1)^{(m)} \alpha_m,
      \ldots, x_{2n}^{(1)}\alpha_1 + x_{2n}^{(m)} \alpha_m).
    \end{eqnarray*}
    Let $s \in \F_p$. As $C^{\prime}(D,G)$ is a linear code, and $\F_{p^m}$ is
    a $\F_p$-linear vector space,
    \begin{eqnarray*}
      & s\cdot x+y = ([s\cdot (a_1 x_1)^{(1)} + (a_1 y_1)^{(1)}] \alpha_1 +
      \cdots + \\ 
      & [s \cdot (a_1 x_1)^{(m)} + (a_1 y_1)^{(m)}] \alpha_m, \ldots, [s\cdot
      x_{2n}^{(1)} + y_{2n}^{(1)}] \alpha_1 + [s\cdot x_{2n}^{(m)} +
      y_{2n}^{(m)}] \alpha_m)
    \end{eqnarray*}
    is in $C^{\prime}(D,G)$ and therefore 
    \begin{eqnarray*}
      s \cdot {\cal X} + {\cal Y}
      & = & (s\cdot (a_1 x_1)^{(1)} + (a_1 y_1)^{(1)},\ldots, s\cdot (a_1
      x_1)^{(m)} + (a_1 y_1)^{(m)},\\
      && \ldots, s\cdot x_{2n}^{(1)} +
      y_{2n}^{(1)}, \ldots, s\cdot x_{2n}^{(m)} + y_{2n}^{(m)}) \in \code(D,G).
    \end{eqnarray*}
    
    Therefore the projected code is a linear code satisfying the symplectic
    inner product and defines a quantum stabilizer code.
\end{proof}

\begin{figure}
  \centering
  \includegraphics[scale=0.5]{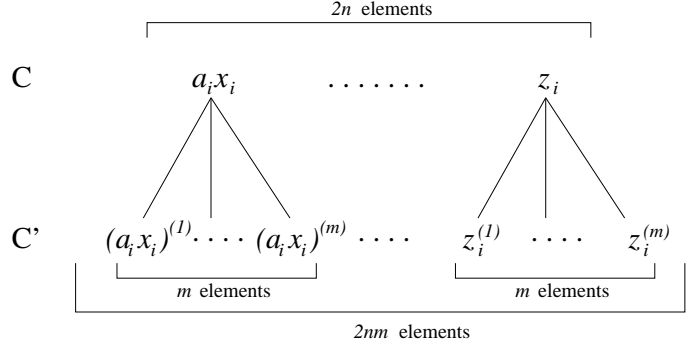}
  \caption{Projection from a code with codewords of length $2n$ over
  $\F_{p^m}$ onto codewords of length $2nm$ over the base field $\F_p$.}
  \label{projection}
\end{figure}
Figure \ref{projection} illustrates how the codewords split with respect to
the self-dual basis. Note that this $p$-ary linear code cannot necessarily be
written as a code of the form
\begin{displaymath}
  {\cal C}^{\prime}(D,G) \cdot \diag(a_1^{(1)},\ldots,a_n^{(m)},1,\ldots,1)
\end{displaymath}
like in the $p^m$-ary case. This is possible in some special cases shown in
the next section.

\subsection{Special Case: The Weights $a_i \in \F_p$ for all $i$}

The following special case allows us to write our projected code in the form
\begin{displaymath}
  {\cal C}^{\prime}(D,G) \cdot \diag(a_1^{(1)},\ldots,a_n^{(m)},1,\ldots,1).
\end{displaymath}

\begin{figure}
  \centering
  \includegraphics[scale=0.5]{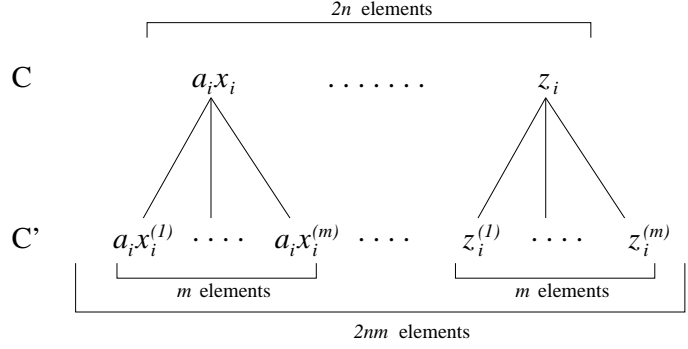}
  \caption{Projection of a codeword of the code $C$ over $\F_{p^m}$ onto the
  base field $\F_p$ if all $a_i$ are in $\F_p$.}
  \label{projection2}
\end{figure}

A way to get an easy projection is the case that all $a_i$ are elements of
$\F_p$, because the trace is $\F_p$-linear, i.e. $\trace(a_i) = a_i
\trace(1)$. So we get for all codewords

\begin{eqnarray*}
  0 = \trace{\sipa{x}{y}}
  & = & \trace(\sum_{i=1}^n a_i (x_i y_{n+i} - x_{n+i} y_i))\\
  & = & \trace(\sum_{i=1}^n a_i \sum_{j=1}^m \sum_{k=1}^m  (x_i^{(j)}
  y_{n+i}^{(k)}\alpha_j \alpha_k - x_{n+i}^{(j)} y_i^{(j)}\alpha_j
  \alpha_k))\\
  & = & \sum_{i=1}^n a_i \sum_{j=1}^m \sum_{k=1}^m  (x_i^{(j)}
  y_{n+i}^{(k)} - x_{n+i}^{(j)}
  y_i^{(j)})\underbrace{\trace(\alpha_j \alpha_k)}_{= \delta_{jk}}\\
  & = & \sum_{i=1}^n \sum_{j=1}^m a_i (x_i^{(j)} y_{n+i}^{(k)} - x_{n+i}^{(j)}
  y_i^{(j)}),\\
\end{eqnarray*}
and our new code can be written as
\begin{displaymath}
  C^{\prime} = \set{(a_1 x_1^{(1)},
    \ldots,a_nx_n^{(m)} |
    z_{1}^{(1)},\ldots,z_{n}^{(m)})\; : \; (a_1 x_1,\ldots,a_n
    x_n|z_{1},\ldots,z_{n}) \in C}.
\end{displaymath}
Figure \ref{projection2} shows how the coefficients $a_i$ split with respect
to the self-dual basis. The figure is just slightly different from the general
case of the previous section.

\section{Examples}


\subsection{Curves with Many Rational Points}


An example of a curve with many rational points is given in \cite[Theorem
6.14]{stepanov}. We will analyse the case over $\F_{p^m}$ where $m$ is odd.
Let $y^2 = f(x)$ with
\begin{eqnarray*}
  f(x) & = & (x + x^{p^{\frac{m-1}{2}}})(x + x^{p^{\frac{m+1}{2}}}) \qquad \in
  \F_{p^m}[x]
\end{eqnarray*}
be a hyperelliptic curve. This curve does not satisfy the standard definition
of a hyperelliptic curve because it is not square-free, i.e. $x^2$ can be
extracted from the equation. In the following we see that it is still possible
to construct a quantum code, i.e. a quantum code with good parameters.

\begin{claim}
  $f(x)$ has no linear divisors except for $x$. 
\end{claim}

\begin{proof}
  We can rewrite $f(x)$ as
  \begin{eqnarray*}
    f(x) & = & x^2(1 + x^{p^{\frac{m-1}{2}}-1})(1 + x^{p^{\frac{m+1}{2}}-1}).
  \end{eqnarray*}
  It suffices to show that none of the two factors different from $x^2$ is
  divisible by a linear polynomial.

  Let $\alpha \in \F_{p^m}$, then
  \begin{eqnarray*}
    \frac{1 + x^{p^{\frac{m-1}{2}}-1}}{x- \alpha} & = & 
    \sum_{i=0}^{p^{\frac{m-1}{2}}-2} \alpha^i x^{p^{\frac{m-1}{2}}-2-i}
    + \frac{\alpha^{p^{\frac{m-1}{2}}-1}+1}{x- \alpha}.
  \end{eqnarray*}
  This equation can only hold if $\alpha^{p^{\frac{m-1}{2}}-1}+1 = 0$ and
  therefore $\alpha^{p^{\frac{m-1}{2}}-1} = -1$. Then
  $\alpha^{2(p^{\frac{m-1}{2}}-1)} = 1$. This is only possible if
  $2(p^{\frac{m-1}{2}}-1) | p^m - 1$. Assume $2(p^{\frac{m-1}{2}}-1) | p^m -
  1$, then
  \begin{eqnarray*}
    \frac{p^m - 1}{2(p^{\frac{m-1}{2}}-1)} & = &
    2^{-1} p^{\frac{m+1}{2}} + 2^{-1} p + \frac{p-1}{2(p^{\frac{m-1}{2}}-1)}.
  \end{eqnarray*}
  So the assumption is only true if $p-1 = 0$, therefore $p=1$. This
  contradicts that $p$ is an odd prime greater than two. Therefore we have
  shown that there exists no linear polynomial dividing $1 +
  x^{p^{\frac{m-1}{2}}-1}$. Now we have to do the same calculations for the
  second factor: Let again $\alpha \in \F_{p^m}$, then
  \begin{eqnarray*}
    \frac{1 + x^{p^{\frac{m+1}{2}}-1}}{x- \alpha} & = & 
    \sum_{i=0}^{p^{\frac{m+1}{2}}-2} \alpha^i x^{p^{\frac{m-1}{2}}-2-i}
    + \frac{\alpha^{p^{\frac{m+1}{2}}-1}+1}{x- \alpha}
  \end{eqnarray*}
  This equation can only hold if $\alpha^{p^{\frac{m+1}{2}}-1}+1 = 0$ and
  therefore $\alpha^{p^{\frac{m+1}{2}}-1} = -1$. If there exists a solution,
  then $2(p^{\frac{m+1}{2}}-1) | p^m - 1$. Assume this equation is true, then
  \begin{eqnarray*}
    \frac{p^m - 1}{2(p^{\frac{m+1}{2}}-1)} & = &
    2^{-1} p^{\frac{m-1}{2}} +
    \frac{p^{\frac{m-1}{2}}-1}{2(p^{\frac{m+1}{2}}-1)}.
  \end{eqnarray*}
  So the assumption is only true if $p^{\frac{m-1}{2}}-1 = 0$, but the
  calculations take place in $\F_{p^m}$ which is a field of characteristic $p$
  and therefore $p^{\frac{m-1}{2}}=0$ which leads to the contradiction $0=
  -1$.

  We conclude that there exists no $\alpha \ne 0$ such that $x-\alpha$ divides
  $f(x)$. Therefore $f(x)$ has no linear divisors except for $x$.
\end{proof}

Corollary 6.15 in \cite{stepanov} shows that it is possible to actually
calculate the number of rational points
\begin{eqnarray*}
  N_{p^m} & = & 2p^m-1
\end{eqnarray*}
As the only linear polynomial that divides $f(x)$ is $x$, there are
$\frac{2p^m-4}{2} = p^m - 2$ pairs that can be used for a quantum stabilizer
code. This follows immediately from Proposition \ref{hypercurvefield} that the
only rational places are those which are zeros of $f(x)$ and $P_{\infty}$.

Now we will apply our construction of Section \ref{qagcode}, use all rational
pairs, and look at the asymptotics in the limit of large $m$.

\begin{enumerate}
  
\item First let us calculate the ratio of rational places versus genus of the
  curve:
  \begin{eqnarray*}
    g_{p^m} 
    & = & \frac{\deg f(x) - 1}{2}\\
    & = & \frac{p^{\frac{m-1}{2}}(p+1) - 1}{2}\\
  \end{eqnarray*}
  Hence the ratio of $N_{p^m}$ and $g_{p^m}$ for $m$ to infinity is given by
  \begin{eqnarray*}
    \lim_{m \rightarrow \infty} \frac{N_{p^m}}{g_{p^m}}
    & = & \lim_{m \rightarrow \infty} 2 \cdot \frac{2p^m -
    1}{p^{\frac{m-1}{2}}(p+1) - 1}\\
    & = & \lim_{m \rightarrow \infty} \frac{4}{p+1} \cdot p^{\frac{m+1}{2}} =
    + \infty \\
    & > & 0.
  \end{eqnarray*}
  This ratio is not bounded and goes to infinity. Therefore it is good.

\item The parameters of our family of codes $C_m(D_m,G_m)$ are given by
  \begin{eqnarray*}
    n_m & = & p^m - 2,\\
    k_m &\ge& r_m,\\
    d_m &\ge& \frac{n_m - g_m +1 - r_m}{2}.
  \end{eqnarray*}

\item Setting $r_m = \floor{R \; n_m} \le n_m - g_m$, we obtain
  \begin{eqnarray*}
    \lim_{m \rightarrow \infty} \frac{k_m}{n_m}
    &\ge& \lim_{m \rightarrow \infty} \frac{r_m}{n_m}\\
    & = & \lim_{m \rightarrow \infty} \frac{\floor{R \; n_m}}{n_m}\\
    &\ge& R\\
    & > & 0,
  \end{eqnarray*}
  and for the distance
  \begin{eqnarray*}
    \lim_{m \rightarrow \infty} \frac{d_m}{n_m}
    &\ge& \lim_{m \rightarrow \infty} \frac{\frac{n_m - g_m +1 -
	r_m}{2}}{n_m}\\
    & = & \lim_{m \rightarrow \infty} \frac{1}{2} \left( \frac{n_m}{n_m} -
    \frac{g_m}{n_m} + \frac{1}{n_m} - \frac{\floor{R \; n_m}}{n_m} \right)\\
    &\ge& \frac{1}{2} - R\\
    & > & 0
  \end{eqnarray*}
  for $R < \frac{1}{2}$.
\end{enumerate}

This gives good quantum codes as long as we do not project it onto the base
field, because the down projection will not necessarily enlarge the distance
of the code and therefore the ratio cannot be bounded from zero. The
disadvantage of these codes is that their alphabet size tends to infinity.

\subsection{Magma Calculations}

In this section we explicitly construct some quantum AG codes with the help of
Magma \cite{magma}. Magma outputs a classical Goppa code and the residues of
the corresponding differential. With simple calculations and reordering of the
columns we get a quantum error correcting code with respect to the standard
symplectic inner product.

\begin{example}
  This example continues the computations begun in Example \ref{exgoppa} and
  transforms it into a quantum code. The Magma code can be found in Example
  \ref{exgoppa}.

  \begin{verbatim}
    //differential corresponding to W
    k := Canonical(W);
    
    //residues of k at the D[i]'s
    for i:= 1 to #D do
      Residue(k,D[i]);
    end for;
  \end{verbatim}

  If we set $r=1$, where $0 \le r \le n-g$ is the parameter that can be chosen,
  the error correcting code $C$ is equal to the one in Example \ref{exgoppa}.
  \begin{verbatim}
    > C;
    [14, 6] Linear Code over GF(19)
    Generator matrix:
    [ 1  0  0  0  0 13  0 14 15 11 17 15  4 16]
    [ 0  1  0  0  0  6  0  5  6 10 13 15 13  1]
    [ 0  0  1  0  0 10  0  7 12 11  2  6  6 14]
    [ 0  0  0  1  0  9  0 12  4  5 16 12  2 13]
    [ 0  0  0  0  1  1  0  0  4  4  5  5  3  3]
    [ 0  0  0  0  0  0  1  1 17 17  5  5 11 11]
  \end{verbatim}
  Now we have to rearrange the columns with the permutation
  \begin{eqnarray*}
    (x_1, z_1, x_2, z_2, \ldots, x_n, z_n) & \mapsto &
    (x_1,x_2,\ldots,x_n \,|\, z_1, z_2,\ldots,z_n),
  \end{eqnarray*}
  and we get the generator matrix
  \begin{eqnarray*}
    \G_1 & = & \left(
    \begin{array}{ccccccc|ccccccc}
      1&0&0&0&15&17&4 & 0&0&13&14&11&15&16\\
      0&0&0&0&6&13&13 & 1&0&6&5&10&15&1\\
      0&1&0&0&12&2&6 & 0&0&10&7&11&6&14\\
      0&0&0&0&4&16&2 & 0&1&9&12&5&12&13\\
      0&0&1&0&4&5&3 & 0&0&1&0&4&5&3\\
      0&0&0&1&17&5&11 & 0&0&0&1&17&5&11
    \end{array}
    \right).
  \end{eqnarray*}
  For the quantum code with respect to the symplectic inner product
  $\sipa{x}{y}= \sum_{i=1}^n a_i (x_i y_{n+i} - x_{n+i} y_i)$ we have to
  determine the elements $a_i$ which are given by the residues of the
  differential
  \begin{verbatim}
    > k;
    ((a + 2)^-1 * (a + 3)^-1 * (a + 4)^-1 * (a + 8)^-1 
    * (a + 12)^-1 * (a + 13)^-1 * (a + 7)^-1 * (b)^-1) d(a)
  \end{verbatim}
  These are given by the vector
  \begin{eqnarray*}
    a & = & (a_1,a_2,a_3,a_4,a_5,a_6,a_7)\\
    & = & (3,11,1,10,14,5,12).
  \end{eqnarray*}
  Therefore we have to multiply the columns with the residues which is the
  same as multiplying $G_1 \cdot D$ where $D = \diag(a_1,\ldots,a_n,1,\ldots,
  1)$. The transformed quantum code is given by the matrix
  \begin{eqnarray*}
    \G_1^{\prime} & = & \left(
    \begin{array}{ccccccc|ccccccc}
      3&0&0&0&1&9&10 & 0&0&13&14&11&15&16\\
      0&0&0&0&8&8&4 & 1&0&6&5&10&15&1\\
      0&11&0&0&16&10&15 & 0&0&10&7&11&6&14\\
      0&0&0&0&18&4&5 & 0&1&9&12&5&12&13\\
      0&0&1&0&18&6&17 & 0&0&1&0&4&5&3\\
      0&0&0&10&10&6&18 & 0&0&0&1&17&5&11
    \end{array}
    \right).
  \end{eqnarray*}

  If we take another $r$ and set $r=2$, we get the code
  \begin{verbatim}
    > C;
    [14, 5] Linear Code over GF(19)
    Generator matrix:
    [ 1  0  0 14  0  6  0 11 14  5 13 12 13  8]
    [ 0  1  0  5  0 13  0  8  7 16 17 18  4  9]
    [ 0  0  1  1  0  0  0  0 16 16 18 18  8  8]
    [ 0  0  0  0  1  1  0  0  4  4  5  5  3  3]
    [ 0  0  0  0  0  0  1  1 17 17  5  5 11 11]
  \end{verbatim}
  
  and therefore the stabilizer matrix
  \begin{eqnarray*}
    \G_2 & = & \left(
    \begin{array}{ccccccc|ccccccc}
      1&0&0&0&14&13&13 & 0&14&6&11&5&12&8\\
      0&0&0&0&7&17&4 & 1&5&13&8&16&18&9\\
      0&1&0&0&16&18&8 & 0&1&0&0&16&18&8\\
      0&0&1&0&4&5&3 & 0&0&1&0&4&5&3\\
      0&0&0&1&17&5&11 & 0&0&0&1&17&5&11
    \end{array}
    \right)
  \end{eqnarray*}
  with respect to the same symplectic inner product $\sipa{x}{y}$ as
  above. The transformation of the code gives us
  \begin{eqnarray*}
    \G_2^{\prime} & = & \left(
    \begin{array}{ccccccc|ccccccc}
      3&0&0&0&6&8&4 & 0&14&6&11&5&12&8\\
      0&0&0&0&3&9&10 & 1&5&13&8&16&18&9\\
      0&11&0&0&15&14&1 & 0&1&0&0&16&18&8\\
      0&0&1&0&18&6&17 & 0&0&1&0&4&5&3\\
      0&0&0&10&10&6&18 & 0&0&0&1&17&5&11
    \end{array}
    \right).
  \end{eqnarray*}
Therefore we constructed two quantum Goppa codes over $\F_{19}$ with parameters
$\qecc{7}{1}{\ge3}$ and $\qecc{7}{2}{\ge2}$.
\end{example}

The following example uses a prime power and works with abstract symbols
instead of ``numbers''.

\begin{example}
  Let us work over $\F_9$ with the hyperelliptic curve
  \begin{eqnarray*}
    y^2 & = & x^5 - 2 \cdot x^3 + x^2 + 1.
  \end{eqnarray*}
  Then we construct with Magma \cite{magma} the following algebraic code:
  \begin{verbatim}
    > C;
    [8, 3, 5] Linear Code over GF(3^2)
    Generator matrix:
    [  1   0   0   w   w   1 w^6 w^5]
    [  0   1   0 w^5   w w^6 w^3 w^5]
    [  0   0   1   1 w^2 w^2 w^3 w^3]
  \end{verbatim}
  and if we order the columns with respect to the conjugated pairs, we get
  \begin{eqnarray*}
    \G_3 & = & \left(
    \begin{array}{cccc|cccc}
      1&0&w&w^6 & 0&w&1&w^5\\
      0&0&w&w^3 & 1&w^5&w^6&w^5\\
      0&1&w^2&w^3 & 0&1&w^2&w^3
    \end{array}
    \right),
  \end{eqnarray*}
  where $w$ is a generator of $\F_9$. In this case the differential $k$ is
  given by
  \begin{verbatim}
    > k;
    ((a + w^3)^-1 * (a + 2)^-1 * (a)^-1 * (a + w)^-1 
    * (a + 1)^-2 * (a + w^2)^-1 * (a + w^6)^-1 * (b)) d(a)
  \end{verbatim}
  and we get the vector of residues
  \begin{eqnarray*}
    a & = & (a_1,a_2,a_3,a_4)\\
    & = & (2,w^2,w^6,1).
\end{eqnarray*}
  Finally the quantum AG code is given by the matrix
  \begin{eqnarray*}
    \G_3^{\prime} & = & \left(
    \begin{array}{cccc|cccc}
      2&0&w^7&w^6 & 0&w&1&w^5\\
      0&0&w^7&w^3 & 1&w^5&w^6&w^5\\
      0&w^2&1&w^3 & 0&1&w^2&w^3
    \end{array}
    \right).
  \end{eqnarray*}
\end{example}

\begin{example}

  In general, we can use the following program to construct codes, if we use a
  curve that has no linear factors over the given field:

  \begin{verbatim}
    constr_field := function(q)
    //as before
    
    constr_curve := function(P2, f)
    //as before

    constr_code := function(K,P2,f,r)

    local X,g,DG,place1,F,D,D3,G,W,C,k,res;

    X,g := constr_curve(P2,f);
    DG := DivisorGroup(X);
    //place1 gives all places of degree 1
    place1 := Places(X,1);
    F<a,b> := FunctionField(X);
    
    //D are the places where we evaluate the elements of G
    D := [];
    for i:= 1 to (Floor((#place1-1)/2)) do      
      D[i] := place1[2*i];
      D[(Floor((#place1-1)/2)) + i] := place1[2*i+1];
    end for;
    D3 := DG! &+D;
    
    //r can be varied to change the dimension of G
    //G is the space of "codewords"
    G := (Floor(#D/2)+g-1-r) * DG!place1[1];
    //W is canonical divisor
    W := - D3 + (#D + (2*g-2)) * DG!place1[1];
    C := AlgebraicGeometricCode(D,G);
    
    
    //differential corresponding to W
    k := Canonical(W);
    
    //residues of k at the D[i]'s
    res := [];	
    for i:= 1 to Floor(#D/2) do
      res[i] := Residue(k,D[i]);
    end for;

    return C, res;

    end function;
  \end{verbatim}
  
  To use this program we first have to define a field
  \begin{verbatim}
    > K<w>,P2<x,y,z> := constr_field(3);
  \end{verbatim}
  and a curve that has no linear factors over K
  \begin{verbatim}
    f := -y^2*z^3 + (x^2 + z^2)*(x^3 + 2*x^2*z + z^3);
  \end{verbatim}
  Then we can follow the code construction by the command
  \begin{verbatim}
    C, residues := constr_code(K,P2,f,1);
  \end{verbatim}
  We obtain
  \begin{verbatim}
    > C;
    [4, 2, 2] Linear Code over GF(3)
    Generator matrix:
    [1 0 1 0]
    [0 1 0 1]    
    > residues;
    [ 2, 1 ]
  \end{verbatim}
  This means that we have obtained a quantum code with respect to the
  symplectic inner product given by the residues. The columns of the generator
  matrix are already in the right order. We just have to multiply the first
  half of the columns component wise with the residues.
\end{example}

\chapter{Conclusions}

In conclusion, we have seen some constructions of quantum Goppa codes over
binary and non-binary fields. First, a paper of Matsumoto about the
construction of good binary codes has been presented in detail. Second, we
have seen that hyperelliptic curves can be used to construct quantum Goppa
codes over arbitrary finite fields. We have presented two different methods to
construct quantum codes from algebraic curves: either we use the CSS
construction or we work with the properties of splitting places on
hyperelliptic curves and generate quantum codes directly. We have illustrated
the constructions by giving concrete examples, most of which have been
computed with the help of the computer algebra system Magma \cite{magma}. An
example of a family of asymptotically good codes has been presented where the
size of the alphabet grows to infinity. Furthermore, we have presented a way
to project codes over prime power fields on their base field. For the reader
who is not familiar with all the basics, introductions to coding theory,
algebraic geometry, and quantum error correction are provided.

Interesting questions for future work include how to use these quantum Goppa
codes over hyperelliptic curves to construct good families of codes. One
possibility to get good families is to think of a hyperelliptic curve as a
Kummer extension. Kummer extensions define infinite function field towers
that provide families of codes. These codes are asymptotically good if we find
a good tower, i.e. a tower with many rational places.

Finally, this thesis provides the first explicit construction of quantum Goppa
codes over non-binary fields that can be applied to all hyperelliptic curves.

\appendix

\chapter{Postulates of Quantum Mechanics}
\label{appendix}

Quantum mechanics is the key to quantum computing and quantum error
correction. Everything is based on the following four postulates that are
cited from $\cite{nielsen}$. We do not need them for the construction of
quantum error correcting codes, but they help us understand the principles of
quantum codes.
\begin{postulate}{1}
  Associated to any isolated physical system is a complex vector space
  $\hilbert$ with inner product (that is, a Hilbert space)
  known as the {\it state space} of the system. The system is completely
  described by its {\it state vector}, which is a unit vector in the system's
  state space.
\end{postulate}
This first postulate explains why we use a Hilbert space and qubits etc. as
basic states of our system.

\begin{postulate}{2}
  The evolution of a closed quantum system is described by a unitary
  transformation. That is, the state $\ket{\psi}$ of the system at time $t_1$
  is related to the state $\ket{\psi^{\prime}}$ of the system at the time
  $t_2$ by a unitary operator $U$ which depends only on the times $t_1$ and
  $t_2$,
  \begin{eqnarray*}
    \ket{\psi^{\prime}} & = & U \ket{\psi},
  \end{eqnarray*}
  i.e., the time evolution of the state of a closed quantum system is
  described by the {\it Schr\"odinger equation},
  \begin{eqnarray*}
    i\hbar \frac{d\ket{\psi}}{dt} & = & H \ket{\psi},
  \end{eqnarray*}
  where $\hbar$ is {\it Planck's constant} and $H$ is the {\it Hamiltonian} of
  the system.
\end{postulate}

\begin{postulate}{3}
  Quantum measurements are described by a collection $\set{M_m}$ of {\bf
  measurement operators}. These are positive hermitean operators acting on the
  state space of the system being measured. The index $m$ refers to the
  measurement outcomes that may occur in the experiment. If the state of the
  quantum system is $\ket{\psi}$ immediately before the measurement then the
  probability that result $m$ occurs is given by
  \begin{eqnarray*}
    p(m) & = & \bra{\psi} M_m^{\dagger} M_m \ket{\psi},
  \end{eqnarray*}
and the state of the system after the measurement is
\begin{displaymath}
  \frac{M_m \ket{\psi}}{\sqrt{\bra{\psi} M_m^{\dagger} M_m \ket{\psi}}}.
\end{displaymath}
The measurement operators satisfy the completeness equation,
\begin{eqnarray*}
  \sum_m M_m^{\dagger} M_m = \identity
\end{eqnarray*}
The completeness equation expresses the fact that probabilities sum to one:
\begin{displaymath}
  1 = \sum_m p(m) = \sum_m \bra{\psi} M_m^{\dagger} M_m \ket{\psi}
\end{displaymath}
\end{postulate}
Everything we have to know of Postulate 3 is that measuring an arbitrary
quantum state will destroy the superposition and project the state with
respect to a chosen basis.

If we define new operators $E_m \equiv M_m^{\dagger} M_m$, we get $\sum_m E_m
= \identity$. These positive operators $E_m$ are called {\bf POVM elements}
and the set $\{ E_m \}$, known as a {\bf POVM}, suffices to determin the
propabilities of the different measurement outcomes. 

If the operators $E_m$ satisfy $E_m \equiv P_m^{\dagger} P_m \equiv P_m$ and
$P_{m^{\prime}}^{\dagger} P_m \equiv \delta_{m^{\prime} m} P_m$, the operators
are knwon as projectors and the measurement is called {\bf projective} or {\bf
orthogonal}.  In this thesis we will assume that orthogonal measurements are
always possible.

\begin{postulate}{4}
  The state space of a composite physical system is the tensor product of the
  state spaces of the component physical systems. Moreover, if we have systems
  numbered 1 through $n$, and system number $i$ is prepared in the state
  $\ket{\psi_i}$, then the joint state of the total system is $\ket{\psi_1}
  \tensor \ket{\psi_2} \tensor \cdots \tensor \ket{\psi_n}$.
\end{postulate}
This postulate describes the existence of quantum registers introduced in
Definition \ref{qregister}.


\begin{thebibliography}{20}

\bibitem[1]{ashknill} A. Ashikhmin and E. Knill, "Nonbinary quantum stabilizer
  codes", IEEE Transactions on Information Theory Vol. 47 No. 7,
  pp. 3065-3072, November 2001.
\bibitem[2]{barg} A. Barg, ``Complexity Issues in Coding Theory'', Handbook
  of Coding Theory (V. Pless and W. C. Huffman, eds.), vol. 1, pp. 649--754,
  Elsevier Science, 1998.
\bibitem[3]{berlekamp} E. Berlekamp, R. McEliece, H. van Tilborg,''On the
  inherent intractability of certain coding problems'', IEEE
  Transactions on Information Theory, vol. 24, no. 3, pp. 384-386, 1978.
\bibitem[4]{farran} J. I. Farr\'an, ``Decoding algebraic geometry codes
  by a key equation'', math.AG/9910151, October 27, 1999.
\bibitem[5]{caldershor} A. R. Calderbank and Peter W. Shor, ``Good
  quantum error-correcting codes exist'', Phys. Rev. A, vol. 54,
  pp. 1098-1105, August 1996.
\bibitem[6]{eisenbud} D. Eisenbud, ``Commutative algebra with a view toward
algebraic geometry'', Springer, 1999
\bibitem[7]{gs-tower} A. Garcia and H. Stichtenoth, ``A tower of
  Artin-Schreier extensions of function fields, attaining the Drinfeld-Vladut
  bound'', Invent. Math., 121(1):211-222, July 1995.
\bibitem[8]{gottesman} D. Gottesman, ``Stabilizer Codes and Quantum Error
  Correction'', Ph.D. thesis, quant-ph/9705052, Pasadena 1997. 
\bibitem[9]{gottesmanft} D. Gottesman, ``Theory of fault-tolerant quantum
  computation'', Physical Review A, vol. 57, no. 1, pp. 127-137, Jan. 1998
\bibitem[10]{gottesmanheisenberg} D. Gottesman, ``The Heisenberg Representation
  of Quantum Computers'', Proceedings of the XXII International Colloquium on
  Group Theoretical Methods in Physics, eds. S. P. Corney, R. Delbourgo, and
  P. D. Jarvis, pp. 32-43, Cambridge, International Press, 1999, and
  quant-ph/9807006, July 1, 1998.
\bibitem[11]{qeclecture} D. Gottesman, Lecture Notes for CO639 ``Quantum
  Error Correction'', University of Waterloo, Winter 2004.\\
  {\texttt www.perimeterinstitute.ca/people/researchers/dgottesman/CO639-2004/}
\bibitem[12]{grasslroetteler} M. Grassl, M. R\"otteler, Thomas Beth,
  ``Efficient Quantum Circuits for Non-Qubit Quantum Error-Correcting Codes'',
  International Journal of Foundations of Computer Science (IJFCS), Vol. 14,
  No. 5, pp. 757-775, 2003, and quant-ph/0211014, 4 November 2002.
\bibitem[13]{hartshorne} R. Hartshorne, ``Algebraic Geometry'', Springer,
1977.
\bibitem[14]{knilafl} E. Knill, R. Laflamme, ``Theory of quantum
  error-correcting codes'', Physical Review A, vol. 55, pp. 900-911, 1997.
\bibitem[15]{laflamme} E. Knill, R. Laflamme, A. Ashikhmin, H. Barnum,
  L. Viola and W. H. Zurek, ``Introduction to Quantum Error Correction'',
  quant-ph/0207170, July 30, 2002.
\bibitem[16]{lang} S. Lang, ``Algebra'', Springer, 2002.
\bibitem[17]{lidl} R. Lidl, H. Niederreiter, ``Finite Fields'', Encyclopedia
  of Mathematics and its Applications, Addison-Wesley Publishing Company, 1983.
\bibitem[18]{macwil} F. MacWilliams, N. Sloane, ``The theory of
  error-correcting codes'', North-Holland Publishing Company, 1988.
\bibitem[19]{magma} The Magma Computational Algebra System for Algebra, Number
Theory and Geometry, V2.11-5, Sydney, 2004.
\bibitem[20]{matsu} R. Matsumoto, ``Improvement of the
  Ashikhmin-Litsyn-Tsfasman Bound for Quantum Codes'', IEEE Transactions on
  Information Theory Vol. 48 No. 7, pp. 2122-2124, July 2002, see also
  ''Algebraic geometric construction of a quantum stabilizer code'',
  quant-ph/0107129, August 8, 2001.
\bibitem[21]{moreno} C. Moreno, ``Algebraic Curves over Finite Fields'',
  Cambridge University Press, 1991.
\bibitem[22]{neukirch} J. Neukirch, ``Algebraische Zahlentheorie'', Springer,
1992.
\bibitem[23]{nielsen} M. Nielsen and I. Chuang, ``Quantum Computation and
  Quantum Information'', Cambridge University Press, 2000
\bibitem[24]{popp} H. Popp, ``Goppa Codes'', Talk at the
  Summer School ``Datensicherheit'' in Mannheim (Germany), August 2003.\\
  {\texttt http://hilbert.math.uni-mannheim.de/Datensicherheit/notes.html}
\bibitem[25]{shor} P. Shor, ``Scheme for reducing decoherence in quantum
  computer memory'', Phys. Rev. A, vol. 52, no. 4, pp. 2493-2496, Oct. 1995
\bibitem[26]{steanecss} A. M. Steane, ``Multiple particle interference
  and quantum error correction'', Proc. Roy. Soc. Lond. A, vol. 452,
  pp. 2551-2577, November 1996.
\bibitem[27]{steane} A. M. Steane, ``Enlargement of Calderbank Shore Steane
  quantum codes'', quant-ph/9802061, March 31, 1998.
\bibitem[28]{stepanov}S. A. Stepanov, ``Codes on Algebraic Curves'',
  Kluwer Academic/ Plenum Publishers, New York, 1999.
\bibitem[29]{s-selfdual}H. Stichtenoth, ``Self-dual Goppa
  Codes'', Journal of Pure and Applied Algebra, vol. 55, pp. 199-211, 1988.
\bibitem[30]{stichtenoth} H. Stichtenoth, ``Algebraic Function
  Fields and Codes'', Springer-Verlag, Berlin, 1993.
\bibitem[31]{sugiyama} K. Sugiyama, ``Algebraic Curves for Coding
  Theory'', Talk at the Summer School ``Datensicherheit'' in Mannheim
  (Germany), August 2003.\\
  {\texttt http://hilbert.math.uni-mannheim.de/Datensicherheit/notes.html}
\bibitem[32]{vardy} A. Vardy, ``Algorithmic complexity in coding theory and
  the minimum distance problem'', STOC '97, pp. 92-109, 1997
\bibitem[33]{xing} Chao-Ping Xing, ``Hyperelliptic function fields and
 codes'', Journal of Pure and Applied Algebra, vol. 74, pp. 109-118, 1991.
\end{thebibliography}
\end{document}